\documentclass[useAMS,usenatbib]{mn2e}
\usepackage{amsmath}
\usepackage{amssymb}
\usepackage{aas_macros}
\usepackage{graphicx}
\usepackage{epstopdf}
\usepackage{epsfig}
\usepackage{threeparttable}
\usepackage{appendix}
\usepackage{rotating}
\usepackage{enumerate}
\usepackage{tabularx}
\usepackage[T1]{fontenc}
\usepackage{aecompl}

\title[Fossil group origins -- VI]
{Fossil group origins -- VI. Global X-ray scaling relations of fossil galaxy clusters}

\author[A. Kundert et al.]
{A. Kundert$^{1}$\thanks{E-mail: kundert@astro.wisc.edu},
F. Gastaldello$^{2}$,
E. D'Onghia$^{1}$\thanks{Alfred P. Sloan Fellow},
M. Girardi$^{3,4}$,
J. A. L. Aguerri$^{5,6}$,
\newauthor
R. Barrena$^{5,6}$,
E. M. Corsini$^{7,8}$,
S. De Grandi$^{9}$,
E. Jim\'{e}nez-Bail\'{o}n$^{10}$,
\newauthor
M. Lozada-Mu\~{n}oz$^{10}$,
J. M\'{e}ndez-Abreu$^{11}$,
R. S\'{a}nchez-Janssen$^{12}$,
E. Wilcots$^{1}$,
\newauthor
and S. Zarattini$^{5,6,7}$\\
$^{1}$ Department of Astronomy, University of Wisconsin-Madison, 475 N. Charter St., Madison, WI 53706, USA \\
$^{2}$ INAF -- IASF Milano, via E. Bassini 15, I-20133 Milano, Italy \\
$^{3}$ Dipartimento di Fisica-Sezione Astronomia, Universit\`{a} degli Studi di Trieste, via Tiepolo 11, I-34143, Trieste, Italy \\
$^{4}$ INAF -- Osservatorio Astronomico di Trieste, via Tiepolo 11, I-34143, Trieste, Italy \\
$^{5}$ Instituto de Astrof\'{i}sica de Canarias, C/ V\'{i}a L\'{a}ctea s/n, E-38200 La Laguna, Tenerife, Spain \\
$^{6}$ Departamento de Astrof\'{i}sica, Universidad de La Laguna, E-38205 La Laguna, Tenerife, Spain \\
$^{7}$ Dipartimento di Fisica e Astronomia `G. Galilei', Universit\`{a} di Padova, vicolo dell'Osservatorio 3, I-35122 Padova, Italy \\
$^{8}$ INAF -- Osservatorio Astronomico di Padova, vicolo dell'Osservatorio 5, I-35122 Padova, Italy \\
$^{9}$ INAF -- Osservatorio Astronomico di Brera, via E. Bianchi 46, I-23807 Merate, Italy \\
$^{10}$ Instituto de Astronom\'{i}a, Universidad Nacional Aut\'{o}noma de M\'{e}xico, Apartado Postal 70-264, 04510-M\'{e}xico DF, Mexico \\
$^{11}$ School of Physics and Astronomy, University of St Andrews, North Haugh, St Andrews, KY16 9SS, UK \\
$^{12}$ NRC Herzberg Institute of Astrophysics, 5071 West Saanich Road, Victoria, BC, V9E 2E7, Canada
}

\date{Accepted 2015 August 12. Received 2015 July 15; in original form 2015 March 16}

\begin{document}

\pagerange{\pageref{firstpage}--\pageref{lastpage}} \pubyear{2015}

\maketitle

\label{firstpage}

\begin{abstract}

We present the first pointed X-ray observations of 10 candidate fossil galaxy groups and clusters. With these {\it Suzaku} observations, we determine global temperatures and bolometric X-ray luminosities of the intracluster medium (ICM) out to $r_{500}$ for six systems in our sample. The remaining four systems show signs of significant contamination from non-ICM sources. For the six objects with successfully determined $r_{500}$ properties, we measure global temperatures in the range $2.8 \leq T_{\mathrm{X}} \leq 5.3 \ \mathrm{keV}$, bolometric X-ray luminosities of $0.8 \times 10^{44} \ \leq L_{\mathrm{X,bol}} \leq 7.7\times 10^{44} \ \mathrm{erg} \ \mathrm{s}^{-1}$, and estimate masses, as derived from $T_{\mathrm{X}}$, of $M_{500} \ga 10^{14} \ \mathrm{M}_{\odot}$. Fossil cluster scaling relations are constructed for a sample that combines our {\it Suzaku} observed fossils with fossils in the literature. Using measurements of global X-ray luminosity, temperature, optical luminosity, and velocity dispersion, scaling relations for the fossil sample are then compared with a control sample of non-fossil systems. We find the fits of our fossil cluster scaling relations are consistent with the relations for normal groups and clusters, indicating fossil clusters have global ICM X-ray properties similar to those of comparable mass non-fossil systems.

\end{abstract}


\defcitealias{girardi2014}{G14}
\defcitealias{khosroshahi2007}{KPJ07}
\defcitealias{santos2007}{S07}
\defcitealias{zarattini2014}{Z14}


\section{Introduction}

Fossil galaxy systems are group and cluster mass objects characterized by extended, relaxed X-ray isophotes and an extreme magnitude gap in the bright end of the optical luminosity function of their member galaxies. Typically, fossils are identified with the criteria of a halo luminosity of $L_{\mathrm{X,bol}}\geq 0.5 \times 10^{42}$ erg s$^{-1}$ and a first ranked galaxy more than 2 $R$-band magnitudes brighter than the second brightest galaxy within half the virial radius \citep{jones2003}. Fossil systems comprise 8-20 per cent of groups and clusters in the same X-ray luminosity regime \citep{jones2003}, and thus determining the origin of the features characterizing these systems is important for understanding the nature and evolution of a significant fraction galaxy groups and clusters.

The features of fossil systems seem to fulfil theoretical predictions that the Milky Way luminosity ($L$*) galaxies in a group will merge into a central bright elliptical in less than a Hubble time, but the time-scale for the cooling and collapse of the hot gas halo is longer \citep{barnes1989,ponman1993}. Indeed the first fossil group discovered, RX J1340.6+4018 \citep{ponman1994}, appeared as a solitary bright elliptical located in the centre of a group-sized X-ray luminous halo. It was thought the central galaxy of this group was the final merger remnant of the former group galaxies, and hence this object was named a `fossil group'. Since then, deeper observations have found this system to consist of galaxies other than the bright central galaxy \citep[BCG;][]{jones2000} and as a result the magnitude gap criterion of fossils has been established. The motivation for this criterion is that over time, an increasingly growing difference between the two brightest galaxies will form as a result of the merging of the most massive galaxies into a single bright central elliptical if no infall occurs. This formation scenario is well suited for group mass fossils where the velocity dispersion is low and the dynamical friction time-scale is short.

A number of objects meeting the fossil criteria have also been observed in the cluster mass regime as well \citep{cypriano2006,khosroshahi2006,voevodkin2010,aguerri2011,harrison2012}. It is possible fossil clusters may form as the result of two systems merging, where one group has had its bright galaxies merge due to dynamical friction, and the other has comparatively fainter galaxies \citep{harrison2012}. Should merging occur between systems with similarly bright galaxies, any previously existing magnitude gaps may become filled in. Therefore, meeting the fossil criteria may only be a transitory phase in the evolution of a group or cluster \citep{vonbendabeckmann2008, dariush2010}.

Numerical and hydrodynamic simulations indicate the large magnitude gaps characterizing fossil groups and clusters are associated with an early formation time: fossil systems have been found to assemble more of their total dynamical mass than non-fossil systems at every redshift \citep{dariush2007}, where half the dynamical mass is assembled by $z \ga 1$ \citep{donghia2005}. Evidence that fossils have formed and evolved in a different manner than normal groups and clusters should then manifest in differences in their respective properties.

The bright central galaxy which dominates the optical output of fossil systems has a number of unique characteristics, although whether this demonstrates a clearly distinct formation scenario from non-fossil BCGs is still uncertain. The BCGs of fossils are more massive in both the stellar component and in total than the central ellipticals in non-fossil systems of the same halo mass \citep{harrison2012}. \citet{mendezabreu2012} find fossil BCGs are consistent with the Fundamental Plane of non-fossil BCGs, but show lower velocity dispersions and higher effective radii when compared to non-fossil intermediate-mass elliptical BCGs of the same $K_\mathrm{s}$-band luminosity. These results suggest the fossil BCG has experienced a merger history of early gas-rich dissipational mergers, followed by gas-poor dissipationless mergers later.

On the global scale, the scaling relations of fossil systems remain a point of contention due to limited data and inhomogeneities between studies. \citet[hereafter KPJ07]{khosroshahi2007} performed a comprehensive analysis of a sample of group mass fossil systems and found their sample fell on the same $L_{\mathrm{X}}$--$T_{\mathrm{X}}$ relation as non-fossils. However, the fossil groups were found to have offset $L_{\mathrm{X}}$ and $T_{\mathrm{X}}$ for a given optical luminosity $L_{\mathrm{opt}}$ or velocity dispersion $\sigma_v$ when compared to normal groups, which was interpreted as an excess in the X-ray properties of fossil systems for their mass. In a comparable study, \citet{proctor2011} found similar deviations between fossils and non-fossils. This offset, however, was interpreted as fossils being underluminous in the optical which is supported by their large mass-to-light ratios. These features would not result from galaxy-galaxy merging in systems with normal luminosity functions, and thus this analysis calls into question the formation scenario commonly attributed to generating the characteristic large magnitude gap of fossil systems. Later studies, such as \citet{harrison2012} and \citet[hereafter G14]{girardi2014}, find no difference in the $L_{\mathrm{X}}$--$L_{\mathrm{opt}}$ relation of fossil systems and non-fossils. Even so, most recently \citet{khosroshahi2014} present a sample of groups, one of which qualifies as a fossil, that lies above the $L_{\mathrm{X}}$--$L_{\mathrm{opt}}$ relation of non-fossil systems, reopening the debate on fossil system scaling relations.

In this paper we have undertaken an X-ray study of 10 candidate fossil systems, never previously studied with detailed pointed observations in the X-ray regime. Using {\it Suzaku} data, we present the first measurements of intracluster medium (ICM) temperatures, bolometric X-ray luminosities, and estimates of the $M_{500}$ masses of our systems. This work comprises the sixth instalment of the FOssil Group Origins (FOGO) series. The FOGO project is a multiwavelength study of the \citet{santos2007} candidate fossil system catalogue. In FOGO I \citep{aguerri2011}, the FOGO project is described in detail and the specific goals of the collaboration are outlined. FOGO II \citep{mendezabreu2012} presents a study of the BCG scaling relations of fossil systems and the implications for the BCG merger history. Global optical luminosities of our FOGO sample are measured in FOGO III \citepalias{girardi2014} and used to construct the global $L_{\mathrm{X}}$--$L_{\mathrm{opt}}$ relation which reveals no difference between the fossil and non-fossil fits. Deep $r$-band observations and an extensive spectroscopic database were used to redetermine the magnitude gaps of the FOGO sample and reclassify our fossil candidate catalogue in FOGO IV \citep[hereafter Z14]{zarattini2014}. In FOGO V \citep{zarattini2015}, the correlation of the size of the magnitude gap and the shape of the luminosity function is investigated. In this work (FOGO VI) we advance the characterization of the X-ray properties of fossil systems and constrain the global scaling relations of these objects.

The details and observations of our {\it Suzaku} sample are described in Sections~\ref{sec:sample} and~\ref{sec:observations}. A discussion on how non-ICM sources may contribute to the observed emission of our systems follows in Section~\ref{sec:nonicm}. Tests to determine the contribution of these non-ICM sources are presented in Sections~\ref{sec:imageanalysis} and \ref{sec:spectralanalysis}. Measurements of the global ICM properties of the thermally dominated subset of our sample are recorded in Section~\ref{sec:globalprop}. Global scaling relations and their implications are presented in Section~\ref{sec:scalrel}. For our analysis, we assume a $\Lambda$CDM cosmology with a Hubble parameter $H_0$=70 km s$^{-1}$ Mpc$^{-1}$, a dark energy density parameter of $\Omega_{\Lambda}$=0.7, and a matter density parameter $\Omega_{\mathrm{M}}$=0.3.


\section{The Sample}\label{sec:sample}

Our sample of 10 observed galaxy groups and clusters was selected from the \citet[hereafter S07]{santos2007} catalogue of candidate fossil systems. The \citetalias{santos2007} catalogue was assembled by first identifying luminous $r<$19 mag red galaxies in the luminous red galaxy (LRG) catalogue \citep{eisenstein2001}, and selecting only those galaxies associated with extended X-ray emission in the \textit{ROSAT} All-Sky Survey (RASS). Sloan Digital Sky Survey (SDSS) Data Release 5 was then used to spatially identify companion galaxies to these bright galaxies. Group or cluster membership was assigned to galaxies identified within a radius of 0.5 $h_{70}^{-1}$ Mpc from one of the bright LRGs and with a redshift consistent with that of the LRG. While spectroscopic redshifts were used when available, galaxy membership was primarily determined using photometric redshifts. Groups and clusters with more than a 2 $r$-band magnitude difference between the brightest and second brightest member galaxies within the fixed 0.5 $h_{70}^{-1}$ Mpc system radius were then selected, and those with an early-type BCG were identified as fossils.

\citetalias{zarattini2014} observed the \citetalias{santos2007} fossil candidate list with the Nordic Optical Telescope, the Isaac Newton Telescope, and the Telescopio Nazionale Galileo to obtain deeper $r$-band images and spectroscopic redshifts for candidate group members allowing for improved system membership. Additionally, the search radius for galaxy system members was extended to the virial radius of the system as calculated from the RASS X-ray luminosity. The \citetalias{zarattini2014} study confirms 15 targets out of 34 \citetalias{santos2007} candidates are fossil galaxy systems. According to this characterization, our sample contains five confirmed fossil systems and five non-confirmed or rejected fossil systems (see Table~\ref{table:obslog}).


\section{Observations and Data Reduction}\label{sec:observations}

\begin{table*}
\centering
\caption{Summary of observations.}
\begin{threeparttable}
\begin{tabular}{c c c c c c c}
\hline
Object &Sequence number &RA &Dec. &Start date & Exposure [ks] &Type$^{\dagger}$ \\ [0.5ex]
\hline
FGS03		&807052010	&07:52:44.2	&+45:56:57.4	&2012 Oct 28 18:39:14	&14.3		&F \\
FGS04		&807053010	&08:07:30.8	&+34:00:41.6	&2012 May 06 16:24:20	&10.1		&NC \\
FGS09		&807050010	&10:43:02.6	&+00:54:18.3	&2012 May 30 05:18:38	&\phantom{0}9.9			&NC \\
FGS14		&807055010	&11:46:47.6	&+09:52:28.2	&2012 May 29 17:06:08	&12.4		&F \\
FGS15		&807057010	&11:48:03.8	&+56:54:25.6	&2012 May 26 17:58:41	&13.6		&NF \\
FGS24		&807058010	&15:33:44.1	&+03:36:57.5	&2012 Jul 28 08:10:10	&13.2		&NF \\
FGS25		&807049010	&15:39:50.8	&+30:43:04.0	&2012 Jul 28 18:06:02	&10.6		&NF \\
FGS26		&807054010	&15:48:55.9	&+08:50:44.4	&2012 Jul 29 02:05:54	&\phantom{0}8.6		&F \\
FGS27		&807056010	&16:14:31.1	&+26:43:50.4	&2012 Aug 05 07:14:36	&10.6		&F \\
FGS30		&807051010	&17:18:11.9	&+56:39:56.1	&2012 May 02 11:43:31	&14.0		&F \\
\hline
\end{tabular}
\begin{tablenotes}
\item [$\dagger$] The fossil status column contains the \citetalias{zarattini2014} updated fossil characterizations of the \citetalias{santos2007} catalogue. In the fossil status column, `F' is a confirmed fossil, `NF' is a rejected fossil, and `NC' is not confirmed as either a fossil or non-fossil according to \citetalias{zarattini2014} and remains a fossil candidate.
\end{tablenotes}
\label{table:obslog}
\end{threeparttable}
\end{table*}

\begin{figure*}
\begin{minipage}{.315\textwidth}
\centering
	\includegraphics[width=\linewidth]{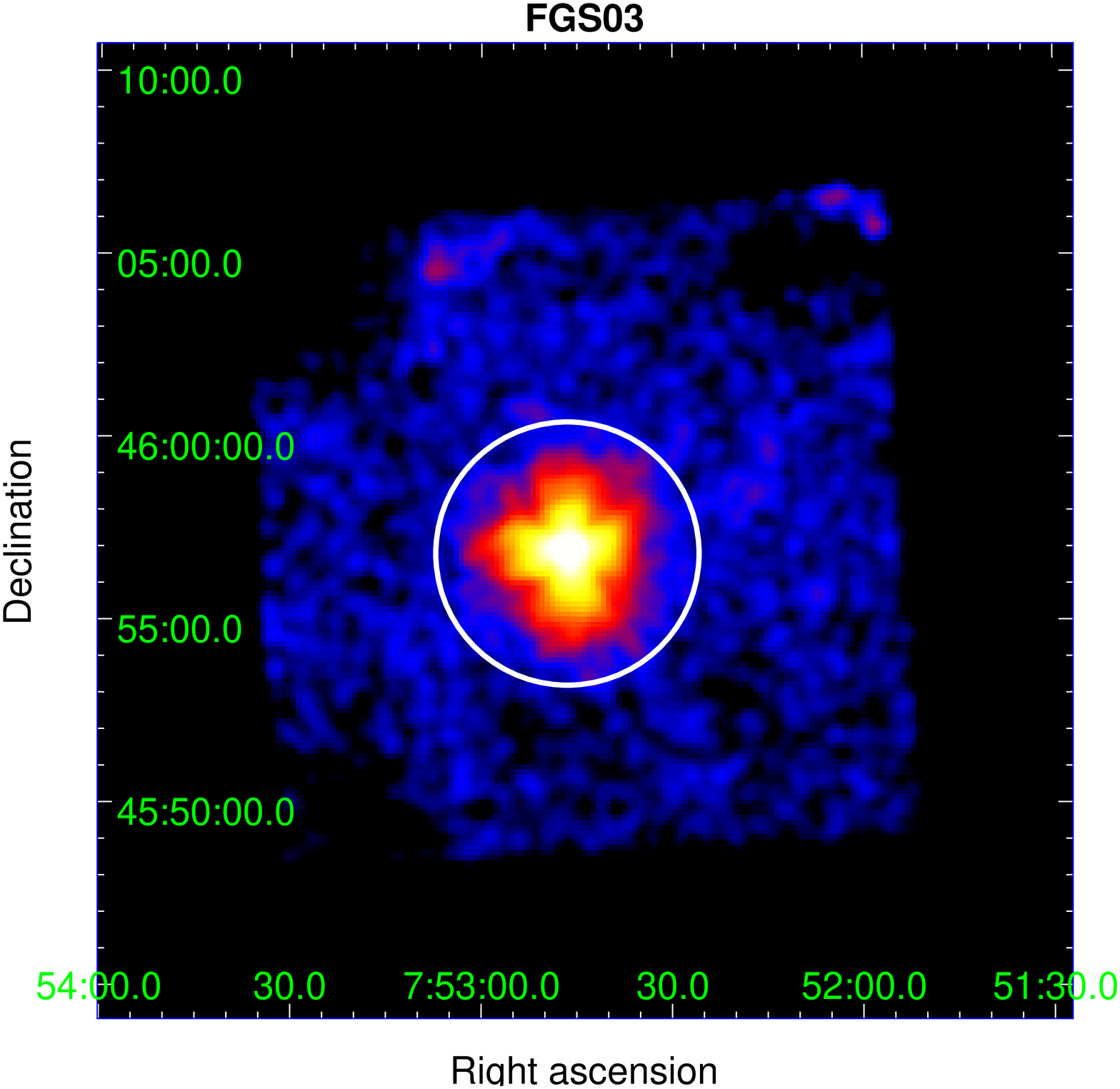}
\end{minipage}
\begin{minipage}{.315\textwidth}
\centering
	\includegraphics[width=\linewidth]{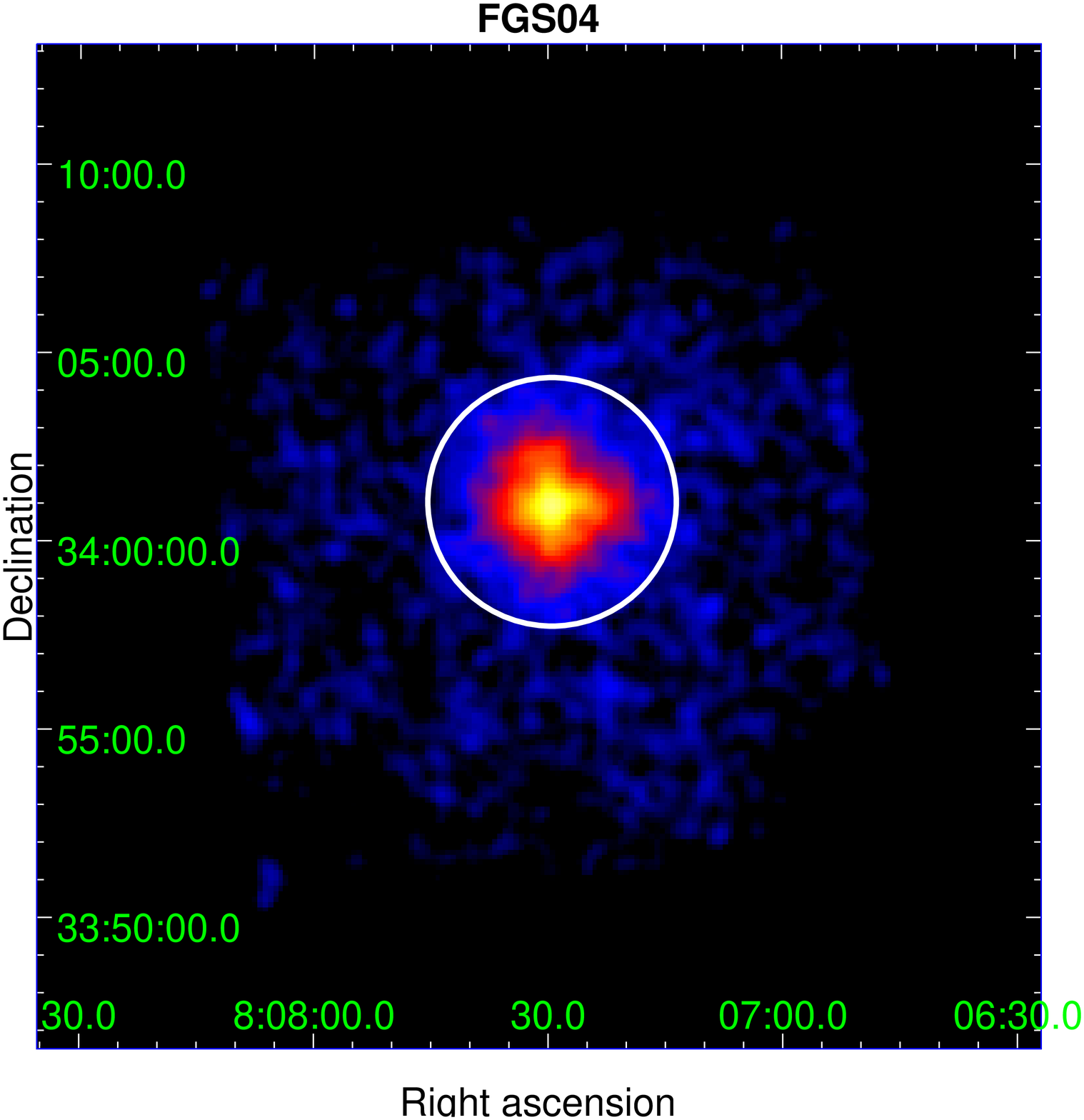}
\end{minipage}
\begin{minipage}{.315\textwidth}
\centering
	\includegraphics[width=\linewidth]{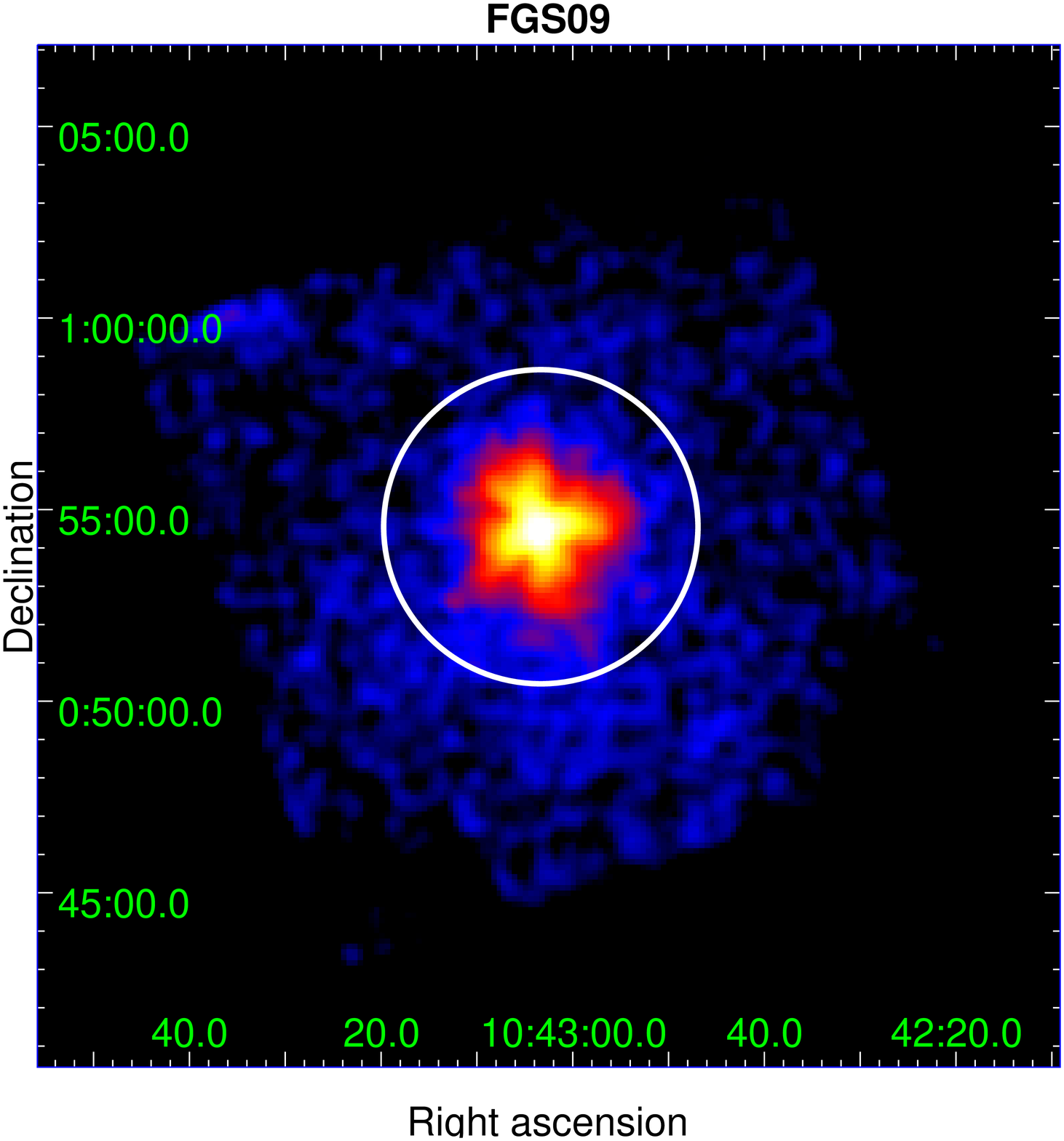}
\end{minipage}
\begin{minipage}{.315\textwidth}
\centering
	\includegraphics[width=\linewidth]{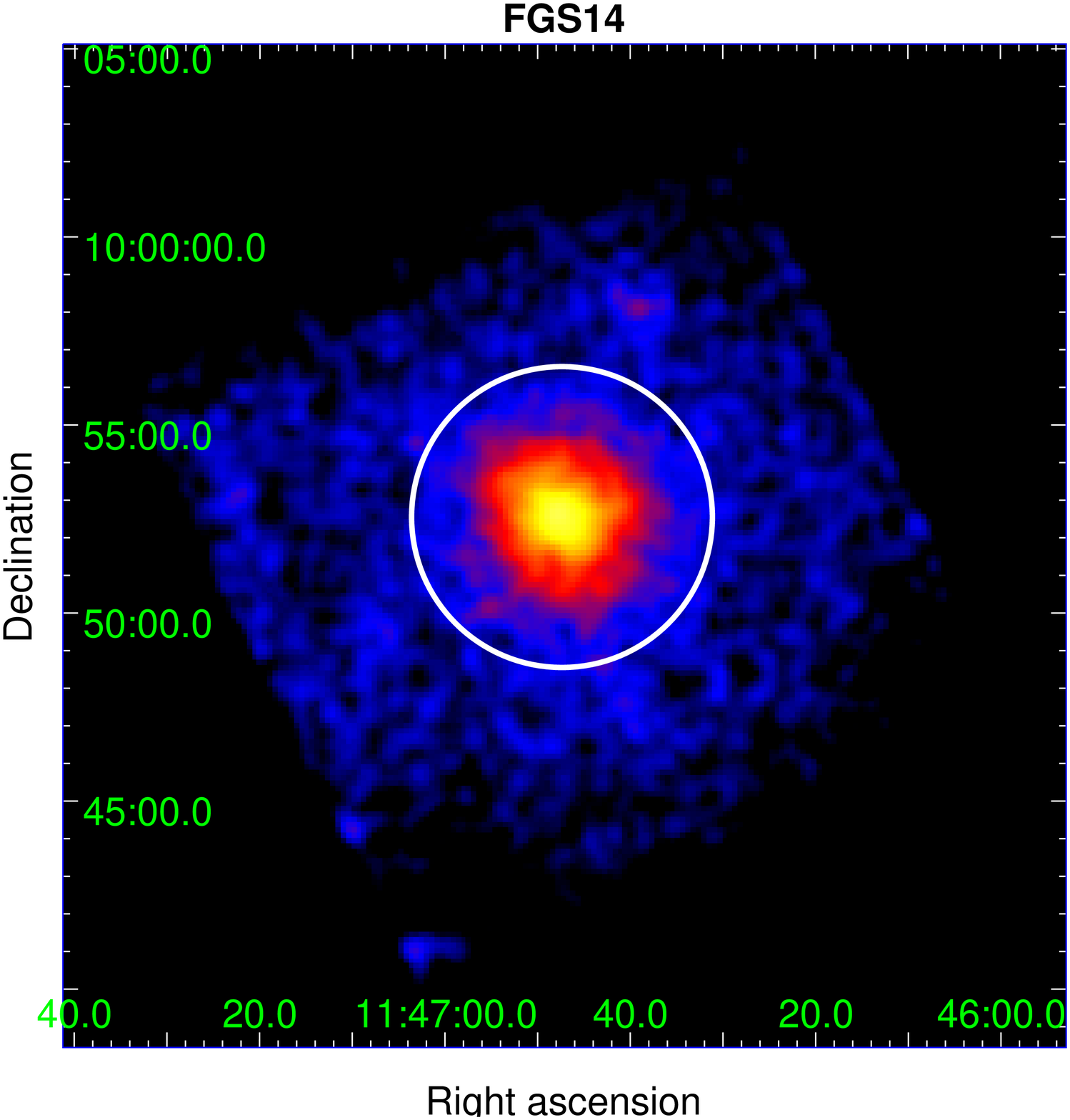}
\end{minipage}
\begin{minipage}{.315\textwidth}
\centering
	\includegraphics[width=\linewidth]{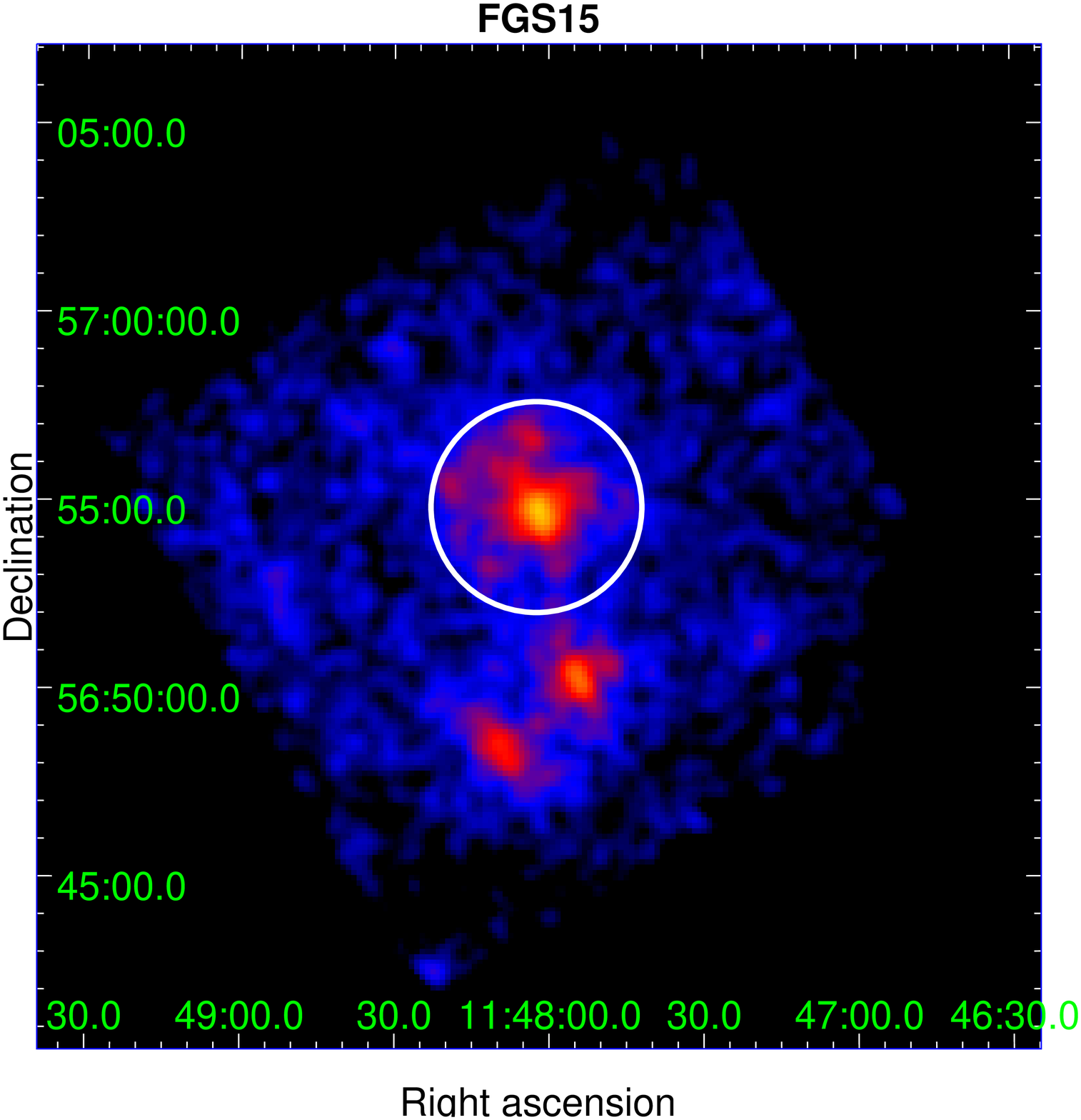}
\end{minipage}
\begin{minipage}{.315\textwidth}
\centering
	\includegraphics[width=\linewidth]{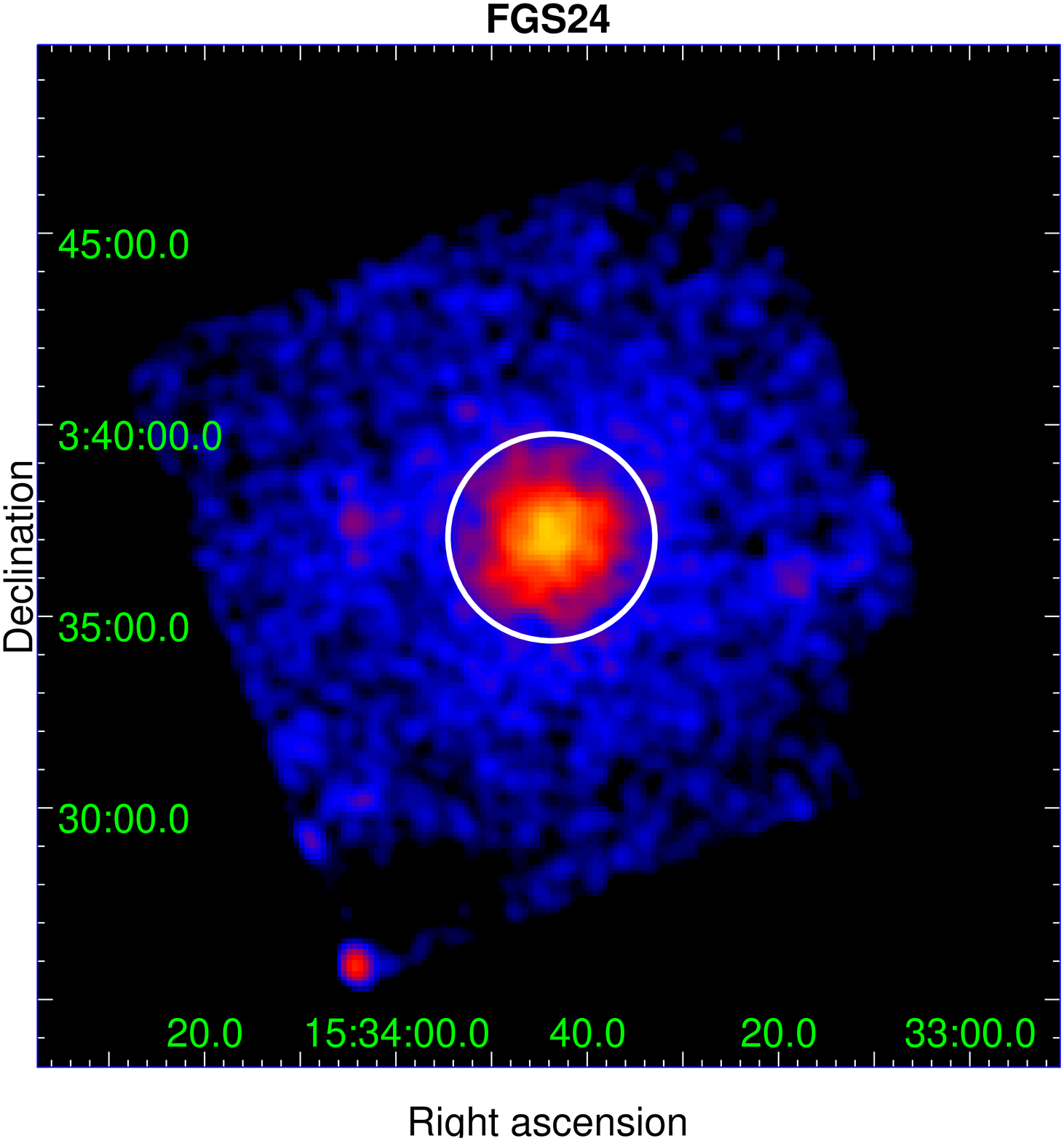}
\end{minipage}
\begin{minipage}{.315\textwidth}
\centering
	\includegraphics[width=\linewidth]{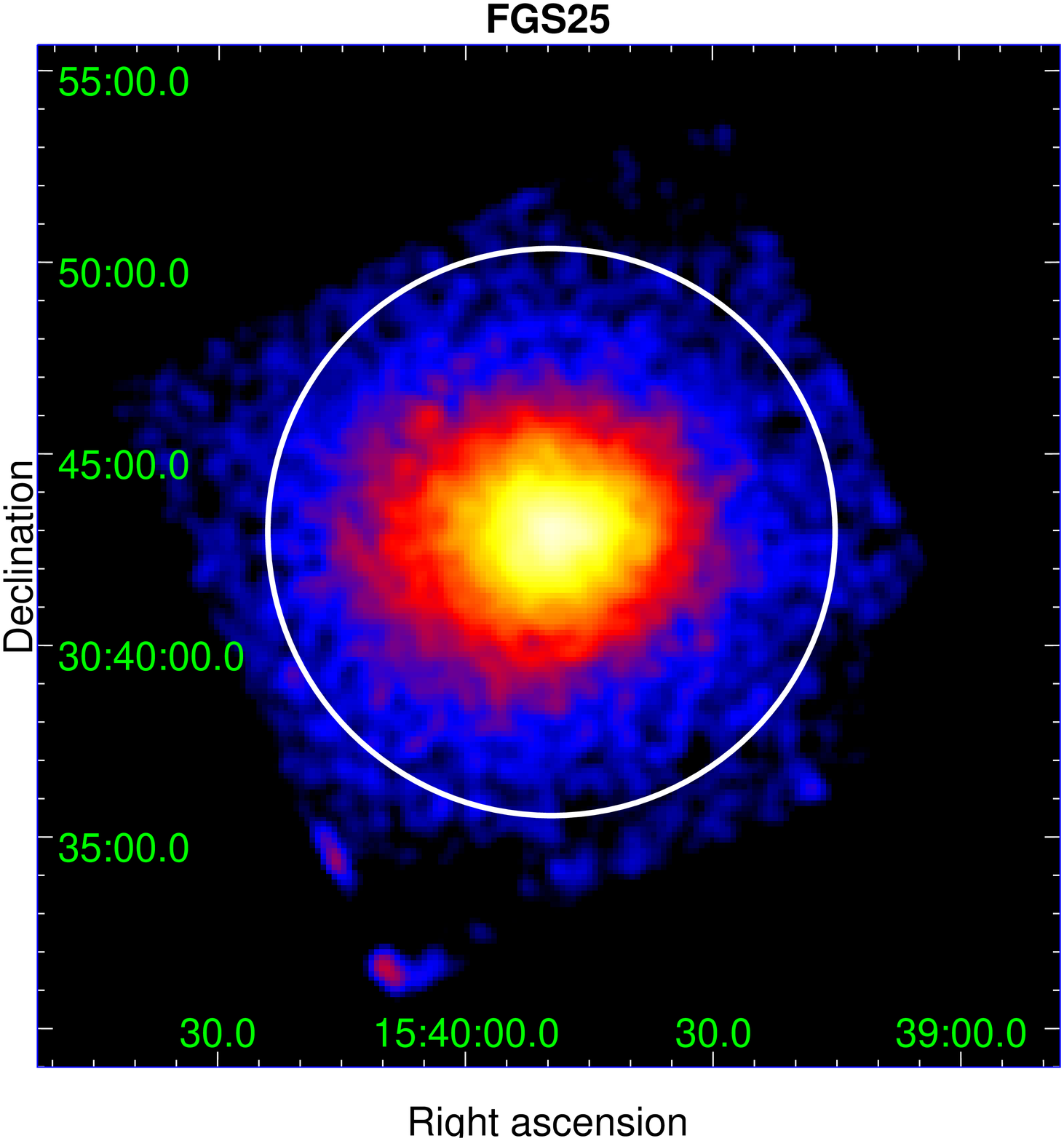}
\end{minipage}
\begin{minipage}{.315\textwidth}
\centering
	\includegraphics[width=\linewidth]{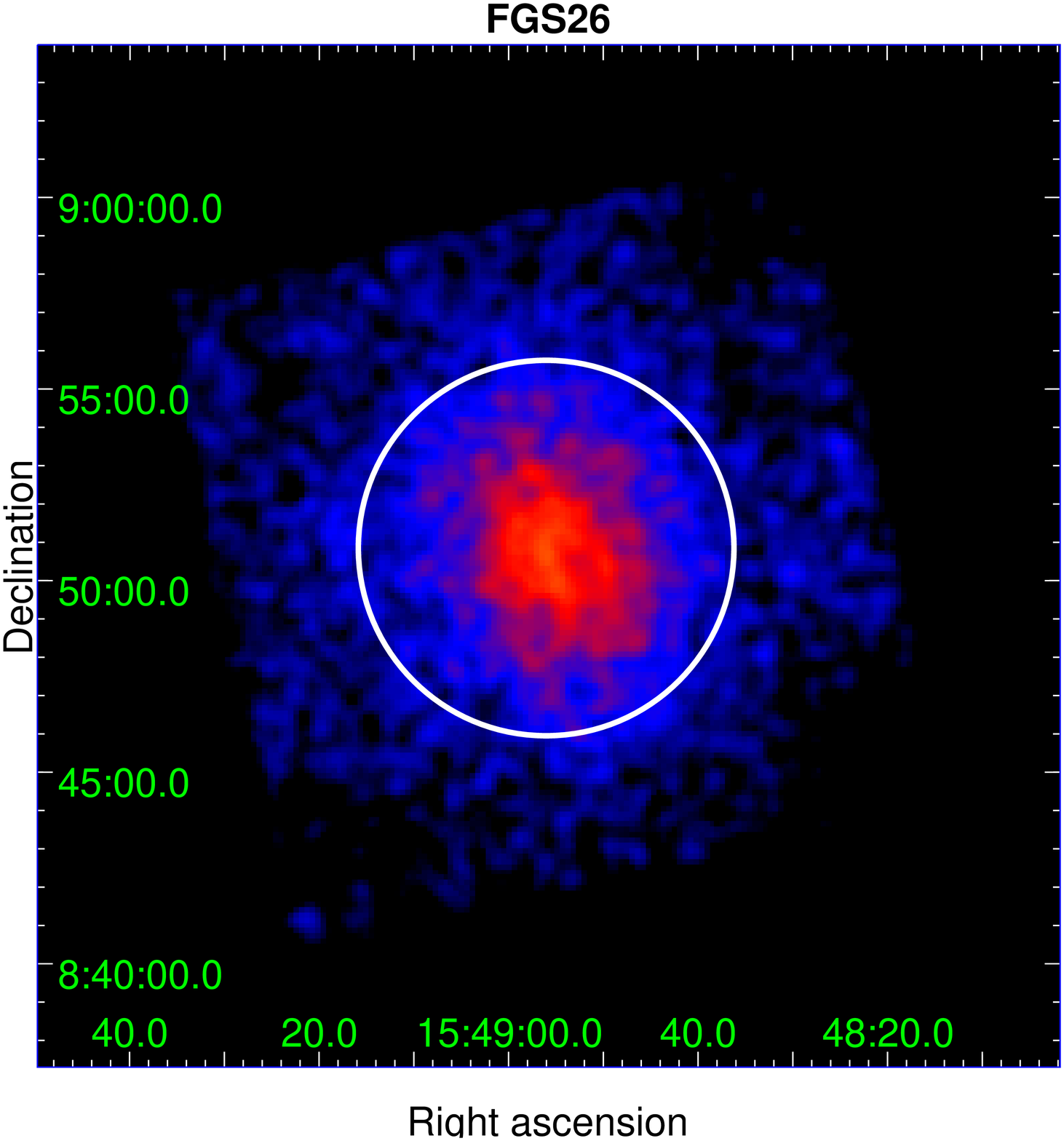}
\end{minipage}
\begin{minipage}{0.315\textwidth}
\centering
	\includegraphics[width=\linewidth]{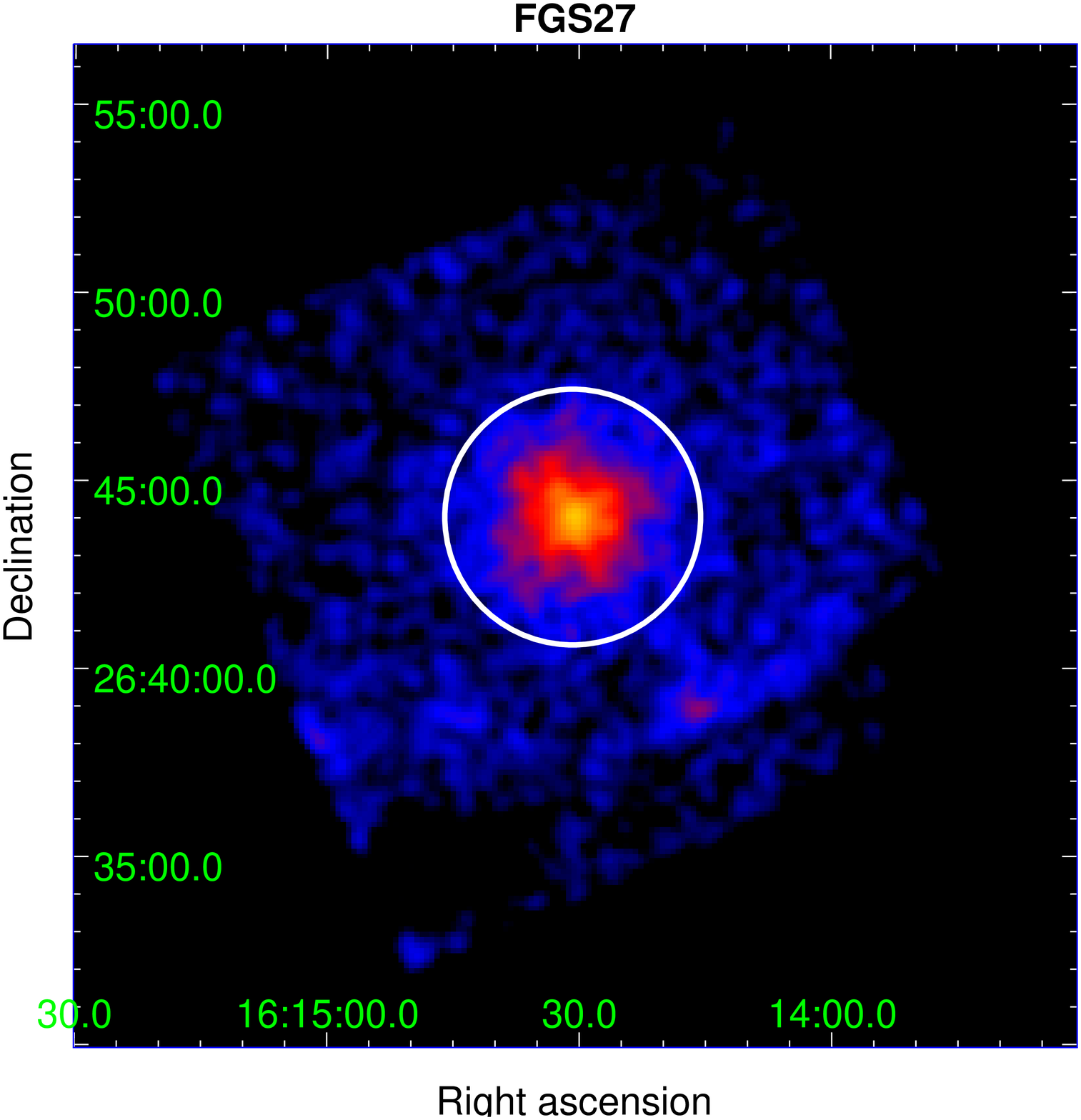}
\end{minipage}
\begin{minipage}{0.315\textwidth}
\centering
	\includegraphics[width=\linewidth]{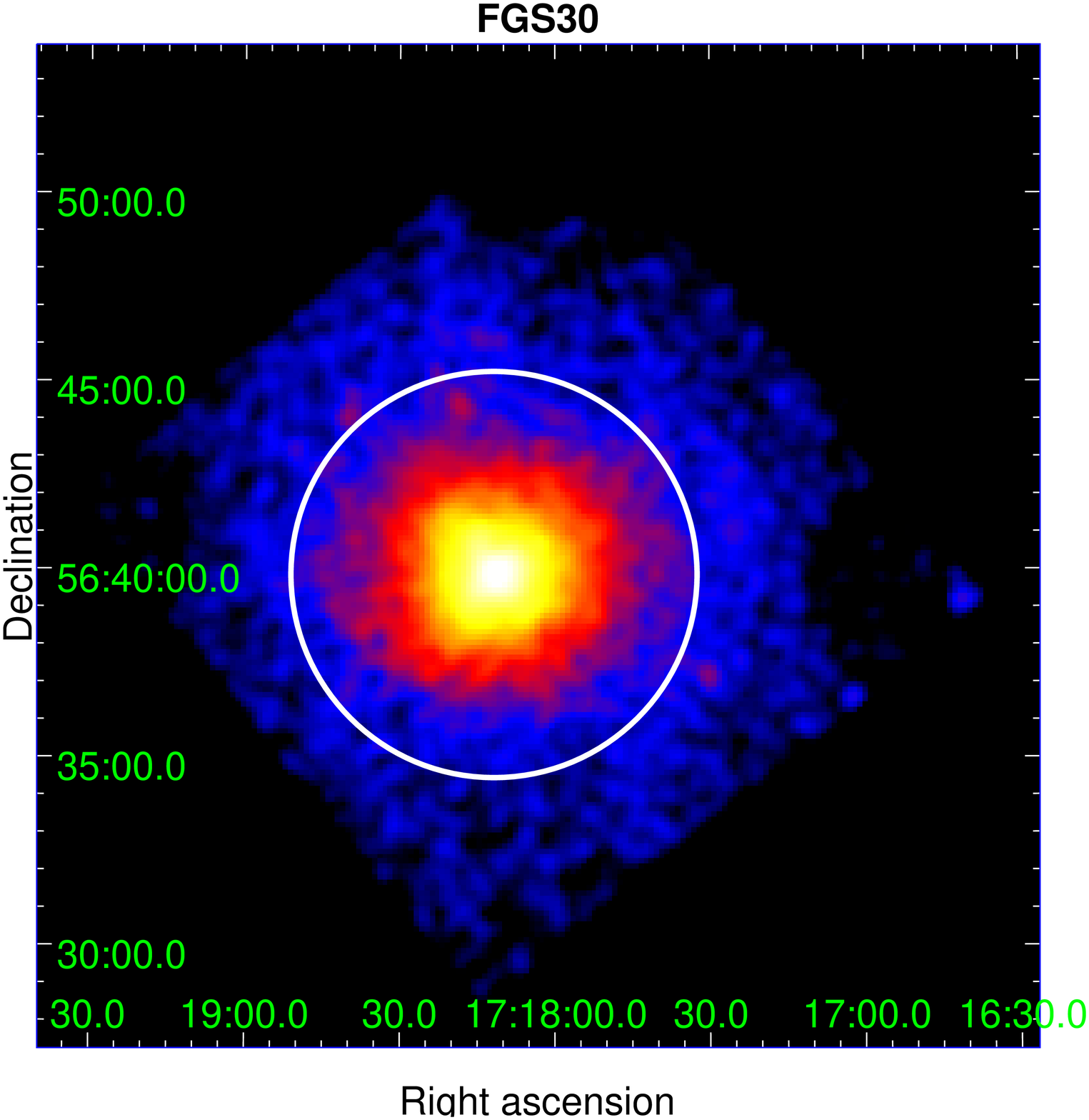}
\end{minipage}
\begin{minipage}{0.63\textwidth}
\caption{The {\it Suzaku} combined raw counts XIS0+XIS1+XIS3 images in the 0.5--10 keV band. The image is Gaussian smoothed with $\sigma$ = 0.42 arcmin. White circles demarcate the initial spectral extraction region $r_{\mathrm{ap,src}}$ defined to encircle the source-dominated region ($r_{\mathrm{ap,src}}$ values in Table~\ref{table:bfp}). $^{55}$Fe calibration source events have been removed.}\label{fig:x013rawim}
\end{minipage}
\end{figure*}

The 10 systems in our sample were observed with the {\it Suzaku} X-ray telescope between 2012 May and October (Table~\ref{table:obslog}). Our analysis uses the data from {\it Suzaku}'s three X-ray Imaging Spectrometers (XIS) sensitive to the 0.5--10 keV band. Our single-pointing observations were taken with a normal clocking mode, and an editing mode of 3$\times$3 or 5$\times$5 which were combined when both were available. The stacked XIS0+XIS1+XIS3 raw count images of the sample are shown in Fig.~\ref{fig:x013rawim}.

The analysis of our study was conducted using the {\sc HEASOFT} version 6.15 software library with the calibration database {\sc CALDB} XIS update version 20140520. Spectra were extracted using {\sc XSELECT} version 2.4c and fit using {\sc XSPEC} version 12.8.1g. The event files were reprocessed using {\tt aepipeline} with the {\sc CALDB} XIS update 20140203 using the default settings with an additional criterion of COR$>$6. In our spectral analysis, emission from the $^{55}$Fe calibration sources, located in the corners of each XIS detector, was removed. Additionally, the XIS0 damaged pixel columns caused by micro-meteorites were masked.

A Redistribution Matrix File (RMF) was created for all spectral extraction regions with {\tt xisrmfgen}. For each RMF, two Ancillary Response Files (ARFs) were created with {\tt xissimarfgen}, one to be convolved with the background spectral model, and the other to be convolved with the source model following the method of \citet{ishisaki2007}. Background ARFs were created out to a radius of 20 arcmin using a uniform emission source mode. For the source ARFs, an image of the stacked XIS field-of-view (FOV) was used to model the emission.

\section{Treatment of non-ICM emission}\label{sec:nonicm}

High fidelity measurements of the ICM temperature and luminosity require careful consideration of non-ICM sources of emission during our analysis.

\subsection{Background and foreground sources}\label{sec:bkg}

The standard {\it Suzaku} XIS background consists of a non-X-ray particle background \citep[NXB;][]{tawa2008}, the cosmic X-ray background \citep[CXB;][]{fabian1992}, and foreground Galactic emission from the Local Hot Bubble (LHB) and the Milky Way Halo \citep[MWH;][]{kuntz2000}.

The contribution of the NXB for each object was assessed using the night earth database within 150 days of the observation using the FTOOL {\tt xisnxbgen} \citep{tawa2008}. Our XIS1 observations were taken in a charge injection mode of CI = 6 keV which increases the NXB. Accordingly, the {\tt nxbsci6 } calibration file was used as input for XIS1 to counteract this.

The contribution of the galactic foreground to a XIS spectrum is well described by two thermal plasma models: {\sc apec}$_{\mathrm{LHB}}$+({\sc wabs} $\times$ {\sc apec}$_{\mathrm{MWH}}$) where $z_{\mathrm{LHB}} = z_{\mathrm{MWH}} = 0$, $Z_{\mathrm{LHB}} = Z_{\mathrm{MWH}}=1 \ Z_{\odot}$, and $kT_{\mathrm{LHB}}$ = 0.1 keV \citep{kuntz2000}. The CXB was modelled by an absorbed power-law: {\sc wabs} $\times$ {\sc powerlaw}$_{\mathrm{CXB}}$ with $\Gamma=1.412$ \citep{kushino2002}. During spectral analyses, the summed background and foreground model: {\sc apec}$_{\mathrm{LHB}}$+{\sc wabs}({\sc apec}$_{\mathrm{MWH}}$+{\sc powerlaw}$_{\mathrm{CXB}})$ was convolved with the uniform emission ARF.


\subsection{Solar wind charge exchange}\label{sec:swcx}

The interaction of ions in the solar wind with neutral atoms in the heliosphere and in Earth's atmosphere can produce $E<1$ keV photons in the X-ray regime \citep{cravens2000,fujimoto2007}. To check for contamination from solar wind charge exchange (SWCX), proton flux light curves with a sampling frequency of 90 s were obtained from the NASA WIND-SWE database over the time span of each observation. The intensity of proton flux has been found to be related to the strength of geocoronal SWCX contaminating photons, where flux levels above $4\times10^8 \ \mathrm{protons} \ \mathrm{cm}^{-2} \ \mathrm{s}^{-1}$ commonly indicate potentially significant contamination to X-ray spectra from charge exchange \citep{yoshino2009}. Following \citet{fujimoto2007}, 2700 s were added to the time points in the WIND-SWE light curve to account for the travel time between the WIND satellite, located at the L1 point, and Earth, where the geocoronal SWCX emission is produced.

Much of the FGS24 observation occurs during an elevated period of proton flux; however, the light curve of FGS24 displays no significant duration flares. Furthermore, as a check, we have performed our spectral analysis on the time windows where the proton flux was less than $4\times10^8 \mathrm{cm}^{-2} \mathrm{s}^{-1}$ and found the results were consistent with the spectral analysis of the full baseline. We therefore consider the effects of SWCX to be small and have recorded the results of the analysis of the full observation in the main text and include the FGS24 light curve and shortened exposure time analysis in Appendix~\ref{appendix:swcxappendix}.


\subsection{Point source contamination}\label{sec:pointsrc}

Our {\it Suzaku} observations are the first pointed X-ray observations of the objects in our sample. Consequently we must assess point source contamination primarily relying on the {\it Suzaku} data alone. Visual inspection of the XIS images (Fig.~\ref{fig:x013rawim}) reveal two obvious point sources in the FGS15 FOV which we are able to exclude in our analysis using circular regions of radius 2.5 arcmin. Additionally, FGS03 and FGS09 show diffraction spikes from a strong point-like sources near the peak of the X-ray emission. However, the large 2 arcmin half-power diameter (HPD) of the {\it Suzaku} X-ray Telescope \citep[XRT;][]{serlemitsos2007} inhibits the exclusion of these sources and the robust identification of other point sources.

Optical and radio studies of the objects in our sample have found a number of active galactic nuclei (AGN) in spatial proximity to our galaxy systems. Especially concerning are the radio-loud AGN, located near the projected location of the BCGs, found in 7 out of the 10 objects in our sample \citep{hess2012}. To determine if these radio-loud AGN, and other optical and radio AGN in the FOV, are significant contributors to the source emission in the X-ray regime, we perform image (Section~\ref{sec:imageanalysis}) and spectral (Section~\ref{sec:spectralanalysis}) analyses. In the 0.5--10 keV range of the XIS, the strength of AGN emission increases towards the harder energies of the spectrum. As a result, the harder photons from an AGN may falsely boost the measured temperature of the ICM if only a thermal model is used to fit the spectrum. Assessing AGN contribution is therefore a crucial step in determining the properties of the ICM.


\subsection{Implementation of RASS data}

Because most of our objects extend over the entire single {\it Suzaku} pointing, a local {\it Suzaku} background region is not consistently available to assess the background contamination in our source regions. To aid in constraining the LHB, MWH, and CXB, we employ RASS background spectra sensitive to the 0.1--2.4 keV X-ray regime. RASS spectra were obtained through the High Energy Astrophysics Science Archive Research Center (HEASARC) X-ray background tool \footnote{ http://heasarc.gsfc.nasa.gov/cgi-bin/Tools/xraybg/xraybg.pl} in an annulus of inner radius 0.5 degrees and outer radius 1 degree centred on each of our sources. The size of this annulus is sufficient to minimize contamination from the source itself where the largest $r_{500}$ radius found for an object in our sample only extends to $\sim$20 per cent of the inner radius of the annular RASS background region.


\section{Image Analysis}\label{sec:imageanalysis}

\subsection{Determination of the source aperture for the spectral analysis region}\label{sec:rsrc}

The region of our initial spectral analysis for each object was established to encircle where the emission from the source dominates the emission from the background, enabling the parameters describing the source spectrum to be determined in a high signal-to-noise ratio (S/N) region. We determine this source region using vignetting and exposure corrected images of the source as well as simulated images of the background estimated from RASS spectra.

For each {\it Suzaku} pointing, an exposure map was created with {\tt xisexpmapgen} and a flat-field using {\tt xissim}. The flat-field was simulated over the XIS 0.5--10 keV energy range at a monochromatic photon energy of 1 keV for a uniform sky out to 20 arcmin.

An image of the NXB particle background for each pointing was produced with {\tt xisnxbgen} over the same energy range. This image was estimated from night Earth observations within 150 days of the {\it Suzaku} observation date. The NXB image was uniformly corrected by dividing the count rates by the exposure time.

Emission from the CXB, LHB, and MWH was estimated from RASS background spectra. These spectra were fit with the background model: {\sc apec}$_{\mathrm{LHB}}$+{\sc wabs}({\sc apec}$_{\mathrm{MWH}}$+{\sc powerlaw}$_{\mathrm{CXB}})$. Because the RASS background spectrum consists of only 7 data points, only the normalizations of the three background components were allowed to vary; the other parameters were fixed at the standard literature values as described in Section~\ref{sec:bkg}. The \textit{ROSAT} PSPC response matrix provided by the background tool was implemented for the fit. In calculating the background photon flux in the {\it Suzaku} XIS 0.5--10 keV energy range, the {\sc XSPEC} {\tt dummyrsp} command was used to extrapolate beyond the \textit{ROSAT} PSPC sensitivity range of 0.1--2.4 keV.

An image of the estimated CXB+LHB+MWH emission was produced with {\tt xissim} out to a radius of 20 arcmin from the coordinates of the X-ray centre of the systems. The emission was modeled with the best-fitting spectral model and photon flux of the RASS background data. Because of the low count rate of CXB+LHB+MWH photons over the exposure time for each object, the exposure time was increased by a factor of 10, and corrected later, to improve the statistics of the surface brightness profile of the resulting image following the method of \citet{kawaharada2010}.

An image of the source could then be created from the images constructed during this procedure. Because the NXB background is not affected by vignetting, the exposure corrected image of the NXB was subtracted from the exposure corrected image of the XIS detector. The resulting image was then vignetting corrected with the flat-field and the vignetting and exposure corrected image of the CXB+LHB+MWH was subtracted to obtain the estimated vignetting corrected image of source emission.

\begin{figure}
\centering
\includegraphics[width=\linewidth]{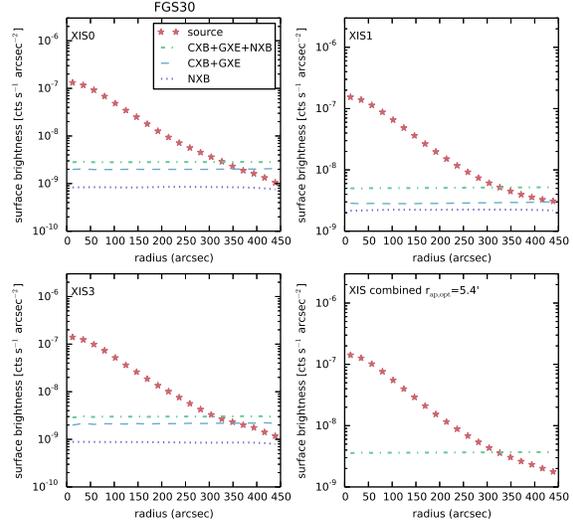}
\caption{An example of the estimated source and background surface brightness profiles for FGS30. The bottom right-hand panel shows the average source and background profile for the three XIS detectors.}
\label{fig:rsrcsbp}
\end{figure}

Surface brightness profiles were created using ds9 for the vignetting corrected source, NXB, and CXB+LHB+MWH images as shown for example in Fig.~\ref{fig:rsrcsbp}. The coordinates of peak X-ray emission (Table~\ref{table:geninfo}) were used as the centre of the surface brightness profile. The profile was constructed from 20 uniformly spaced circular annuli out to the radius of the largest circle that could be inscribed within the XIS FOV from the centre coordinates. The source and combined background profiles were then averaged for the three XIS detectors and the radius at which the source and background emission are equal was identified. We find that within this radius the source contributes on average $\sim$80 per cent of the total counts, with no less than a $\sim$70 per cent source contribution for all objects in our sample. It is this radius, the source radius $r_{\mathrm{ap,src}}$, which we have used to define our region of initial source spectral analysis.

\begin{table}
\begin{center}
\caption{General information.}
\begin{threeparttable}
\begin{tabular}{c c c c c}
\hline
FGS$^{a}$			&\multicolumn{2}{c}{Coordinates of Peak X-ray}$^{b}$		&$z^{c}$		&$n_{\mathrm{H}}$$^{d}$	\\	\cline{2-3}
			&RA		&Dec.		&		&$[10^{20} \mathrm{cm}^{-2}]$ \\ [0.5ex]
\hline
03$^*$				&07:52:46.48	&+45:56:48.40		&0.052		&5.06		\\
04$^{\phantom{*}}$		&08:07:29.47	&+34:01:02.95		&0.208		&4.27		\\
09$^{\phantom{*}}$		&10:43:03.33	&+00:54:33.26		&0.125		&3.88		\\
14$^*$				&11:46:47.37	&+09:52:33.38 		&0.221		&2.89		\\
15$^{\phantom{*}}$		&11:48:02.43	&+56:54:49.57		&0.105		&0.998		\\
24$^{\phantom{*}}$		&15:33:43.74	&+03:37:03.74		&0.293		&3.65		\\ 	
25$^{\phantom{*}}$		&15:39:49.57 	&+30:42:58.40 		&0.097		&2.29		\\
26$^*$				&15:48:56.03	&+08:50:51.27		&0.072		&3.14		\\	
27$^*$				&16:14:30.77	&+26:44:02.18		&0.184		&3.61		\\
30$^*$				&17:18:11.79	&+56:39:51.33		&0.114		&2.21		\\
\hline
\end{tabular}
\begin{tablenotes}
\item [a] [SMS2007] ID
\item [b] Coordinates determined from the stacked XIS0+XIS1+XIS3 raw count image in the 0.5--10 keV band
\item [c] Spectroscopic redshift of the central bright galaxy in the fossil cluster \citepalias{santos2007}
\item [d] Weighted average galactic hydrogen column density in the direction of the target \citep{kalberla2005}
\item [*] Confirmed fossil system
\end{tablenotes}
\label{table:geninfo}
\end{threeparttable}
\end{center}
\end{table}

\subsection{Surface brightness analysis}\label{sec:sbp}

\begin{figure*}
\begin{minipage}{.3\textwidth}
\centering
	\includegraphics[width=\linewidth]{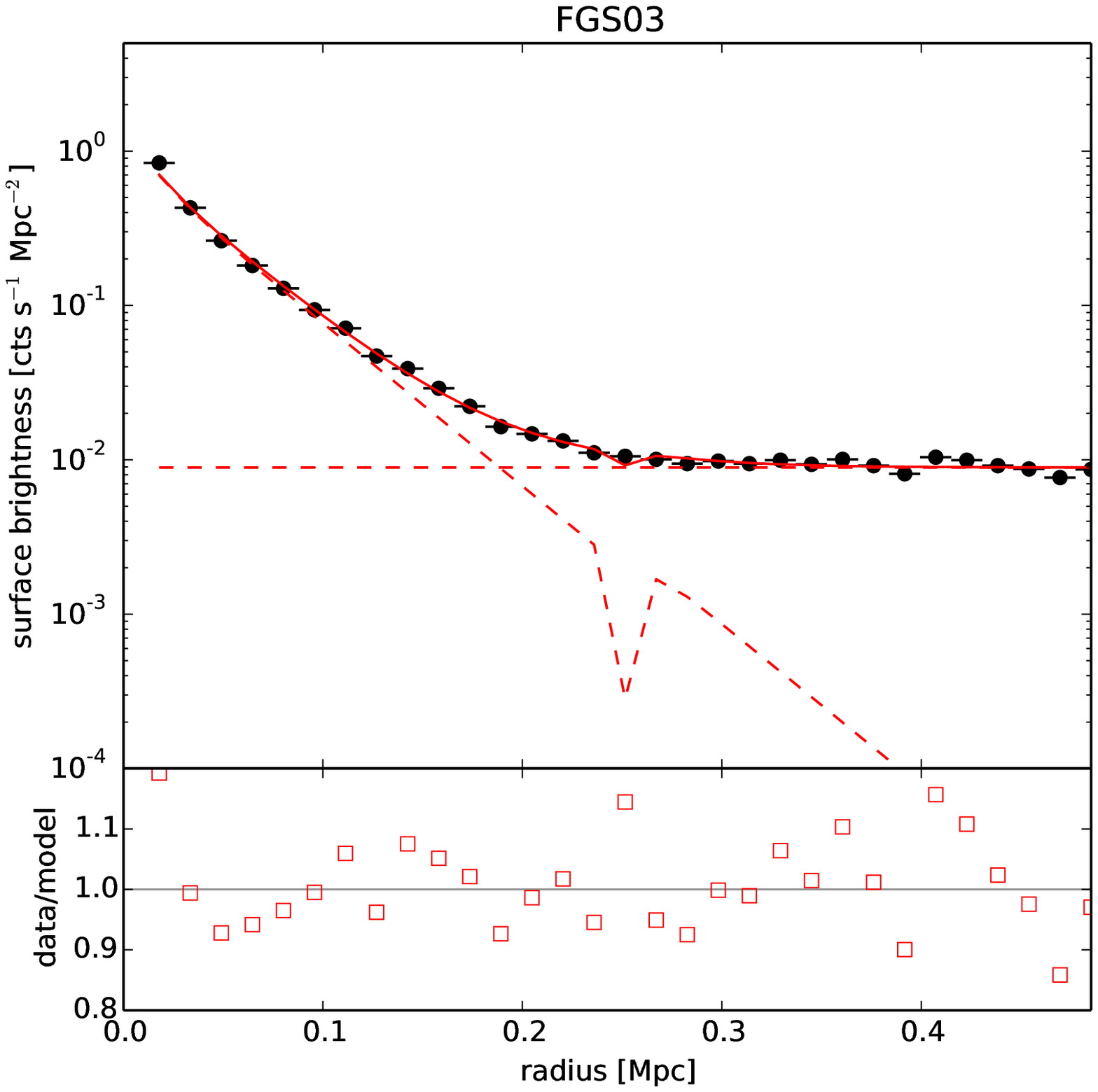}
\end{minipage}
\begin{minipage}{.3\textwidth}
\centering
	\includegraphics[width=\linewidth]{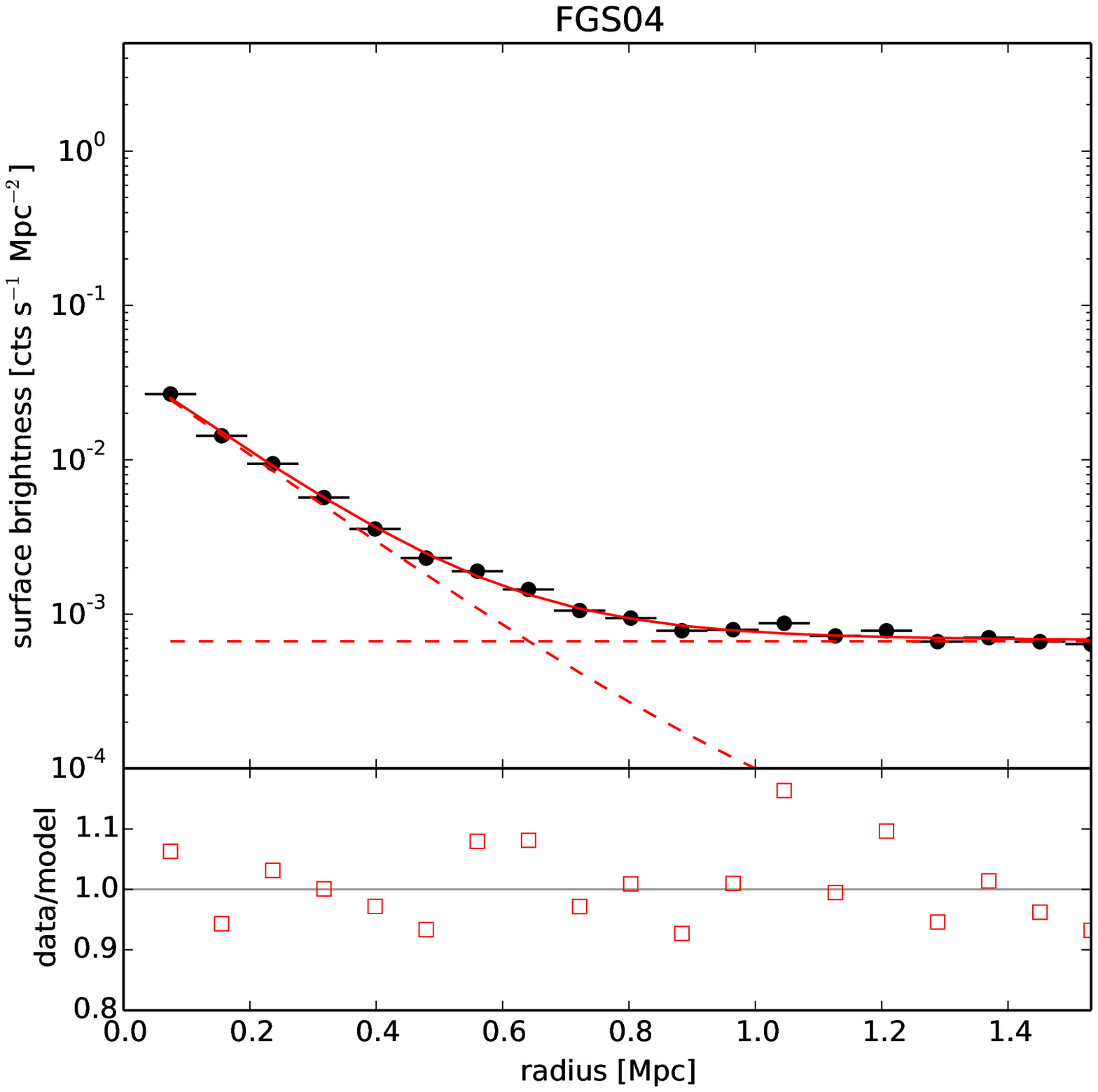}
\end{minipage}
\begin{minipage}{.3\textwidth}
\centering
	\includegraphics[width=\linewidth]{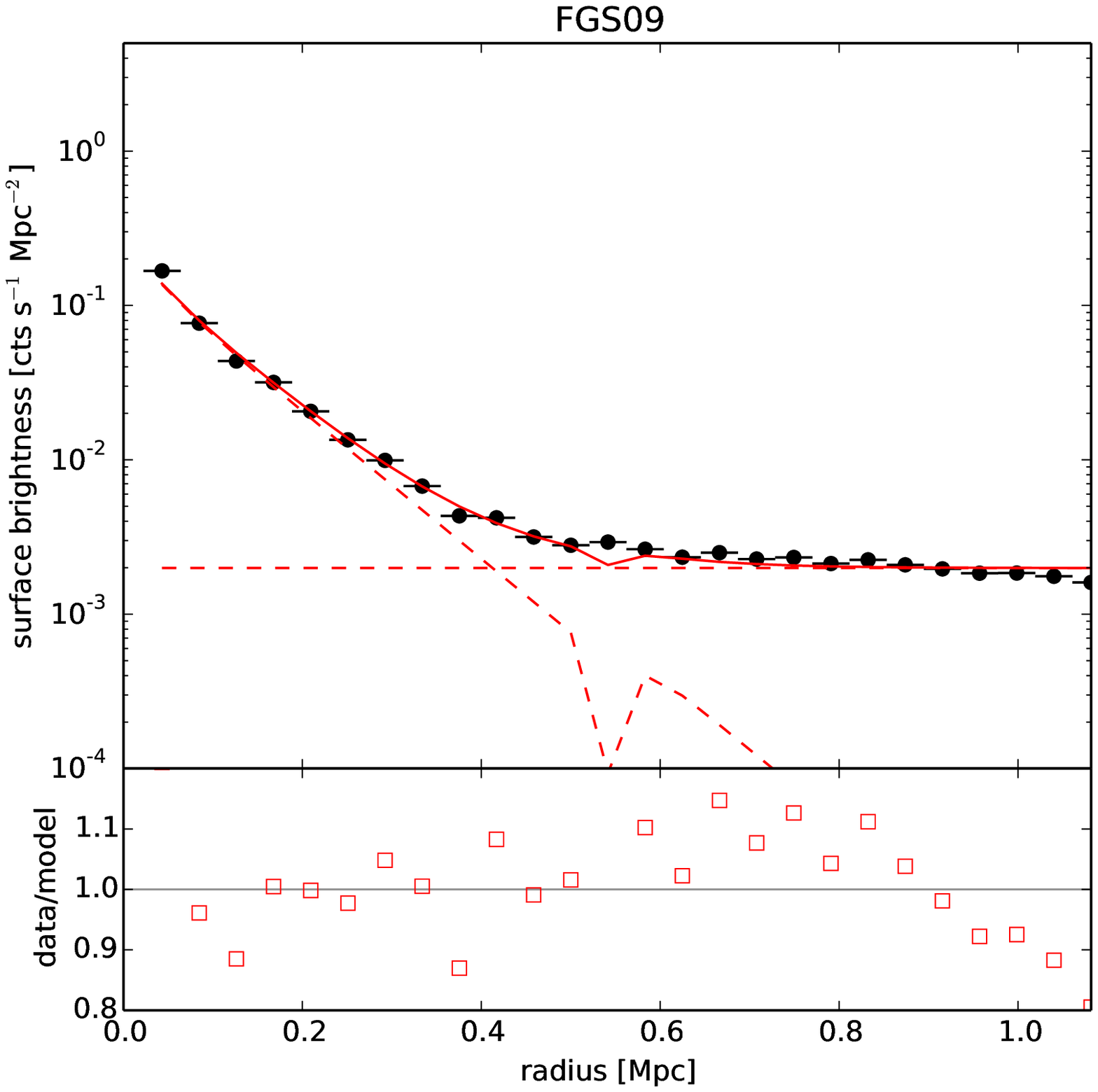}
\end{minipage}
\begin{minipage}{.3\textwidth}
\centering
	\includegraphics[width=\linewidth]{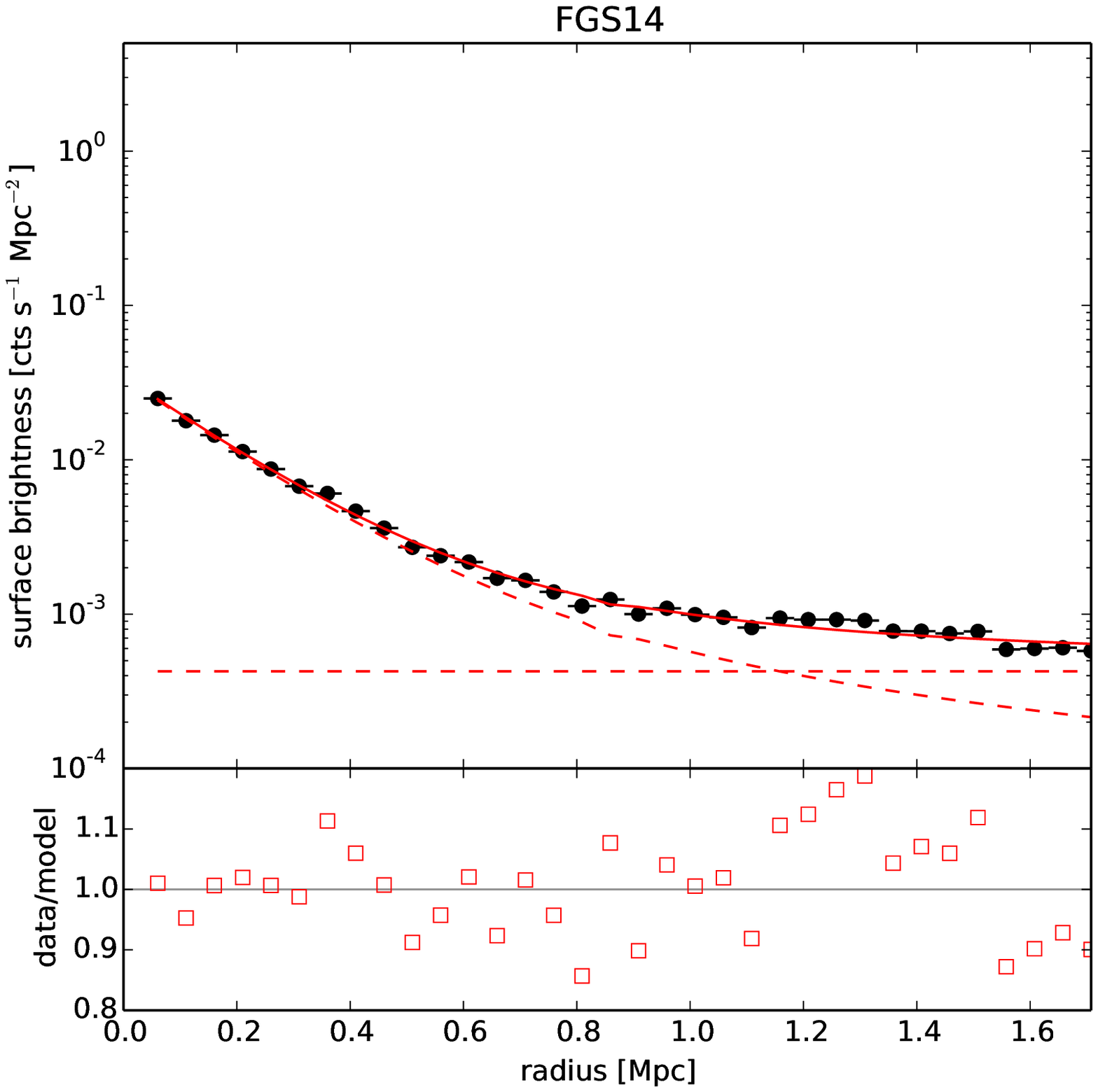}
\end{minipage}
\begin{minipage}{.3\textwidth}
\centering
	\includegraphics[width=\linewidth]{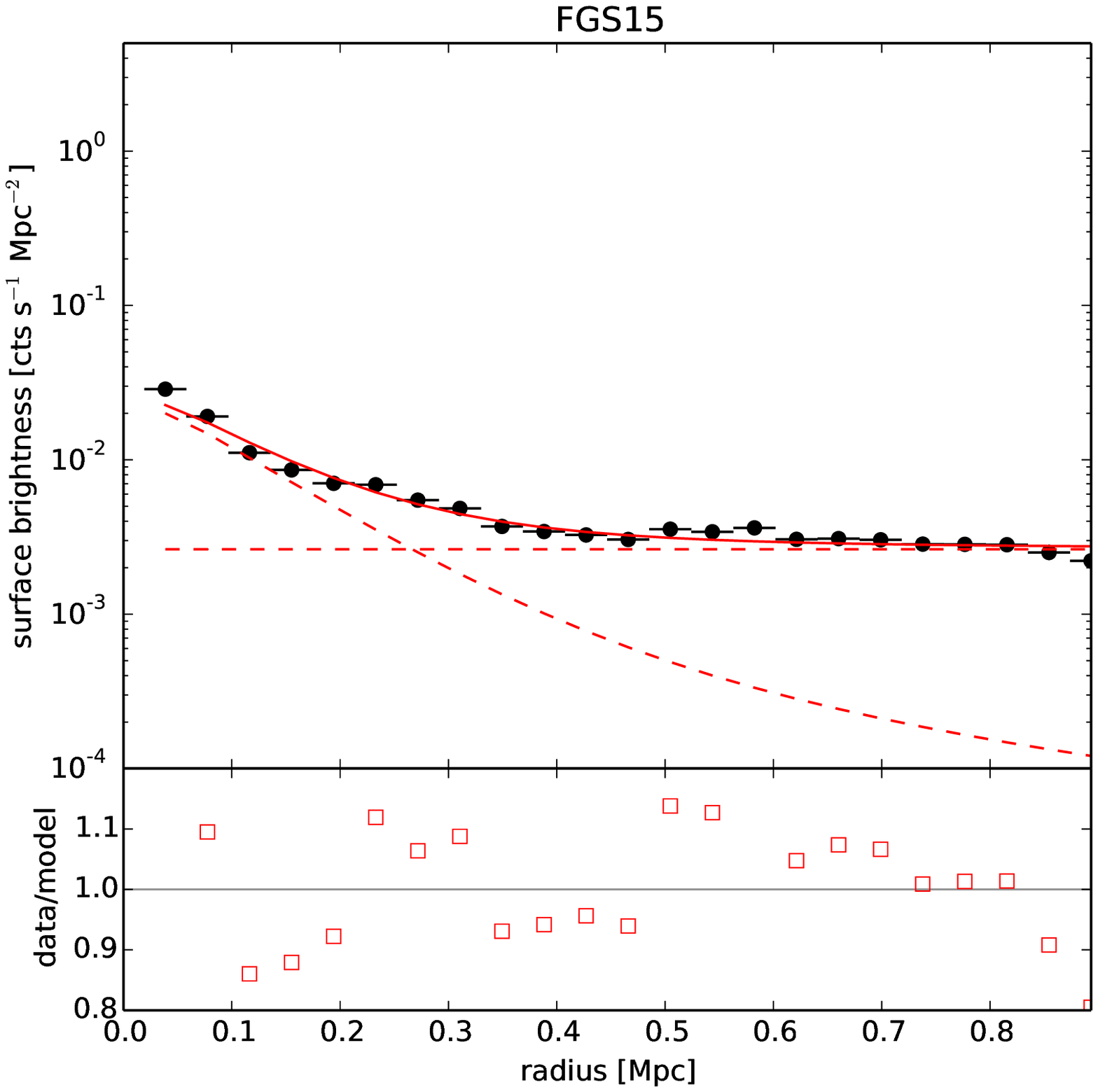}
\end{minipage}
\begin{minipage}{.3\textwidth}
\centering
	\includegraphics[width=\linewidth]{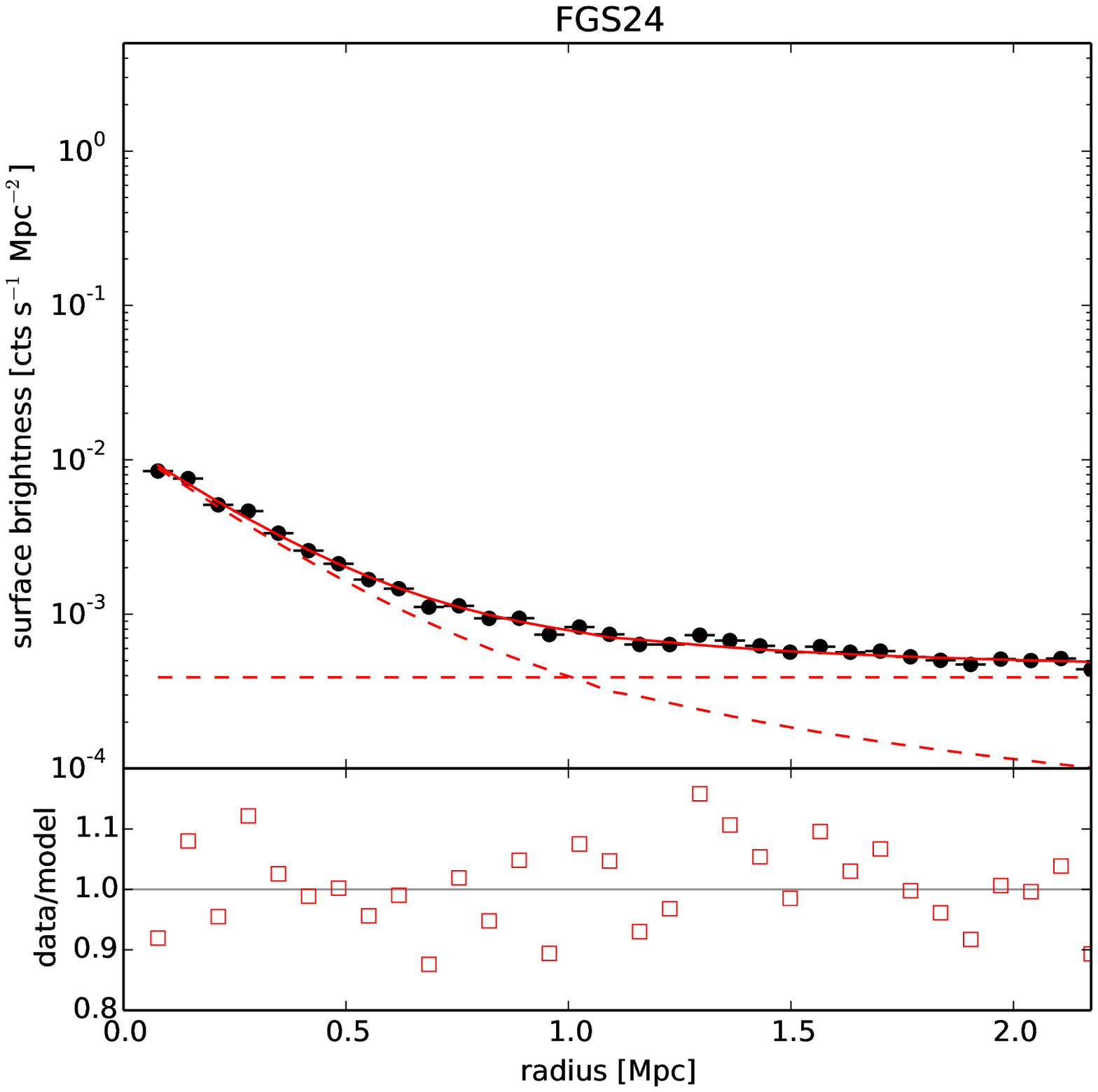}
\end{minipage}
\begin{minipage}{.3\textwidth}
\centering
	\includegraphics[width=\linewidth]{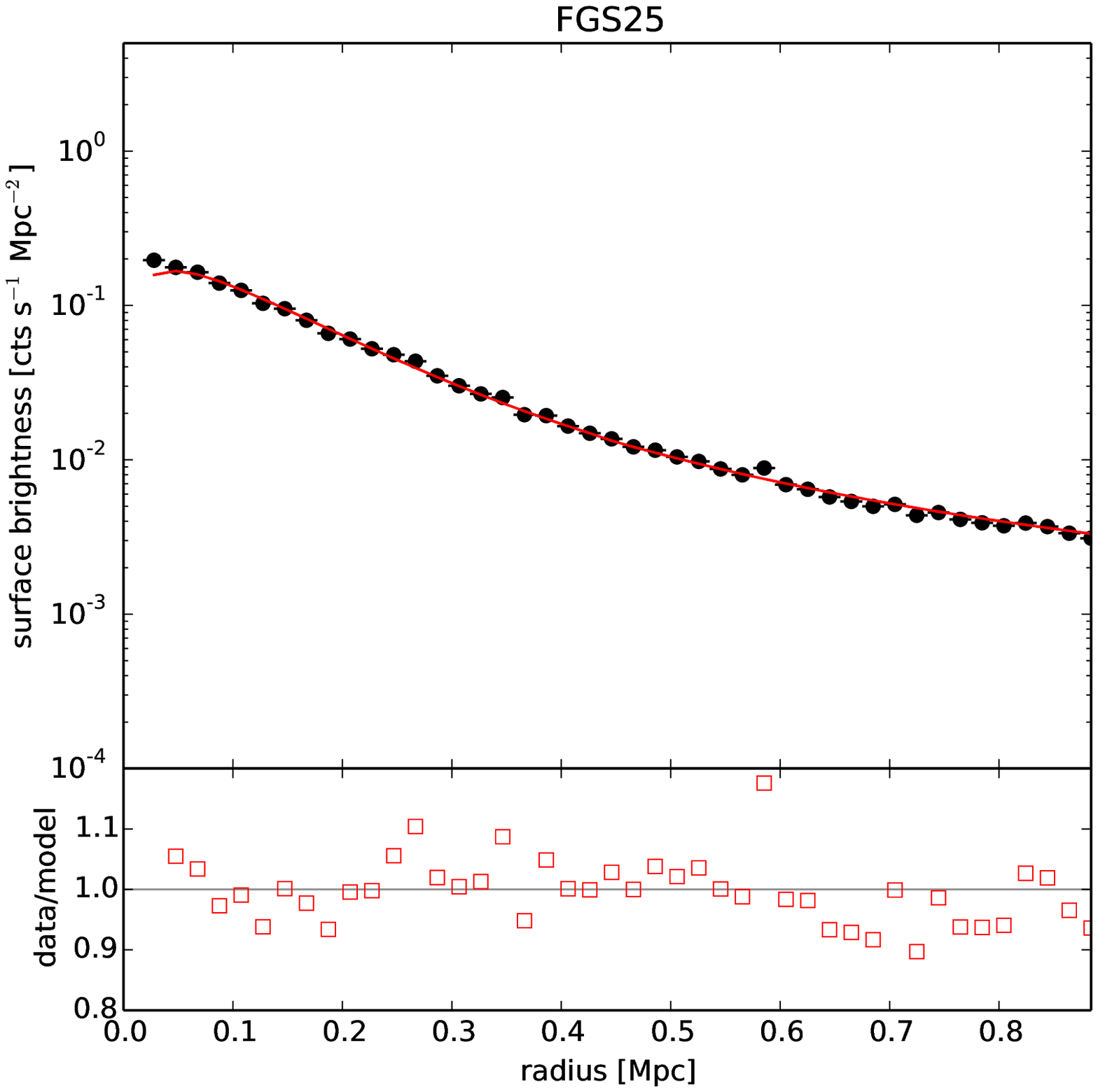}
\end{minipage}
\begin{minipage}{.3\textwidth}
\centering
	\includegraphics[width=\linewidth]{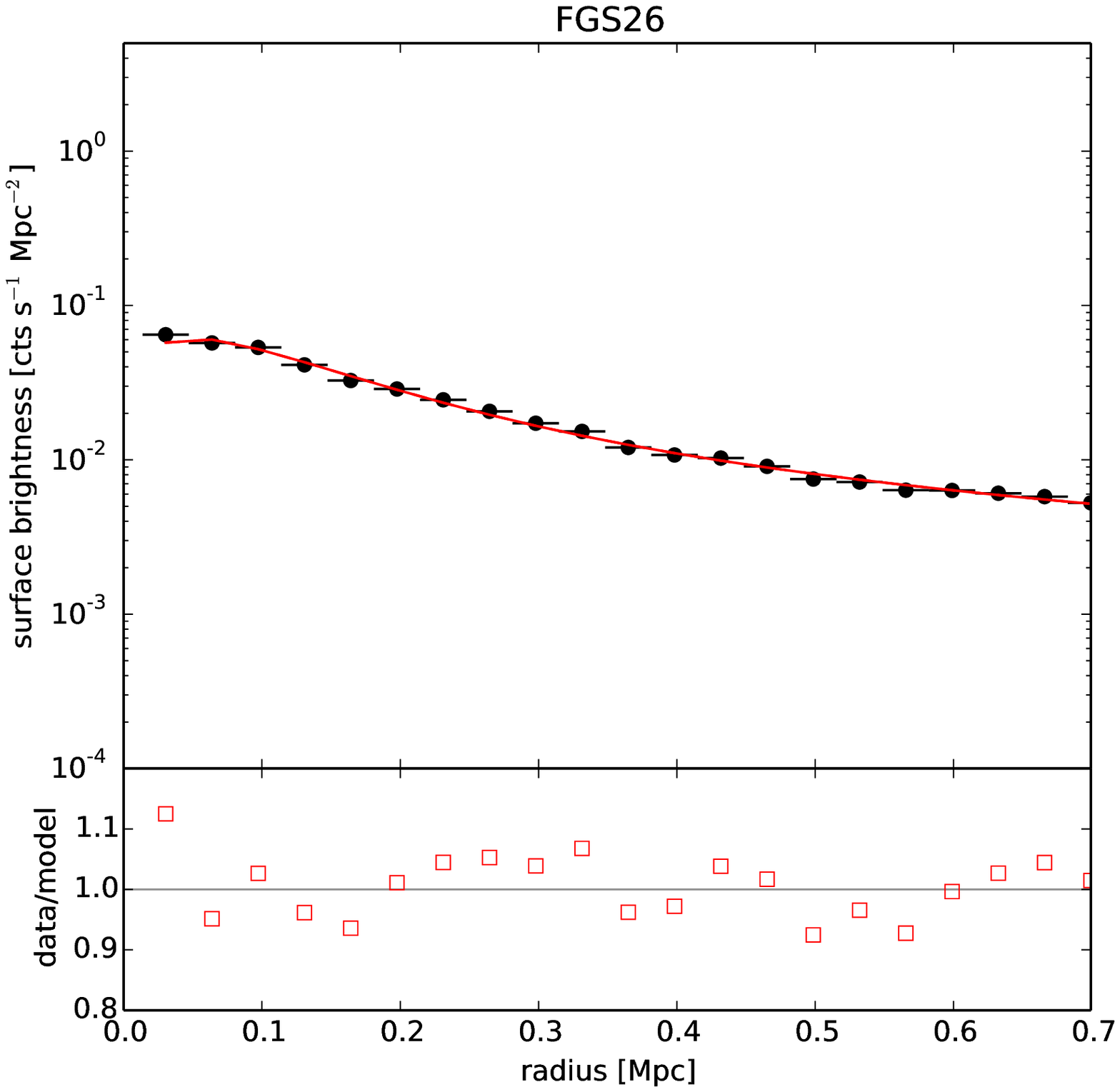}
\end{minipage}
\begin{minipage}{.3\textwidth}
\centering
	\includegraphics[width=\linewidth]{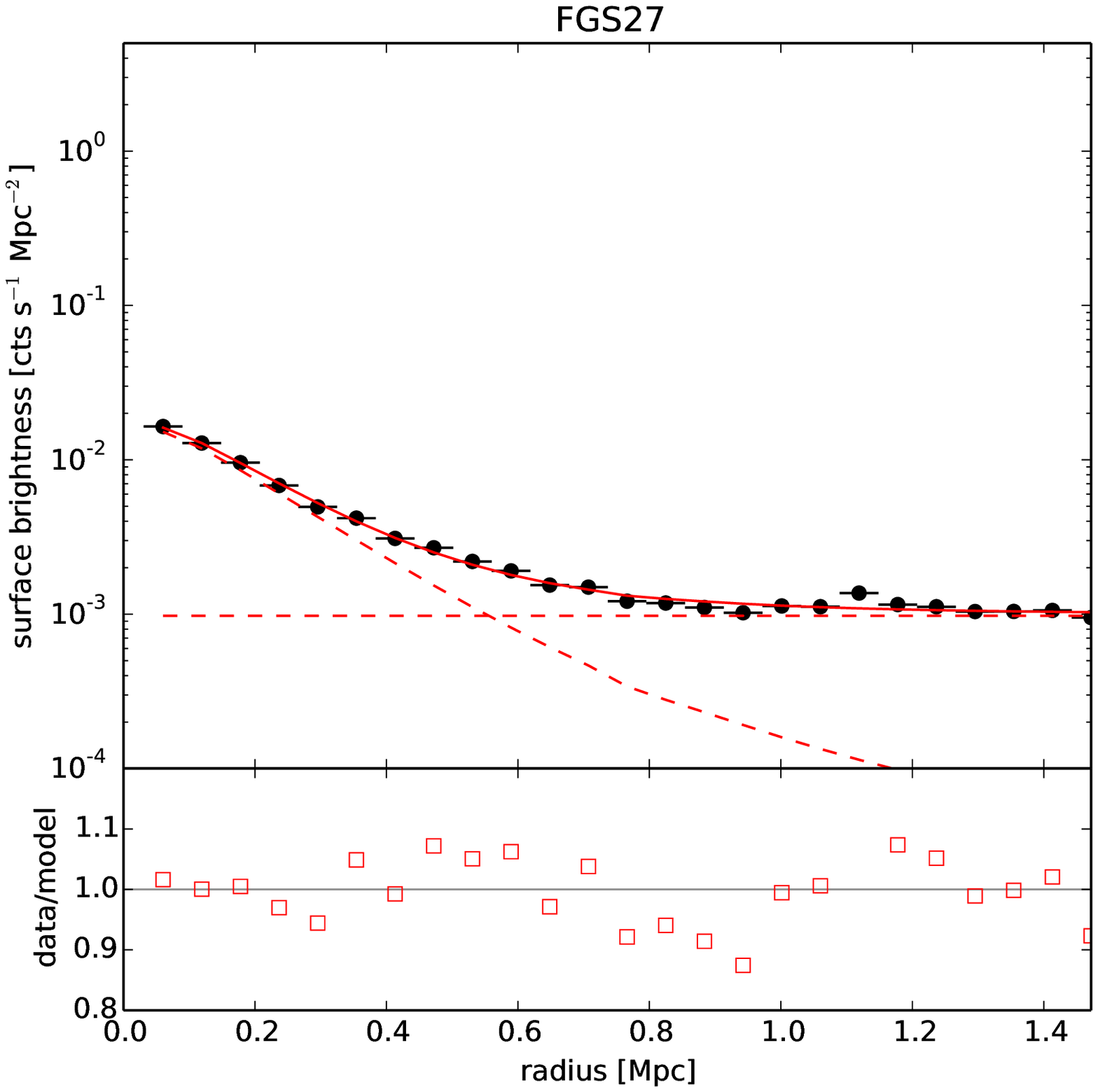}
\end{minipage}
\begin{minipage}{0.3\textwidth}
\centering
	\includegraphics[width=\linewidth]{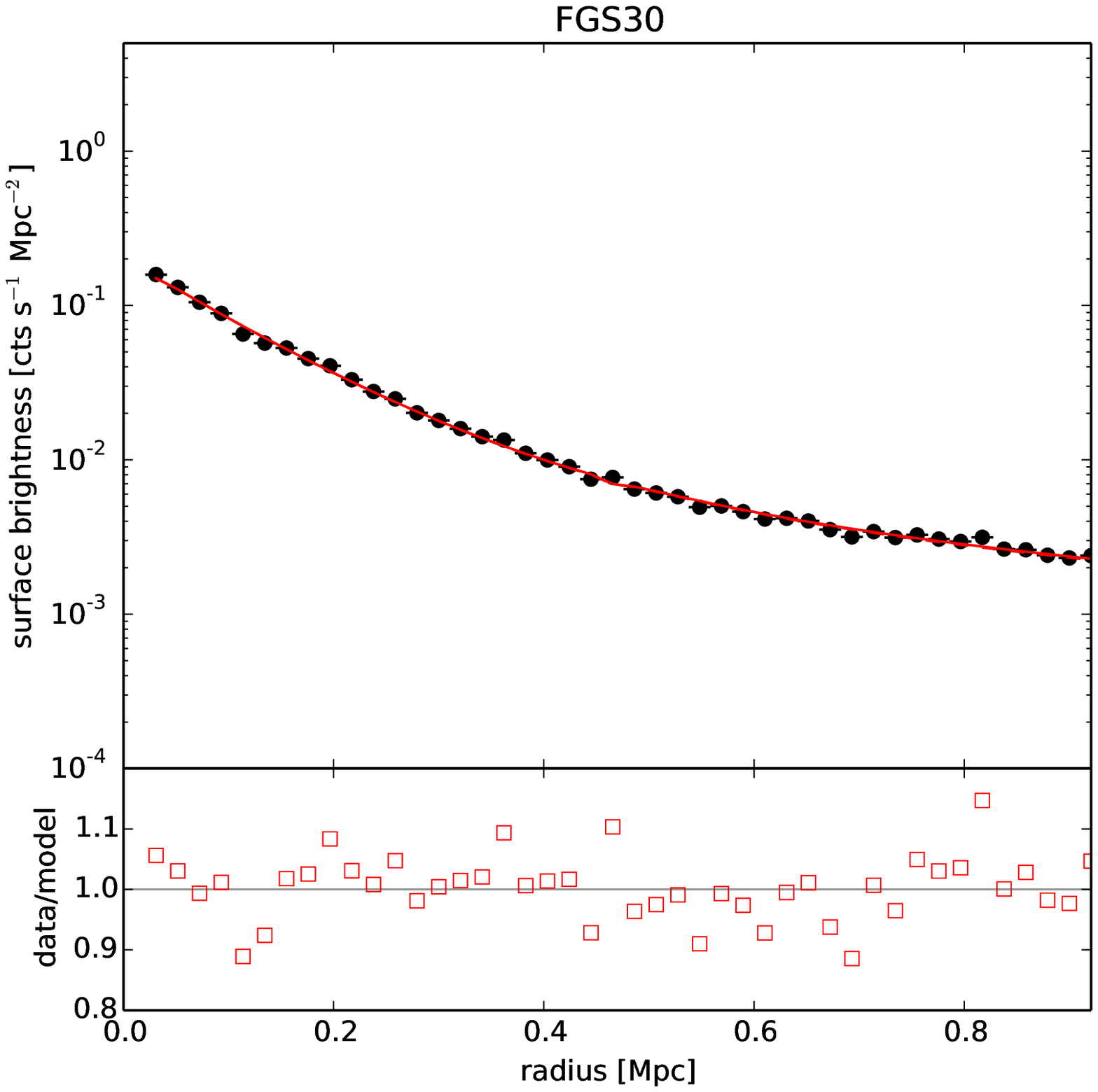}
\end{minipage}
\begin{minipage}{0.6\textwidth}
\caption{Surface brightness profiles of the stacked XIS image in the 0.5--10 keV band. The best-fitting convolved $\beta$-model is plotted in solid red; dashed lines represent the components to the model. Residuals for the $\beta$-model are plotted as squares.}\label{fig:pltsbp}
\end{minipage}
\end{figure*}

Radial surface brightness profiles were constructed for each object using stacked 0.5--10 keV XIS0+XIS1+XIS3 observed images. For the purpose of this profile analysis, we apply an additional satellite attitude correction to the event files used to create the images. {\it Suzaku} XIS images can contain up to a 1 arcmin position error as a result of a recurrent offset between the XRT optical axis and the satellite attitude \citep{uchiyama2008}. With the application of a corrected attitude file, the XIS images can thus be sharpened. This correction was performed by generating corrected attitude files with {\tt aeattcor}, and then applying these corrected attitude files to our cleaned event files using {\tt xiscoord}. The new corrected event files are used to produce the images used in our brightness profile analysis, the brightness profiles of which are shown in Fig.~\ref{fig:pltsbp}. The number of annuli for each profile was determined such that each annulus had at minimum 225 counts, which, assuming Poissonian noise, requires the number of counts to be 15 times the error.

The brightness profile of a spherically symmetric and isothermal ICM in hydrostatic equilibrium will follow a $\beta$-model \citep{cavaliere1976,cavaliere1978}. These are appropriate assumptions for virialized and relaxed groups and clusters. Disparity between the data and the single $\beta$-model can therefore result from processes such as merger asymmetries, multiple thermal components, and non-thermal emission, for example, as produced by an AGN. Our initial fit of the profiles consists of a $\beta$-model plus a background constant:
\begin{equation}
S(r)=S_0(1+(r/r_{\mathrm{c}})^2)^{-3\beta+1/2}+k,
\end{equation}
where $S_0$ is the central surface brightness, $r_{\mathrm{c}}$ is the core radius, and $k$ is the background surface brightness. In this model, the $\beta$-model component was convolved with a radial model of the {\it Suzaku} XRT PSF (see Appendix~\ref{appendix:psffwhm}). Fits were performed with the Sherpa Python module \citep{doe2007}.

The returned best-fitting parameters are recorded in Table~\ref{table:sbp} and the convolved best-fitting model is shown in Fig.~\ref{fig:pltsbp}. We note that FGS03, FGS09, FGS15 have $\chi^2_r >3$ indicating the $\beta$-model poorly describes the observed emission. For these objects, we test adding to the original model a point-like component consisting of a $\delta$ function convolved with the PSF model. This additional point-source component does not offer an improvement in $\chi^2_r$ compared to the original $\beta$-model fits. Nevertheless, the emission from these three objects seems to indicate that either the ICM is not relaxed, or there is some significant source of non-ICM emission.

Because the annuli used are smaller than the {\it Suzaku} XRT PSF and, additionally, discrepancy from a $\beta$-model could be attributed to multiple phenomena, we consider the results as merely suggestive and to be used and interpreted in conjunction with our spectral analysis.

\begin{table}
\caption{Best-fitting parameters of the surface brightness profiles.}
\begin{threeparttable}
\scriptsize
\begin{tabular}{c | c c c c c }
\hline
FGS & \multicolumn{5}{c}{$\beta$-model + background constant} \\
\cline{2-6}
& $S_0^{\dagger}$ & $r_{\mathrm{c}}$ & $\beta$ & $k^{\dagger}$ & $\chi^2$/d.o.f. ($\chi^2_{\mathrm{r}}$) \\
& [10$^{-2}$] & [kpc] & & [10$^{-4}$] & \\ [0.5ex]
\hline
03$^*$ & 299.9$_{-37.3}^{+32.9}$ & 14$_{-2}^{+2}$ & 1.00$_{-0.05}^{+\infty}$ & 89.2$_{-1.3}^{+1.3}$ & 90/27 (3.3) \\
04$^{\phantom{*}}$ & 26.7$_{-5.2}^{+\infty}$ & 48$_{-2}^{+7}$ & 0.64$_{-0.02}^{+0.02}$ & 6.7$_{-0.2}^{+0.2}$ & 25/15 (1.6) \\
09$^{\phantom{*}}$ & 50.4$_{-4.5}^{+4.3}$ & 38$_{-5}^{+5}$ & 1.00$_{-0.05}^{+\infty}$ & 19.9$_{-0.3}^{+0.3}$ & 126/22 (5.7) \\
14$^*$ & 8.3$_{-1.4}^{+2.3}$ & 28$_{-10}^{+11}$ & 0.41$_{-0.02}^{+0.02}$ & 4.3$_{-0.7}^{+0.6}$ & 65/30 (2.2) \\
15$^{\phantom{*}}$ & 28.7$_{-5.8}^{+\infty}$ & 16$_{-1}^{+3}$ & 0.49$_{-0.02}^{+0.02}$ & 26.3$_{-0.8}^{+0.7}$ & 99/19 (5.2) \\
24$^{\phantom{*}}$ & 2.7$_{-0.5}^{+0.8}$ & 38$_{-17}^{+19}$ & 0.40$_{-0.03}^{+0.03}$ & 3.9$_{-0.5}^{+0.4}$ & 49/28 (1.8) \\
25$^{\phantom{*}}$ & 34.4$_{-0.4}^{+2.8}$ & 56$_{-2}^{+3}$ & 0.45$_{-0.00}^{+0.01}$ & 0.0$_{-\infty}^{+1.4}$ & 80/40 (2.0) \\
26$^*$ & 12.7$_{-0.9}^{+0.9}$ & 47$_{-5}^{+7}$ & 0.37$_{-0.00}^{+0.01}$ & 0.0$_{-\infty}^{+5.8}$ & 23/17 (1.4) \\
27$^*$ & 3.0$_{-0.3}^{+0.4}$ & 88$_{-20}^{+22}$ & 0.55$_{-0.05}^{+0.06}$ & 9.8$_{-0.4}^{+0.4}$ & 37/21 (1.8) \\
30$^*$ & 80.0$_{-15.7}^{+16.0}$ & 11$_{-2}^{+3}$ & 0.39$_{-0.00}^{+0.01}$ & 0.4$_{-2.7}^{+2.5}$ & 60/40 (1.5) \\
\hline
\end{tabular}
\begin{tablenotes}
\item [$\dagger$] Units of counts s$^{-1}$ Mpc$^{-2}$
\item [*] Confirmed fossil system
\end{tablenotes}
\label{table:sbp}
\end{threeparttable}
\end{table}

\section{Spectral Analysis}\label{sec:spectralanalysis}

Our spectral analysis consists of measuring spectral properties within a region of high S/N (Section~\ref{sec:optreg}) and using these results to classify these objects as thermally dominated or AGN contaminated (Section~\ref{sec:speccomp}). The results of this section will then be used to measure or estimate the global properties of the ICM-dominated systems within $r_{500}$ (Section~\ref{sec:globalprop}).


\subsection{Spectral fitting in the source region}\label{sec:optreg}

In order to disentangle ICM emission from potential contaminating point source emission, we perform our analysis on the source aperture region where the source emission is more than half of the total emission from the object. By determining this source aperture radius, $r_{\mathrm{ap,src}}$ as described in Section~\ref{sec:rsrc}, we make no assumptions on the type of source emission. Extracting a spectrum from this region therefore improves the spectral analysis of any type of source over the background whether the source is dominated by thermal emission from the ICM or non-thermal emission from an AGN.

The results of our surface brightness profile analysis indicate some objects in our sample may have a strong non-thermal point-like component to the total emission. As a result, we compare the fit of three source models to our spectra:
\begin{enumerate}[(i)]
\item an absorbed thermal plasma model, {\sc wabs}$\times${\sc apec}, to model the ICM;
\item an absorbed power-law, {\sc wabs}$\times${\sc powerlaw}, to model an AGN;
\item an absorbed combined thermal and power-law model, {\sc wabs}({\sc apec}+{\sc powerlaw}), to describe contribution from both the ICM and an AGN;
\end{enumerate}
\noindent where the {\sc wabs} absorption component accounts for galactic absorption in all three models.

The background and foreground sources consist of the NXB, LHB, MWH, and CXB. The NXB spectrum was used as the background file for the extracted $r_{\mathrm{ap}}$ region to be subtracted directly during the spectral fit. The CXB, LHB, and MWH were accounted for through modelling as described in Section~\ref{sec:bkg}.

The XIS spectra were grouped with {\tt grppha} such that each bin had a minimum of 25 counts. The binned {\it Suzaku} XIS0, XIS1, and XIS3 spectra were fit simultaneously with the RASS background spectrum. The {\it Suzaku} spectra were fit with the source and background model while the RASS spectra were fit only with the background model. The RASS best-fitting parameters were tied to that of the {\it Suzaku} spectra with a scaling factor to account for the difference in the angular size of the spectral extraction regions. Bad channels were ignored for all spectra. The {\it Suzaku} XIS0 and XIS3 spectra were fit over 0.7--10 keV (Section~\ref{sec:softexcess}), the XIS1 spectra over 0.7--7 keV, and the RASS spectra over the range 0.1--2.4 keV.

In all three models, the neutral hydrogen column density was assigned the weighted average galactic value in the direction of the source \citep{kalberla2005}. The redshifts of our systems were taken to be the spectroscopic redshifts of the bright central galaxies as determined by \citetalias{santos2007}. During the fit, the column density and redshift were always fixed. The metal abundance $Z$ component of the {\sc apec} model was calculated using the abundance tables of \citet{anders1989}. The photon index of the {\sc powerlaw} model was constrained to be within $\Gamma=1.5-2.5$ \citep{ishibashi2010}. Initially, all other parameters were left free to be fit.
However, if during the fit convergence on an {\sc apec} or {\sc powerlaw} parameter within the physically reasonable limits did not occur or the parameter was returned with infinite error bars, the fit was performed again with that parameter fixed. In all further tables, quantities presented without error bars have been fixed to a reasonable value.

\begin{figure*}
\begin{minipage}{.32\textwidth}
\centering
	\includegraphics[angle=-90,width=\linewidth]{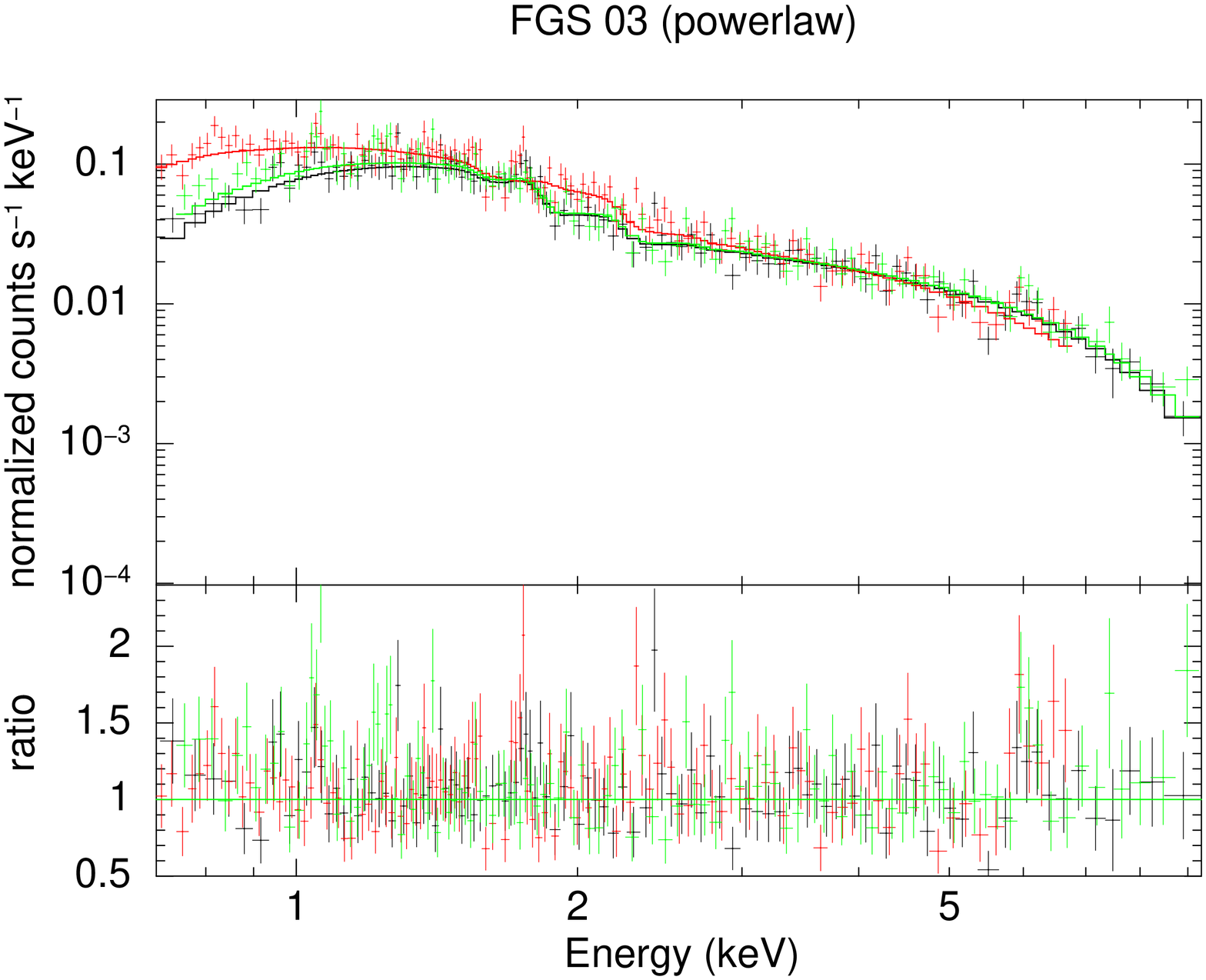}
\end{minipage}
\begin{minipage}{.32\textwidth}
\centering
	\includegraphics[angle=-90,width=\linewidth]{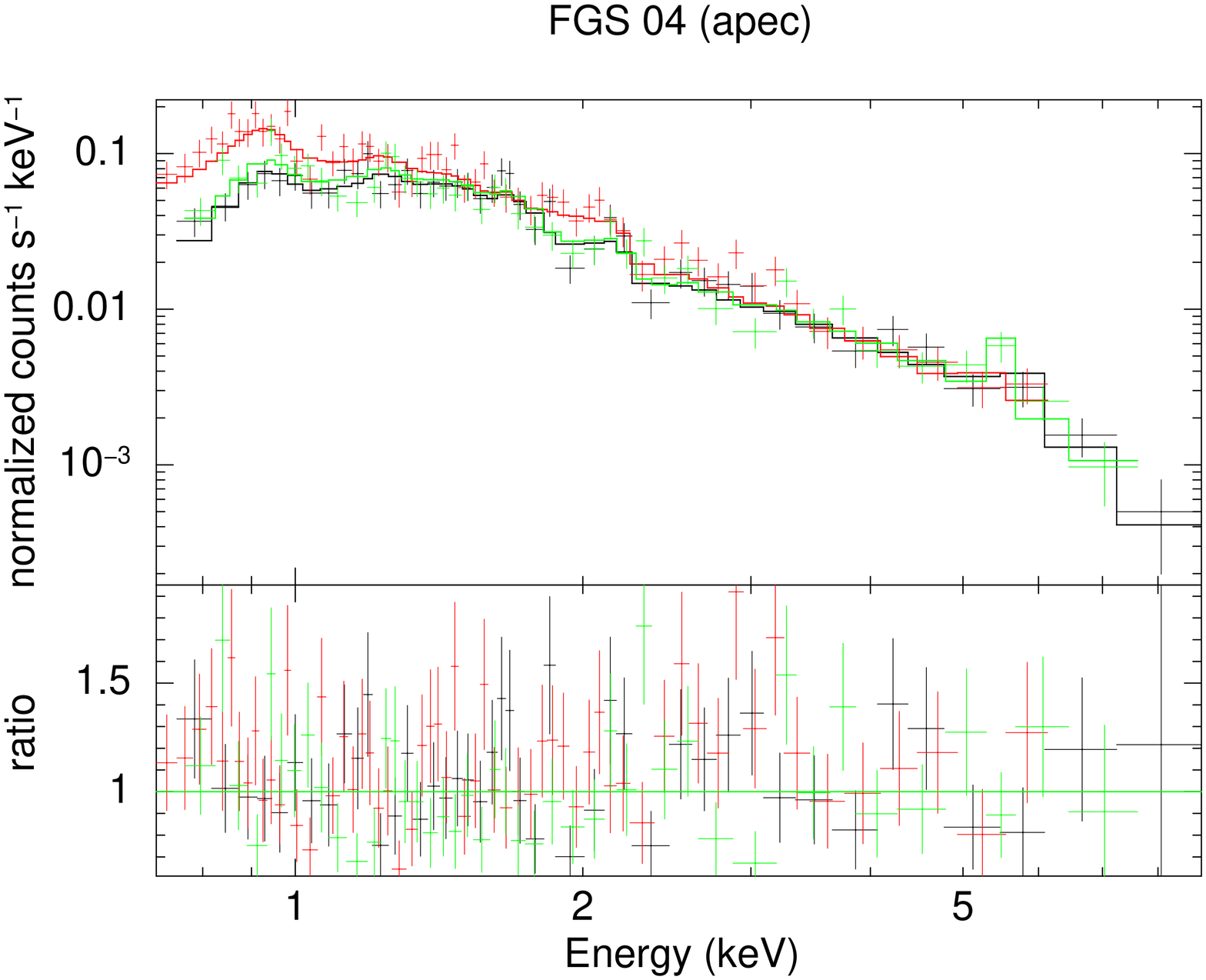}
\end{minipage}
\begin{minipage}{.32\textwidth}
\centering
	\includegraphics[angle=-90,width=\linewidth]{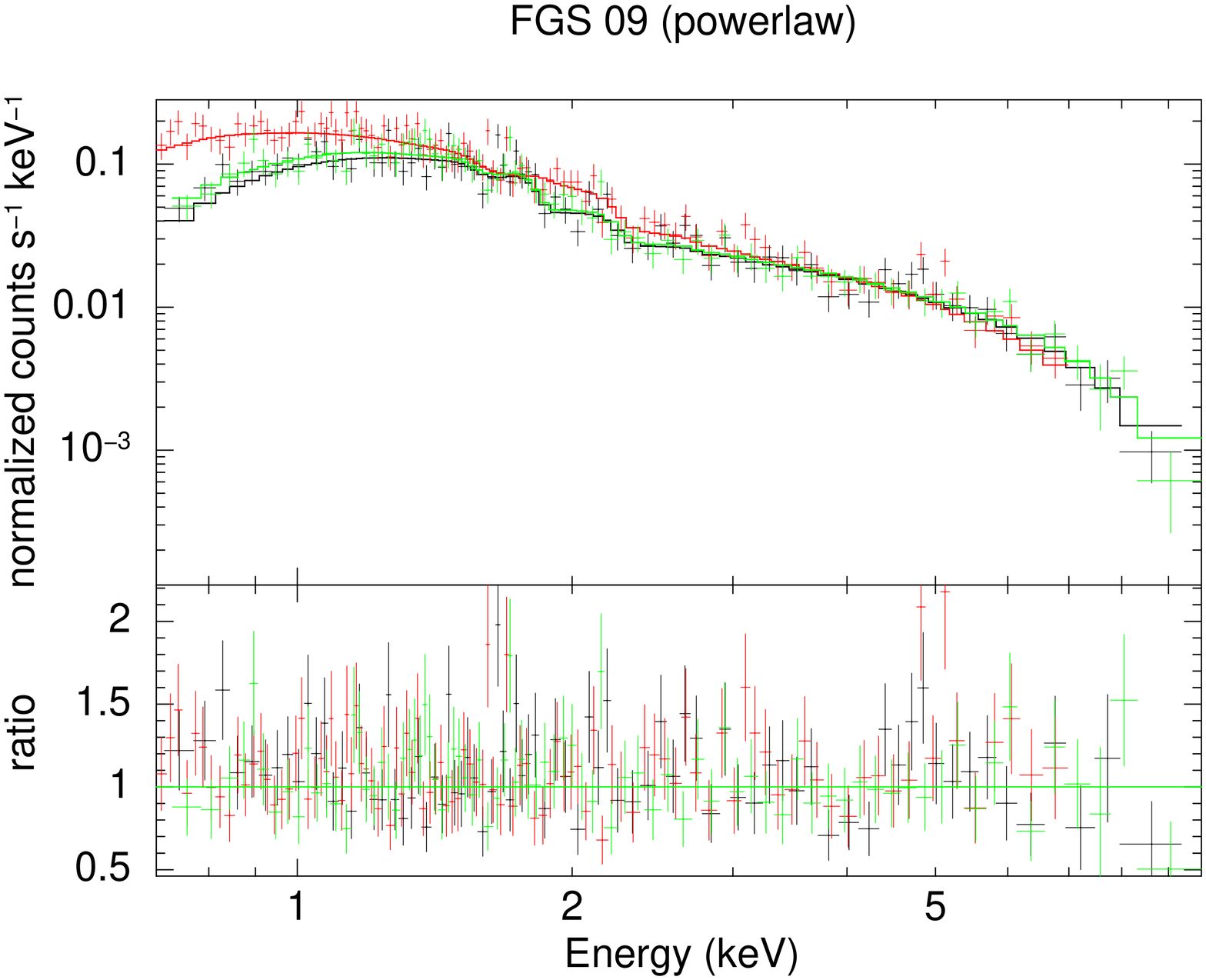}
\end{minipage}
\begin{minipage}{.32\textwidth}
\centering
	\includegraphics[angle=-90,width=\linewidth]{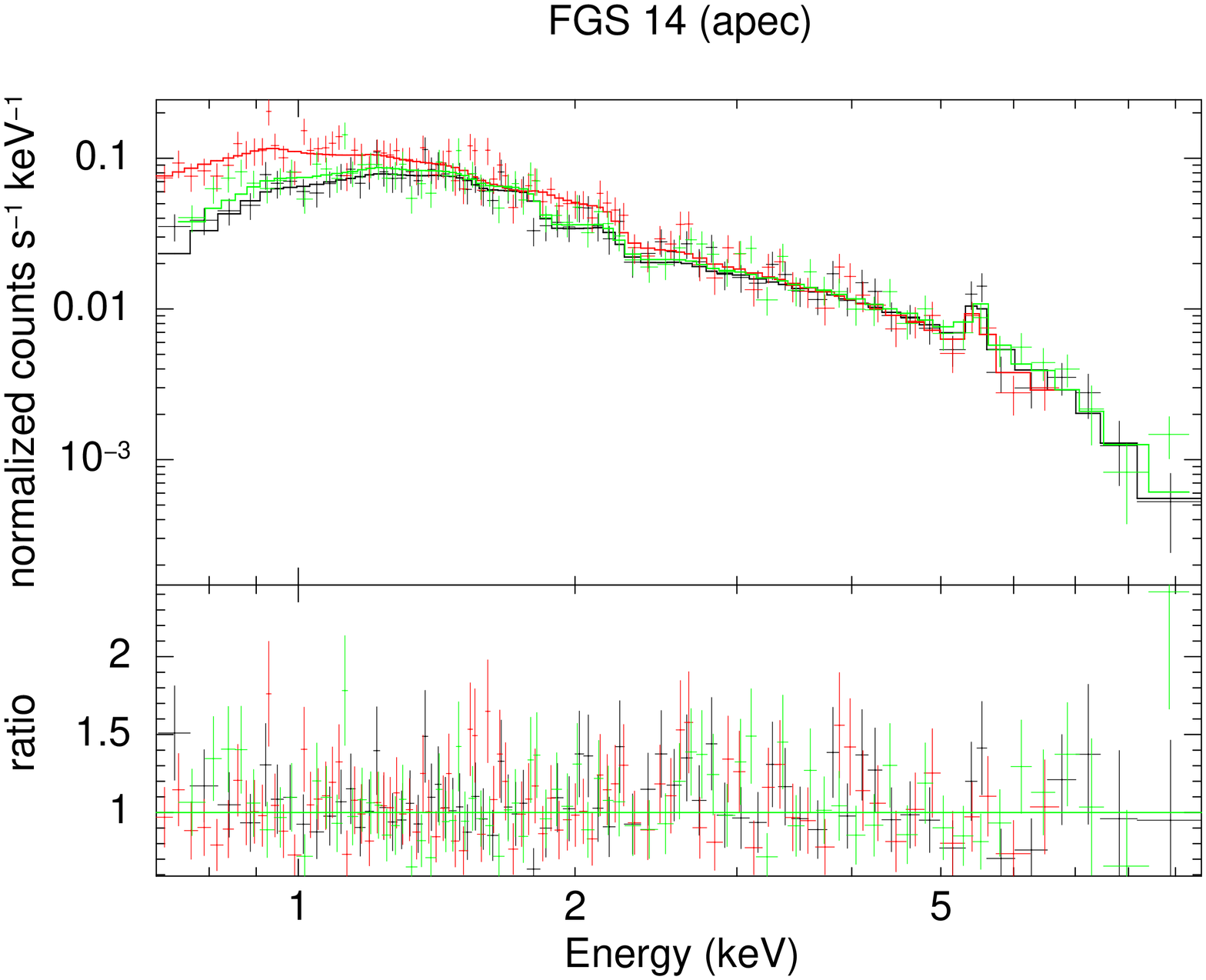}
\end{minipage}
\begin{minipage}{.32\textwidth}
\centering
	\includegraphics[angle=-90,width=\linewidth]{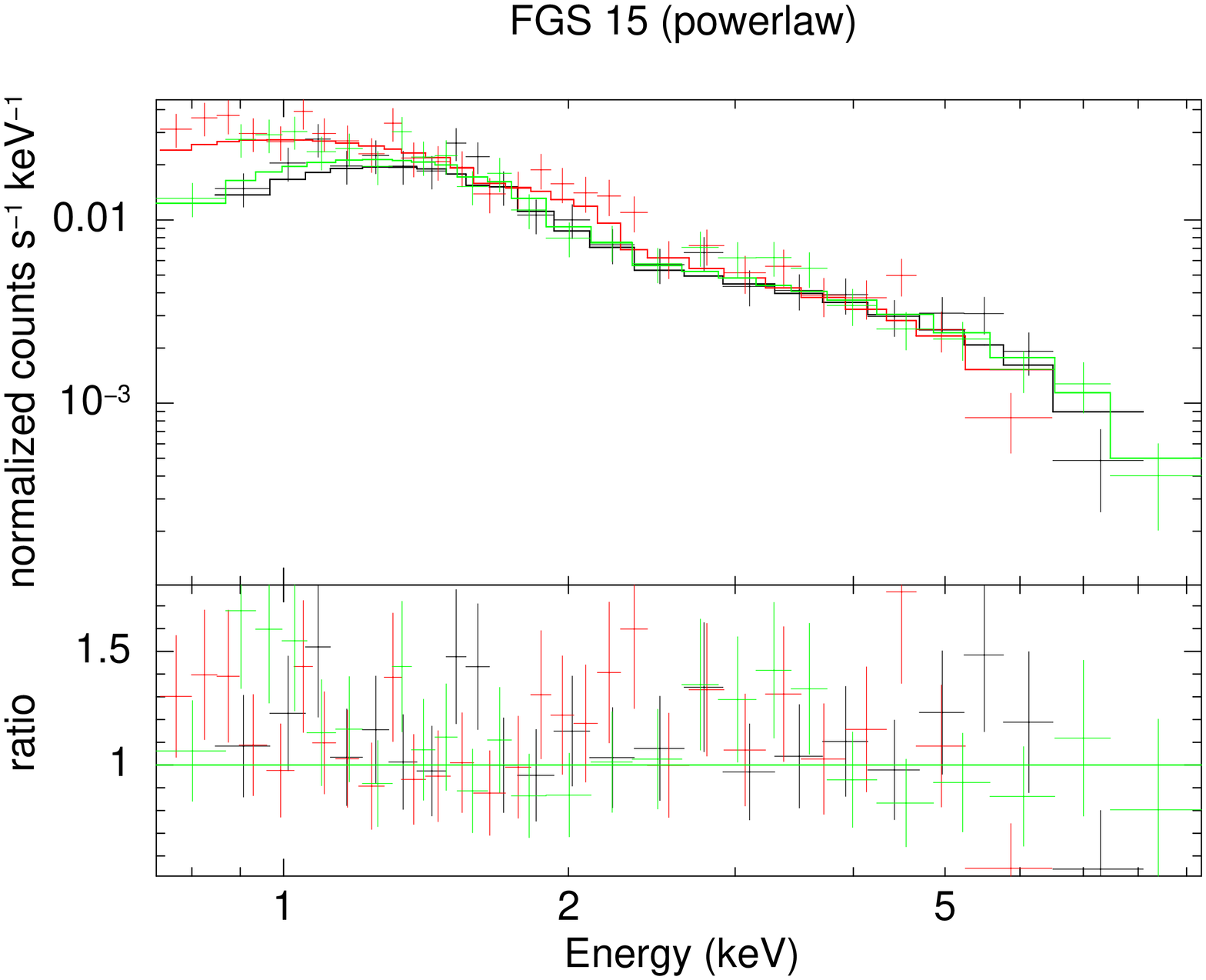}
\end{minipage}
\begin{minipage}{0.32\textwidth}
\centering
	\includegraphics[angle=-90,width=\linewidth]{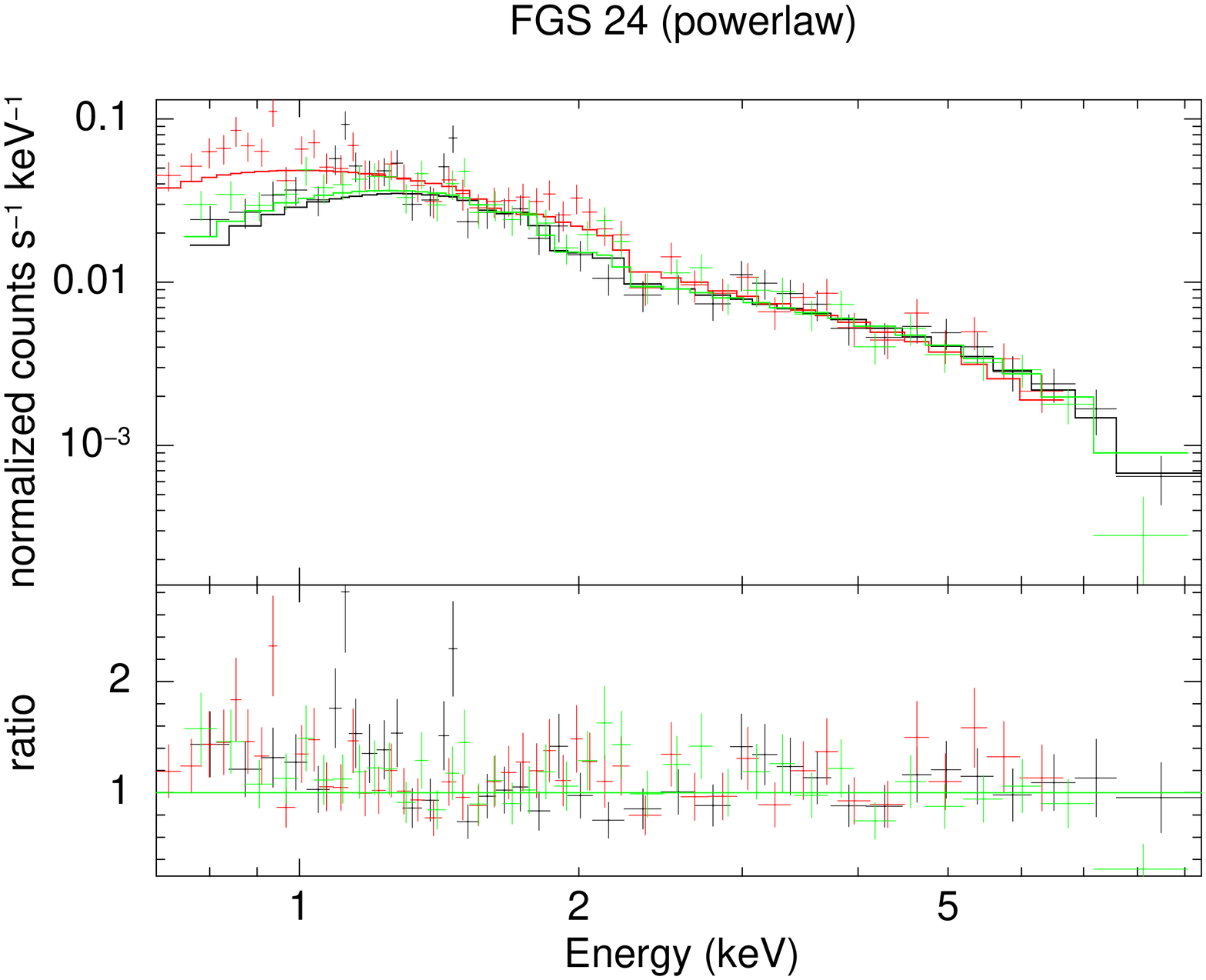}
\end{minipage}
\begin{minipage}{.32\textwidth}
\centering
	\includegraphics[angle=-90,width=\linewidth]{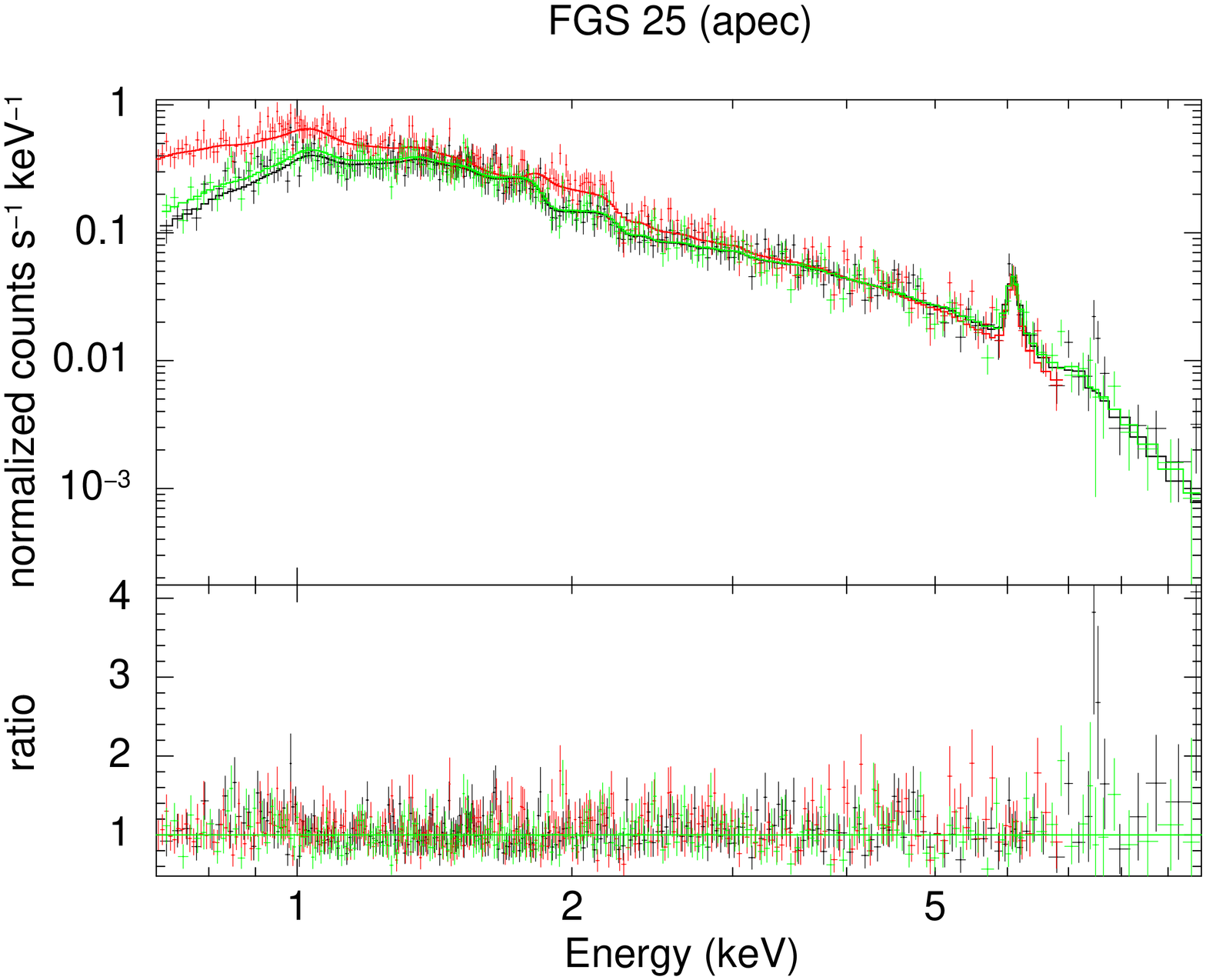}
\end{minipage}
\begin{minipage}{.32\textwidth}
\centering
	\includegraphics[angle=-90,width=\linewidth]{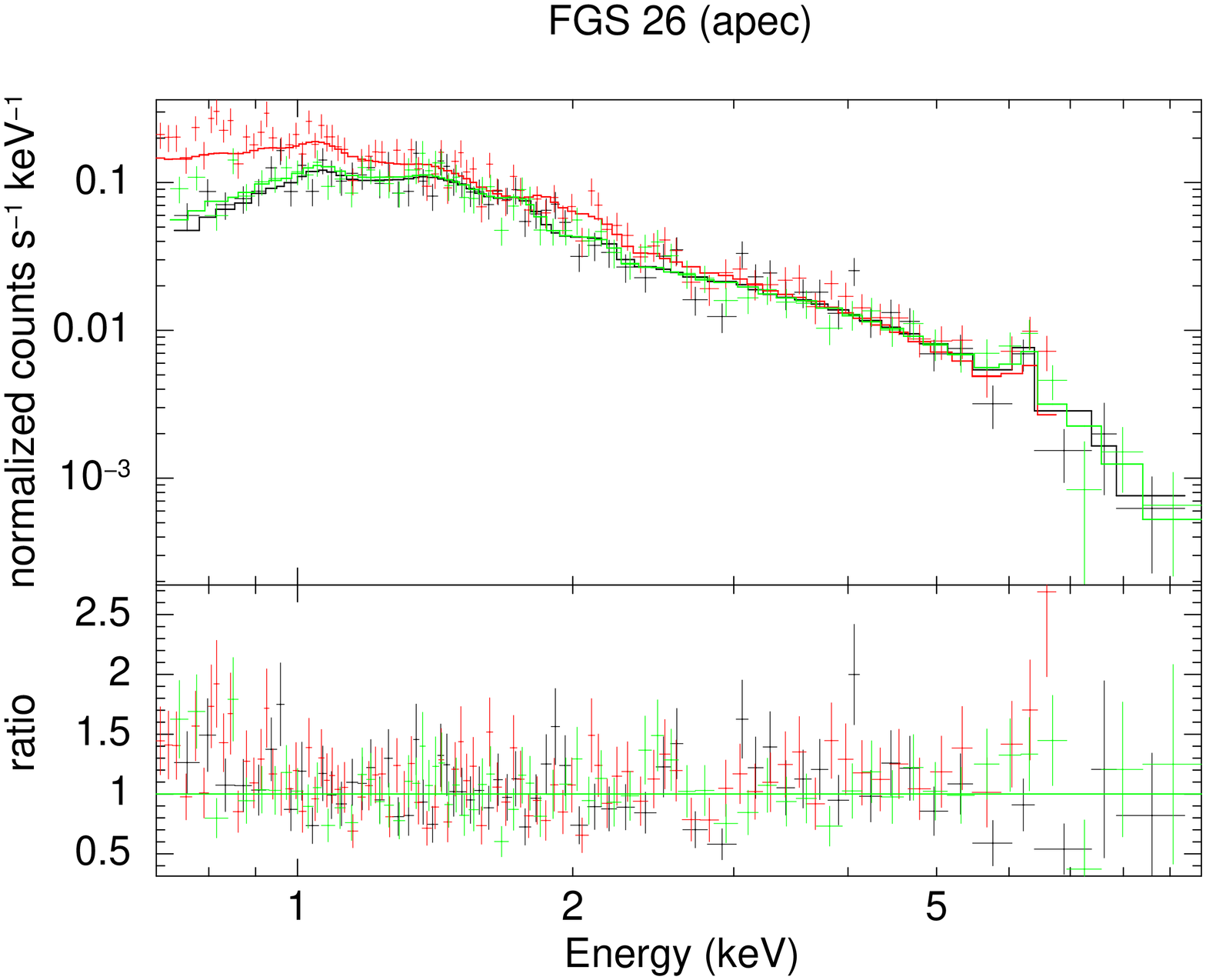}
\end{minipage}
\begin{minipage}{.32\textwidth}
\centering
	\includegraphics[angle=-90,width=\linewidth]{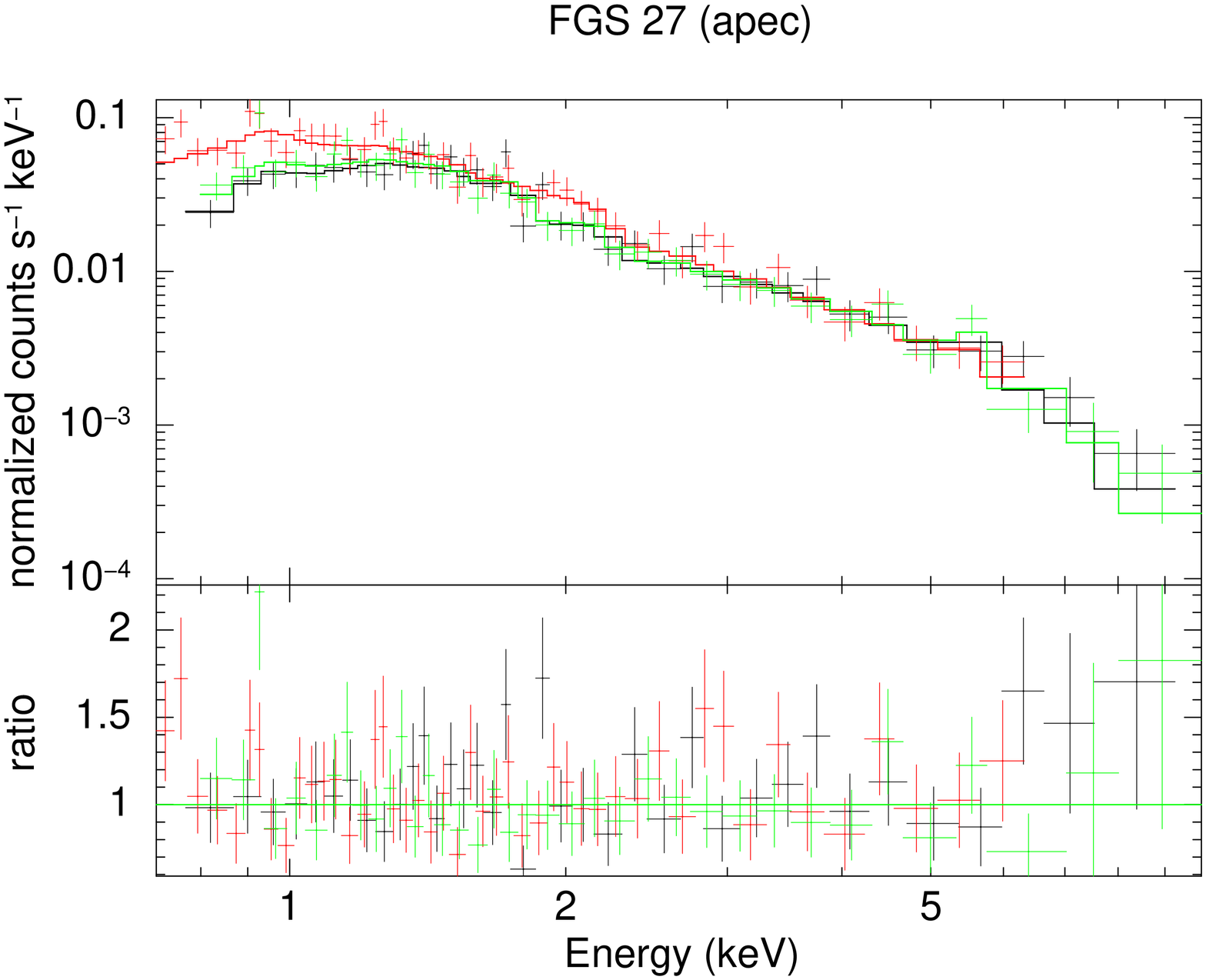}
\end{minipage}
\begin{minipage}{.32\textwidth}
\centering
	\includegraphics[angle=-90,width=\linewidth]{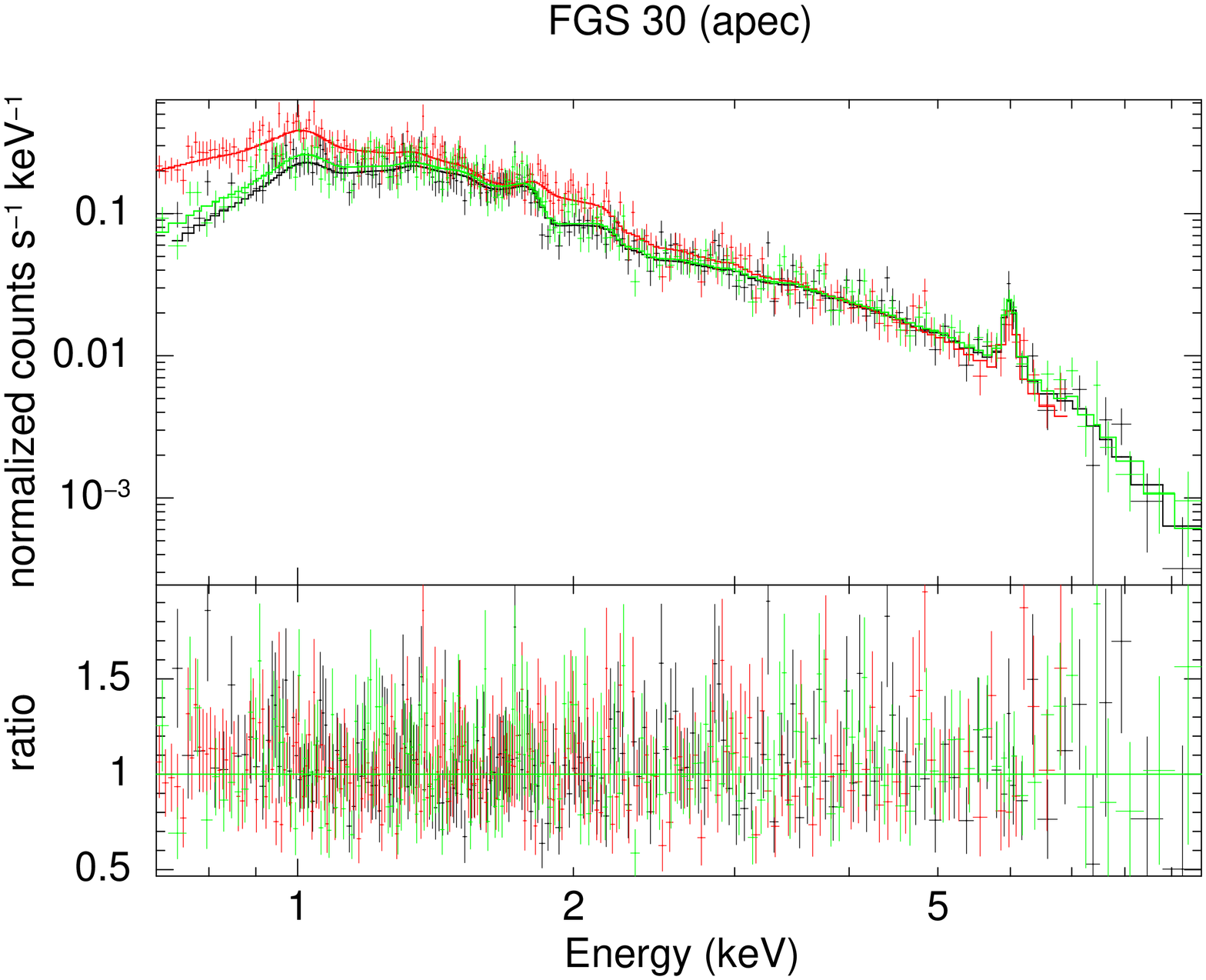}
\end{minipage}
\begin{minipage}{0.64\textwidth}
\caption{The XIS0 (black), XIS1 (red), and XIS3 (green) spectra for the source regions $r_{\mathrm{ap,src}}$ determined in Section~\ref{sec:rsrc}. The best-fitting model to the observed spectra, as determined by the $\chi^2$ values in Table~\ref{table:bfp}, is plotted in a solid line. The RASS spectra that were simultaneously fit with the {\it Suzaku} background model are not shown.}\label{fig:fitspec}
\end{minipage}
\end{figure*}

The resulting best-fitting parameters are listed in Table~\ref{table:bfp} and the best-fitting models to the spectra are shown in Fig.~\ref{fig:fitspec}. The background parameters resulting from each of the model fits were consistent with each other within 1$\sigma$ errors.


\subsubsection{A soft energy excess}\label{sec:softexcess}

While the XIS is sensitive to photons with energy as low as 0.5 keV, we have excluded the $E < 0.7$ keV energy channels from our spectral analysis. In the majority of our observations, an apparent excess in counts was found in the 0.5--0.7 keV range when compared to the fit of the {\sc apec} or {\sc powerlaw} models in the $E>0.7$ keV range.

Potential origins of this soft excess include a second thermal component in the ICM, an AGN, calibration issues, SWCX, or statistical fluctuations. Adding a second thermal model to the ICM model did not improve the fit. If an AGN were the origin of the excess, removing the softest energies should not greatly deter detecting the presence of its emission in the spectra because an AGN will contribute most strongly to the harder energies of the spectrum. Calibration issues with proportional removal of flickering pixels from observations of the source and the NXB may also contribute to energy channels below 0.6 keV. Additionally, it is possible there is some contribution from SWCX in the soft energy regime, although the solar wind proton flux light curves of most of our sample are of a low intensity indicating geocoronal SWCX emission is unlikely to be a significant contaminant (see Section~\ref{sec:swcx}).

Because the origin of this excess is uncertain and thus cannot be appropriately modelled in the spectra, and furthermore the excess only affects the first few low energy channels in the spectrum, we exclude this softest energy regime from our fits. This has little effect on the returned best-fitting parameters and in general the reduced $\chi^2$ of the fits improves with the exclusion of the soft excess energy channels.


\subsection{Comparison and interpretation of the model fits}\label{sec:speccomp}

In comparing the fits of the three models, the {\sc apec}+{\sc powerlaw} model does not appear to significantly improve the characterization of the spectra over the individual {\sc apec} and {\sc powerlaw} fits. Indeed in the combined fit, the {\sc apec} and {\sc powerlaw} components are never simultaneously constrained. As a result, while some {\sc apec}+{\sc powerlaw} fits return $\chi^2_r$ with values slightly less than that for the less complex fits of {\sc apec} or {\sc powerlaw} only, we decide to choose the simpler model that has all parameters constrained.

By the $\chi^2_r$ values, the {\sc powerlaw} model provides a better fit over the thermal {\sc apec} model for FGS03, FGS09, FGS15, and FGS24. We consider these four objects to be dominated by a non-ICM source and with our current observations, we cannot disentangle the ICM and non-ICM emission. Further discussion on these objects is provided in Appendix~\ref{appendix:notes}.

For the remainder of our analysis, we focus on those objects in our sample
with spectra that are best fit by the {\sc apec} model and are thus galaxy systems dominated by ICM emission.

\begin{table}
\begin{turn}{90}
\begin{minipage}[c][80mm]{235mm}
\caption{Best-fitting spectral parameters in the $r_{\mathrm{ap,src}}$ region.}
\begin{threeparttable}
\scriptsize
\begin{tabular}{c c | c c c c | c c c | c c c c c c}
\hline
FGS & r$_{\mathrm{ap,src}}$ & \multicolumn{4}{c}{{\sc apec}} & \multicolumn{3}{c}{{\sc powerlaw}} & \multicolumn{6}{c}{{\sc apec}+{\sc powerlaw}} \\ 
& & $kT_{\mathrm{apec}}$ & $Z_{\mathrm{apec}}$ & norm$_{\mathrm{apec}}^{a}$ & $\chi^2$/d.o.f ($\chi^{2}_{r})$ & $\Gamma_{\mathrm{PL}}$ & norm$_{\mathrm{PL}}^{b}$ & $\chi^2$/d.o.f ($\chi^{2}_{r}$) & $kT_{\mathrm{apec}}$ & $Z_{\mathrm{apec}}$ & norm$_{\mathrm{apec}}^{a}$ & $\Gamma_{\mathrm{PL}}$ & norm$_{\mathrm{PL}}^{b}$ & $\chi^2$/d.o.f ($\chi^{2}_{r}$) \\
& ['(Mpc)] & $[$keV$]$ & $[$Z$_{\odot}]$ & $[10^{-3}]$ & & & [$10^{-4}$] & & $[$keV$]$ & $[$Z$_{\odot}]$ & $[10^{-3}]$ & & [$10^{-4}$] & \\ [0.5ex]
\hline
03$^*$			& 3.6(0.22) & 6.71$_{-0.47}^{+0.91}$ & 0.14$_{-0.14}^{+0.08}$ & 4.8$_{-0.1}^{+0.1}$ & 409/351 (1.17) & 1.72$_{-0.03}^{+0.03}$ & 11.9$_{-0.4}^{+0.4}$ & 361/352 (1.02) & 2 & 0.3 & 0.3$_{-0.3}^{+0.5}$ & 1.68$_{-0.09}^{+0.07}$ & 10.9$_{-1.6}^{+1.3}$ & 360/351 (1.02) \\
04$^{\phantom{*}}$ 	& 3.3(0.67) & 2.84$_{-0.19}^{+0.18}$ & 0.40$_{-0.10}^{+0.12}$ & 4.3$_{-0.3}^{+0.3}$ & 169/148 (1.14) & 2.30$_{-0.06}^{+0.06}$ & 10.3$_{-0.4}^{+0.4}$ & 212/149 (1.43) & 2.46$_{-0.37}^{+0.32}$ & 0.42$_{-0.13}^{+0.16}$ & 3.4$_{-0.7}^{+0.7}$ & 1.8 & 1.7$_{-1.3}^{+1.3}$ & 165/147 (1.12) \\
09$^{\phantom{*}}$ 	& 4.1(0.55) & 4.73$_{-0.34}^{+0.35}$ & 0.13$_{-0.13}^{+0.07}$ & 5.8$_{-0.2}^{+0.2}$ & 287/265 (1.08) & 1.94$_{-0.04}^{+0.04}$ & 13.6$_{-0.4}^{+0.4}$ & 246/266 (0.92) & 2 & 0.3 & 0.0$_{-0.0}^{+0.6}$ & 1.94$_{-0.07}^{+0.04}$ & 13.6$_{-1.7}^{+0.4}$ & 246/265 (0.93) \\
14$^*$ 			& 4.0(0.86) & 5.23$_{-0.38}^{+0.44}$ & 0.21$_{-0.08}^{+0.08}$ & 4.9$_{-0.2}^{+0.2}$ & 212/237 (0.90) & 1.87$_{-0.04}^{+0.04}$ & 9.8$_{-0.4}^{+0.4}$ & 262/238 (1.10) & 5.23$_{-0.45}^{+0.44}$ & 0.21$_{-0.08}^{+0.09}$ & 4.9$_{-1.1}^{+0.1}$ & 1.8 & 0.0$_{-0.0}^{+2.1}$ & 212/236 (0.90) \\
15$^{\phantom{*}}$ 	& 2.8(0.32) & 5.41$_{-0.80}^{+1.26}$ & 0.3 & 2.5$_{-0.1}^{+0.1}$ & 66/79 (0.83) & 1.79$_{-0.09}^{+0.09}$ & 6.0$_{-0.4}^{+0.4}$ & 63/79 (0.80) & 2 & 0.3 & 0.0$_{-0.0}^{+0.3}$ & 1.8 & 5.9$_{-0.8}^{+0.4}$ & 63/79 (0.80) \\
24$^{\phantom{*}}$ 	& 2.7(0.71) & 5.67$_{-0.76}^{+0.88}$ & 0.21$_{-0.21}^{+0.15}$ & 4.7$_{-0.3}^{+0.3}$ & 174/126 (1.38) & 1.88$_{-0.07}^{+0.07}$ & 8.9$_{-0.5}^{+0.5}$ & 169/127 (1.33) & 4.10$_{-2.82}^{+1.95}$ & 0.3 & 1.7$_{-1.4}^{+1.9}$ & 1.8 & 5.7$_{-3.6}^{+2.6}$ & 167/126 (1.33) \\
25$^{\phantom{*}}$ 	& 7.4(0.80) & 3.91$_{-0.14}^{+0.14}$ & 0.31$_{-0.04}^{+0.04}$ & 12.3$_{-0.3}^{+0.3}$ & 742/777 (0.96) & 2.04$_{-0.02}^{+0.02}$ & 31.5$_{-0.6}^{+0.6}$ & 983/778 (1.26) & 3.76$_{-0.26}^{+0.22}$ & 0.33$_{-0.05}^{+0.06}$ & 11.0$_{-1.5}^{+1.4}$ & 1.8 & 3.0$_{-3.0}^{+3.2}$ & 740/776 (0.95) \\
26$^*$ 			& 4.9(0.40) & 3.39$_{-0.24}^{+0.28}$ & 0.25$_{-0.10}^{+0.10}$ & 5.2$_{-0.3}^{+0.3}$ & 325/220 (1.48) & 2.14$_{-0.06}^{+0.06}$ & 14.2$_{-0.6}^{+0.6}$ & 331/221 (1.50) & 3.10$_{-0.75}^{+0.45}$ & 0.24$_{-0.12}^{+0.13}$ & 4.3$_{-1.1}^{+1.0}$ & 1.8 & 2.3$_{-2.3}^{+3.0}$ & 323/219 (1.48) \\
27$^*$ 			& 3.4(0.63) & 3.42$_{-0.28}^{+0.34}$ & 0.20$_{-0.20}^{+0.13}$ & 4.0$_{-0.3}^{+0.3}$ & 107/121 (0.88) & 2.14$_{-0.07}^{+0.07}$ & 8.9$_{-0.5}^{+0.5}$ & 115/122 (0.94) & 3.06$_{-0.91}^{+0.60}$ & 0.19$_{-0.19}^{+0.16}$ & 3.3$_{-1.0}^{+0.9}$ & 1.8 & 1.6$_{-1.6}^{+2.2}$ & 106/120 (0.88) \\
30$^*$ 			& 5.4(0.67) & 3.43$_{-0.11}^{+0.15}$ & 0.30$_{-0.05}^{+0.05}$ & 8.0$_{-0.2}^{+0.2}$ & 627/599 (1.05) & 2.08$_{-0.03}^{+0.03}$ & 20.0$_{-0.4}^{+0.4}$ & 844/600 (1.41) & 3.26$_{-0.20}^{+0.18}$ & 0.32$_{-0.06}^{+0.06}$ & 7.0$_{-0.8}^{+0.9}$ & 1.8 & 2.2$_{-1.8}^{+1.8}$ & 623/598 (1.04) \\
\hline
\end{tabular}
\begin{tablenotes}
\item [] Note: quantities without errors have been fixed at the listed value
\item [$a$] norm$_{\mathrm{apec}}=\frac{10^{-14}}{4 \pi [D_A (1+z)]^2} \int n_e n_H dV \mathrm{cm}^{-5} $
\item [$b$] norm$_{\mathrm{PL}}$ has units of photons keV$^{-1}$ cm$^{-2}$ s$^{-1}$ arcmin$^{-2}$ at 1 keV
\item [*] Confirmed fossil system
\end{tablenotes}
\label{table:bfp}
\end{threeparttable}
\end{minipage}
\end{turn}
\end{table}


\section{Global ICM temperatures and luminosities}\label{sec:globalprop}

In order to compare the ICM temperatures and luminosities of our fossil systems with those of other groups and clusters, we calculate these properties within the fiducial radius of $r_{500}$, the radius at which the average enclosed density is 500 times the critical density of the Universe. We calculate $r_{500}$, and the spectral properties within this radius, using an iterative procedure.

Using the temperature calculated within some aperture, $T_{\mathrm{ap}}$, we calculate $r_{500}$ using the $r_{500}$--$T_{\mathrm{X}}$ relation of \citet{arnaud2005}:
\begin{equation}\label{eq:arnaudrt}
r_{500}=1.104 \ h_{70}^{-1} \ E(z)^{-1} \left(\frac{kT}{5 \ \mathrm{keV}}\right)^{0.57} \mathrm{Mpc},
\end{equation}
where $h_{70}=H_0/(70 \ \mathrm{km} \ \mathrm{s}^{-1} \ \mathrm{Mpc}^{-1})$ and $E(z) = H(z)/H_0 =\sqrt{\Omega_{\mathrm{M}} (1+z)^3 + \Omega_{\mathrm{k}} (1+z)^2 + \Omega_{\Lambda}}$ \citep{hogg1999}. This value of $r_{500}$ is used as our next radius of extraction to determine a new $T_{\mathrm{ap}}$, and we continue this process until convergence is reached between $r_{500}$ and the temperature, and thus $T_{500}$ has been determined. This analysis is performed on the subset of our sample that is thermally dominated (Section~\ref{sec:speccomp}). The iterative process is begun with the $T_{\mathrm{ap}}$ determined from the {\sc apec} only fit as recorded in Table~\ref{table:bfp}.

For two of our objects, FGS25 and FGS26, the final estimation of $r_{500}$ extends beyond the largest aperture radius that can be inscribed within the XIS FOV. However, our estimated $r_{500}$ is very similar to the largest aperture size that was used to extract spectral parameters, where the ratio between the maximum $r_{\mathrm{ap}}$ and $r_{500}$ is 0.98 and 0.84 for FGS25 and FGS26, respectively. As a result the $T_{\mathrm{ap}}$ values for these two objects should reasonably describe the true global temperature within $r_{500}$. When considering the luminosity, $L_{\mathrm{X},500}$ is estimated from $L_{\mathrm{X,ap}}$ using a surface brightness profile model that well describes the ICM emission. By integrating this surface brightness model over area, the conversion factor between $L_{\mathrm{X},500}$ and $L_{\mathrm{X, ap}}$ is calculated using the relation
\begin{equation}\label{eq:convlum}
\frac{L_{\mathrm{X},500}}{L_{\mathrm{X,ap}}} = \frac{\int_{0}^{r_{500}} S(r) r \ dr}{\int_{0}^{r_{\mathrm{ap}}} S(r) r \ dr}
\end{equation}
where $S$ is an azimuthally averaged radial surface brightness profile. For our surface brightness model, we use the $\beta$-model where $S_0$, $r_{c}$, and $\beta$ have the values recorded in Table~\ref{table:sbp}.

With the global temperature values listed in Table~\ref{table:globalprop}, we estimate the masses within $r_{500}$ for our systems using the $M_{500}$--$T_{\mathrm{X}}$ relation of \citet{arnaud2005}:
\begin{equation}
M_{500}= 3.84 \times10^{14} \ h_{70}^{-1} \ E(z)^{-1} \left(\frac{kT}{5 \mathrm{keV}}\right)^{1.71} \ \mathrm{M}_{\odot}.
\end{equation}
We find our thermally dominated objects have masses consistent with clusters ($M_{500} \ga 10^{14} \ \mathrm{M}_{\odot}$).

\begin{table*}
\begin{center}
\caption{Global properties of the ICM-dominated subsample.}
\begin{threeparttable}
\begin{tabular}{c | c c c c | c c c}
\hline
FGS &$r_{\mathrm{ap}}/r_{500}$ &$kT_{\mathrm{ap}}$ &$Z_{\mathrm{ap}}$ &$L_{\mathrm{X,bol,ap}}$ &$r_{500}$ &$L_{\mathrm{X,bol,r500}}$ &$M_{500}$ \\
& 	 &[keV] &[Z$_{\odot}$] &[10$^{44}$ erg s$^{-1}$] & &[10$^{44}$ erg s$^{-1}$] &[10$^{14}$M$_{\odot}$] \\ [0.5ex]%
\hline
04$^{\phantom{*}}$ &1 	 &2.81$_{-0.19}^{+0.19}$ &0.40$_{-0.11}^{+0.12}$ &5.03$_{-0.19}^{+0.19}$ 	 &3.5' (0.71 Mpc) &5.03$_{-0.19}^{+0.19}$ &1.3$\pm$0.1 \\
14$^*$ &1 	 &5.26$_{-0.39}^{+0.44}$ &0.21$_{-0.08}^{+0.09}$ &7.71$_{-0.29}^{+0.29}$ 	 &4.8' (1.03 Mpc) &7.71$_{-0.29}^{+0.29}$ &3.8$\pm$0.5 \\
25$^{\phantom{*}}$ &0.98 	 &3.92$_{-0.15}^{+0.15}$ &0.28$_{-0.04}^{+0.04}$ &3.80$_{-0.09}^{+0.09}$ 	 &8.5' (0.92 Mpc) &3.84$_{-0.09}^{+0.09}$ &2.4$\pm$0.2 \\
26$^*$ &0.84 	 &3.33$_{-0.30}^{+0.34}$ &0.19$_{-0.08}^{+0.09}$ &0.70$_{-0.04}^{+0.04}$ 	 &10.3' (0.85 Mpc) &0.82$_{-0.05}^{+0.05}$ &1.9$\pm$0.3 \\
27$^*$ &1 	 &3.30$_{-0.31}^{+0.33}$ &0.18$_{-0.18}^{+0.13}$ &3.38$_{-0.16}^{+0.16}$	 &4.3' (0.80 Mpc) &3.38$_{-0.16}^{+0.16}$	 &1.7$\pm$0.3 \\
30$^*$ &1 	 &3.39$_{-0.11}^{+0.15}$ &0.30$_{-0.05}^{+0.05}$ &3.06$_{-0.06}^{+0.06}$ 	 &6.8' (0.84 Mpc) &3.06$_{-0.06}^{+0.06}$ &1.9$\pm$0.1 \\
\hline
\end{tabular}
\begin{tablenotes}
\scriptsize
\item [] Note: $L_{\mathrm{X,bol}}$ is the unabsorbed X-ray luminosity in the 0.1-100 keV energy range
\item [*] Confirmed fossil system
\end{tablenotes}
\label{table:globalprop}
\end{threeparttable}
\end{center}
\end{table*}


\section{Scaling Relations}\label{sec:scalrel}

We combine our newly measured global $L_{\mathrm{X,bol,500}}$ and $T_{\mathrm{X}}$ with previously measured fossil systems properties, to constrain the scaling relations of these objects with the goal of assessing if fossil systems display different scaling relations than those for normal groups and clusters. Our analysis of fossil system scaling relations is distinguished from previous studies through several updates including the fitting of the largest assembled fossil system data set, using recent X-ray and optical data for our control sample of normal groups and clusters, and a substantial effort of homogenizing both the fossil and non-fossil data sets. We furthermore record the best-fitting $L_{\mathrm{X}}$--$L_r$ relation, and for the first time record the slopes and $y$-intercepts of the $L_{\mathrm{X}}$--$T_{\mathrm{X}}$, $L_{\mathrm{X}}$--$\sigma_v$, $T_{\mathrm{X}}$--$\sigma_v$ scaling relation fits for fossil systems.


\subsection{Sample assembly, correction, and fitting}

We have assembled data from a number of studies to investigate how the global X-ray and optical properties of fossil systems compare to non-fossil groups and clusters. To ensure a reliable comparison, we have made an effort to use quantities determined within the same fiducial radius and defined the same way. For our analysis we use bolometric X-ray luminosities $L_{\mathrm{X,bol}}$, temperatures $T_{\mathrm{X}}$, and optical SDSS $r$-band luminosities $L_r$ all calculated within $r_{500}$, and global velocity dispersions $\sigma_v$. While we have removed known fossils from our sample of non-fossil groups and clusters, we do not have information on the magnitude gap between the first and second brightest galaxies in all of the systems making up our control sample. However, the large magnitude gap characterizing fossil systems should be found in only a fraction of $L_{\mathrm{X,bol}} \geq 5 \times 10^{41}$ erg s$^{-1}$ systems \citep{jones2003,milosavljevic2006}. Thus, we expect our control sample is contaminated by at most a few unidentified fossil systems.

To assemble our group sample, we use the $\sigma_v$ of the `G-sample' from \citet{op2004}. Group $T_{\mathrm{X}}$ values are pulled from \citet{rasmussen2007}, \citet{sun2009}, \citet{hudson2010}, \citet{eckmiller2011}, and \citet{lovisari2015}. \citet{lovisari2015} $L_{\mathrm{X},0.1-2.4\mathrm{keV}}$ are transformed to $L_{\mathrm{X,bol}}$ using the conversion tables of \citet{boehringer2004}.

For the cluster sample, we use the \citetalias{girardi2014} $r$-band optical luminosities calculated within $r_{500}$. The corresponding velocity dispersions are taken from \citet{girardi1998,girardi2002}, \citet{girardi2001}, \citet{popesso2007b}, and \citet{zhang2011}. Bolometric X-ray luminosities within $r_{500}$ and temperatures were sourced from \citet{pratt2009} and \citet{maughan2012}, and supplemented with additional $L_{\mathrm{X,bol}}$ from \citet{zhang2011} and $T_{\mathrm{X}}$ from \citet{wu1999} and \citet{hudson2010}.

Taking our sample of fossil systems observed with {\it Suzaku}, we match the global X-ray properties of the systems in Table~\ref{table:globalprop} with the corresponding $L_{r}$ from \citetalias{girardi2014} and $\sigma_v$ from \citetalias{zarattini2014}. For the remainder of the \citetalias{zarattini2014} confirmed fossil catalogue, we supplement the $L_{\mathrm{X,bol}}$ from \citetalias{girardi2014}. For improved consistency with the $L_{\mathrm{X}}$ of our cluster sample, we approximate X-ray luminosities that more closely resemble those computed using the growth curve analysis (GCA) method \citep{boehringer2000} from the \citetalias{girardi2014} luminosities derived from RASS counts (see section 3.3 of \citetalias{girardi2014} for details). These corrected luminosities also show good agreement with the {\it Suzaku} measured $L_{\mathrm{X}}$ for the sample of objects shared between both the \citetalias{girardi2014} study and ours.

We add to the fossil sample with the X-ray luminosities and temperatures from \citetalias{khosroshahi2007} and \citet{miller2012}, matched with the $L_r$ and $\sigma_v$ data from \citet{proctor2011}. The \citetalias{khosroshahi2007} $L_{\mathrm{X,bol,200}}$ are rescaled to $r_{500}$ using their best-fitting $\beta$-model parameters and our luminosity conversion Eq.~\ref{eq:convlum}. To ensure consistency with our {\it Suzaku} sample, the $r_{500}$ of \citetalias{khosroshahi2007} is recalculated from their temperatures using our Eq.~\ref{eq:arnaudrt} and we use this value to estimate $L_{\mathrm{X,bol,500}}$. To rescale the $L_{r,200}$ of \citet{proctor2011} to $r_{500}$, we assume the light follows the mass, which is a good approximation on the global scale of groups and clusters \citep{bahcall2014}. For a NFW density profile with concentration parameter $c=5$, $M_{500}/M_{200}=0.70 $ \citep{nfw1997}. The assumption of $c=5$ was chosen for agreement with the concentrations of normal clusters of similar temperature and mass \citep{pointecouteau2005,pratt2005,vikhlinin2006,buote2007,ettori2010} because the typical concentration parameter for fossil systems is poorly characterized. Thus, we can rescale using $L_{\mathrm{opt},500}/L_{\mathrm{opt},200} \propto 0.70$. Here, the correction relation is applied only to the non-BCG light.

We also implement the fossil catalogue of \citet{harrison2012}. We take their $L_{\mathrm{X,bol,200}}$ and rescale by assuming a $\beta$-model with $r_{\mathrm{c}}$ estimated using the $r_{\mathrm{c}}$--$L_{\mathrm{X}}$ relation of \citet{boehringer2000} and $\beta$=0.67, then correcting to $L_{\mathrm{X,bol,500}}$ using Eq.~\ref{eq:convlum}. The optical luminosities provided are calculated for $r=0.5r_{200} \sim r_{1000}$. By the reasoning described previously, this luminosity is corrected to $L_{r,500}$ using the relation $M_{500}/M_{1000} \propto 1.3$. Because the magnitudes of the BCG were not recorded, we rescale all of the optical light for these objects. The \citet{harrison2012} $\sigma_v$ are also used, and we assign a 0.1 dex error to these values during our fit of the fossil scaling relations.

\begin{figure*}
\centering
\includegraphics[width=\linewidth]{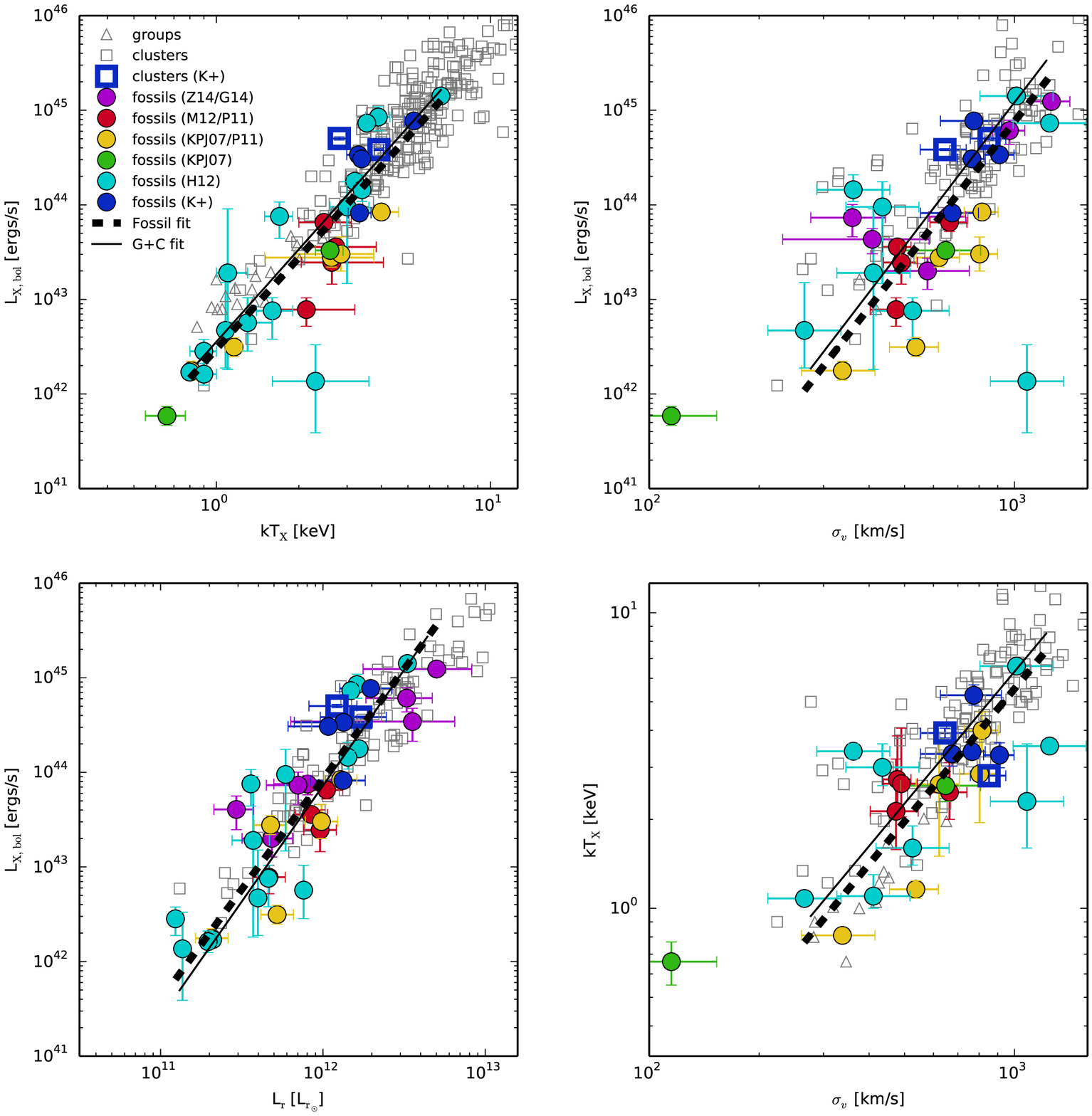}
\caption{$L_{\mathrm{X}}$, $T_{\mathrm{X}}$, $L_r$, $\sigma_v$ scaling relations for fossil and non-fossil samples. We abbreviate this current work as K+, \citet{zarattini2014} as Z14, \citet{girardi2014} as G14, \citet{miller2012} as M12, \citet{proctor2011} as P11, \citet{khosroshahi2007} as KPJ07, and \citet{harrison2012} as H12. The plotted lines are the orthogonal BCES fits to the fossil sample (dashed line) and to the sample of groups and clusters (solid line) in the same parameter range as the fossils.}
\label{fig:scalingrelations}
\end{figure*}

With the above data sets, we have enough data to assemble and quantitatively compare the $L_{\mathrm{X}}$--$T_{\mathrm{X}}$, $L_{\mathrm{X}}$--$\sigma_v$, $L_{\mathrm{X}}$--$L_r$, $T_{\mathrm{X}}$--$\sigma_v$ scaling relations for a sample of fossils and a control sample of normal groups and clusters. We do not investigate the $T_{\mathrm{X}}$--$L_r$ relation due to the small subsample of our control population with both $T_{\mathrm{X}}$ and $L_r$ measurements.

We fit the equation
\begin{equation}\label{eq:sr_fit}
\log(Y)=a+b\log(X)
\end{equation}
to the data using the BCES orthogonal method \citep{akritas1996} which accounts for measurement errors in the data as well as intrinsic scatter in the fitted relation. We choose to compare the fit of the fossil sample to a combined sample of groups and clusters (G+C) in the same parameter range as the fossil sample. For the fossil system data set we exclude NGC 6482 from \citetalias{khosroshahi2007} and XMMXCS J030659.8+000824.9 from \citet{harrison2012} as they are clear outliers.

We cross-checked the results obtained with the BCES method with the IDL Astronomy library tool {\sc LINMIX\_ERR} \citep{kelly2007}, a Bayesian fitting method for linear regression. The plotted scaling relations and BCES fits are shown in Fig.~\ref{fig:scalingrelations} and the best-fitting parameters of the relations are recorded in Table~\ref{table:sr_bfp}. Uncertainties on the BCES best-fitting parameters are estimated using 10,000 bootstrap resamplings. For the {\sc LINMIX\_ERR} fits, the quoted values are the mean and the standard deviation of the posterior distributions for the regression parameters. We investigate changing the pivot point of the fits, i.e. rescaling the $X$ and $Y$ values in Eq.~\ref{eq:sr_fit} by a constant, but no difference is found in the returned fits.

We find the BCES fits to the fossil sample are consistent within error to the combined groups and clusters fit for each scaling relation investigated in this work. The {\sc LINMIX} fossil and non-fossil fits are for the most part consistent within 1$\sigma$; the $y$-intercepts of $L_{\mathrm{X,bol}}$--$T_{\mathrm{X}}$ and the $y$-intercepts and slopes of $L_{\mathrm{X}}$--$\sigma_v$ are consistent within 2$\sigma$. These slight discrepancies in the {\sc LINMIX} fits are most likely due to inhomogeneities in the data or the small sample size of both the fossil and control populations.

The global properties involved in these scaling relations: $L_{\mathrm{X}}$, $T_{\mathrm{X}}$, $L_{\mathrm{opt}}$, $\sigma_v$, are determined predominantly by the shape and depth of the potential well, and are thus well-documented proxies for the total mass of the system. Additional important effects that determine the X-ray properties of the ICM include the entropy structure \citep{donahue2006} and non-gravitational heating and cooling processes, such as can be caused by AGN or mergers. These factors can produce dispersions in scaling relations between X-ray and optical mass proxies. Finding no difference in the scaling relations between fossil and non-fossil groups and clusters thus indicates fossil systems are of similar mass as non-fossils, and on the global scale, the combined effect of mass, ICM entropy, and non-gravitational processes that have occurred in fossil systems are similar to the combined effect of those that have occurred in normal groups and clusters.

\begin{table*}
\begin{center}
\caption{Best fits to the scaling relations.}
\begin{threeparttable}
\begin{tabular}{c c | c c | c c}
\hline
Relation ($Y$-$X$) & Sample & \multicolumn{4}{c}{Fitting Procedure}\\
\cline{3-6}
& & \multicolumn{2}{c}{BCES Orthogonal} & \multicolumn{2}{c}{{\sc LINMIX\_ERR}} \\
& & $a$ & $b$ & $a$ & $b$ \\ [0.5ex]
\hline
$L_{\mathrm{X,bol}}$--$T_{\mathrm{X}}$ & Fossils & 42.48$\pm$0.17 & 3.21$\pm$0.44 & 42.49$\pm$0.13 & 3.39$\pm$0.29 \\
& G+C & 42.55$\pm$0.09 & 3.25$\pm$0.14 & 42.74$\pm$0.07 & 3.03$\pm$0.11 \\
\hline
$L_{\mathrm{X,bol}}$--$\sigma_v$ & Fossils & 30.05$\pm$3.60 & 4.94$\pm$1.29 & 28.34$\pm$3.22 & 5.51$\pm$1.14 \\
& G+C & 29.95$\pm$1.40 & 5.05$\pm$0.49 & 33.30$\pm$0.96 & 3.87$\pm$0.33 \\
\hline
$L_{\mathrm{X,bol}}$--$L_r$ & Fossils & 15.98$\pm$3.18 & 2.33$\pm$0.27 & 17.18$\pm$2.84 & 2.23$\pm$0.24 \\
& G+C & 14.47$\pm$2.03 & 2.45$\pm$0.17 & 17.05$\pm$2.73 & 2.24$\pm$0.22 \\
\hline
$T_{\mathrm{X}}$--$\sigma_v$ & Fossils & -3.73$\pm$2.44 & 1.49$\pm$0.89 & -4.59$\pm$1.67 & 1.79$\pm$0.59 \\
& G+C & -3.65$\pm$0.44 & 1.48$\pm$0.15 & -3.92$\pm$0.27 & 1.58$\pm$0.09 \\
\hline
\end{tabular}
\label{table:sr_bfp}
\end{threeparttable}
\end{center}
\end{table*}


\subsection{Comparison with previous studies}

Our result that fossils share the same $L_{\mathrm{X}}$--$T_{\mathrm{X}}$ relation as non-fossil groups and clusters is consistent with previous studies \citep[\citetalias{khosroshahi2007};][\citetalias{girardi2014}]{proctor2011,harrison2012}. However, the comparison of optical and X-ray properties of fossil and non-fossil systems is a contentious issue in the literature.

The $L_{\mathrm{X}}$--$L_r$, $L_{\mathrm{X}}$--$\sigma_v$, $T_{\mathrm{X}}$--$\sigma_v$ scaling relation fits of our analysis show the relations of fossil systems are consistent within error to normal groups and clusters. This is in good agreement with the findings of \citet{harrison2012} and \citetalias{girardi2014}. \citetalias{girardi2014} recorded the first quantitative values of their fit to the $L_{\mathrm{X}}$--$L_r$ relation and found no difference between fossil systems ($L_{\mathrm{X}}\propto L_{r}^{1.8\pm0.3}$) and a sample of non-fossil clusters ($L_{\mathrm{X}}\propto L_{r}^{1.78\pm0.08}$). While qualitatively we both find no difference in the $L_{\mathrm{X}}$--$L_r$ fossil and non-fossil scaling relations, there are some numerical differences in the returned best-fitting parameters of our study and \citetalias{girardi2014}.

Our fossil fit of $L_{\mathrm{X}}\propto L_{r}^{2.33\pm0.27}$ is consistent within error to \citetalias{girardi2014}, although this is in large part due to the considerable error on both of our slopes. However, our non-fossil fit ($L_{\mathrm{X}}\propto L_{r}^{2.45\pm0.17}$) is not within error of the fit determined by \citetalias{girardi2014}. Differences in the slopes of our fits could be due to multiple reasons: (1) we use bolometric $L_{\mathrm{X}}$ in our fits, while \citetalias{girardi2014} uses $L_{\mathrm{X},\mathrm{0.1-2.4keV}}$; (2) our $L_{\mathrm{X}}$ are defined within $r_{500}$ while the fitted \citetalias{girardi2014} $L_{\mathrm{X}}$ represent a total luminosity that has not been defined within a precise radius; (3) we use different fitting methods; (4) we fit our control sample of non-fossils over a different parameter space (i.e., one defined to match our fossil sample).

We check to see if we can return more consistent results with \citetalias{girardi2014} by repeating our analysis of the $L_{\mathrm{X}}$--$L_r$ relation using $L_{\mathrm{X},\mathrm{0.1-2.4keV}}$ instead of $L_{\mathrm{X},\mathrm{bol}}$ and expanding the fit of our control `G+C' sample to the full parameter space. We find the returned fit to the fossil sample ($L_{\mathrm{X}}\propto L_{r}^{2.11\pm0.26}$) and to the non-fossil sample ($L_{\mathrm{X}}\propto L_{r}^{1.86\pm0.10}$) are both within error of the \citetalias{girardi2014} fits. And again we emphasize that even without the changes made here, although numerically our fits differ from those of \citetalias{girardi2014}, the interpretation is the same: fossil systems follow the same $L_{\mathrm{X}}$--$L_r$ scaling as non-fossil systems, supporting our conclusion that on the global scale, fossil systems have optical and X-ray properties congruent with those of normal groups and clusters.

Accumulation of multiple differences in data and methodology explain the differences in conclusions between our study and those of earlier studies \citep[\citetalias{khosroshahi2007};][]{proctor2011} that find discrepancies in the optical and X-ray scaling relations for fossil and non-fossils. We have compared fossil and non-fossil optical luminosities measured from the same photometric catalogue and band, avoiding the need to make approximative luminosity estimates for comparisons between samples. We have also used optical luminosities defined within the same fiducial radius, thus ensuring a more equal comparison between data pulled from multiple catalogues. Additionally, our large sample size of fossils reduces the effect of noise to ensure a more reliable comparison between the fossil and non-fossil samples.

We note, however, that our best-fitting parameters for both the fossil and non-fossil samples have large errors. Thus, a study of fossil scaling relations could be greatly improved in the future by larger and more homogeneous data sets. Furthermore, our results probe the relations of clusters and high-mass groups, and consequently it is possible differences in the scaling relations exist in the low-mass end \citep{desjardins2014,khosroshahi2014}.


\section{Summary and Conclusions}

We have presented a detailed study of the X-ray properties of 10 candidate fossil galaxy systems using the first pointed X-ray observations of these objects. In particular, {\it Suzaku} XIS data have been used to measure their global X-ray temperatures and luminosities and to estimate the masses of these galaxy clusters. We determine 6 of our 10 objects are dominated in the X-ray by thermal bremsstrahlung emission and thus we are able to measure the global temperatures and luminosities of their ICM. This sample of six objects has temperatures of $2.8 \leq T_{\mathrm{X}} \leq 5.3 \ \mathrm{keV}$, luminosities of $0.8 \times 10^{44} \leq L_{\mathrm{X,bol}} \leq 7.7\times 10^{44} \ \mathrm{erg} \ \mathrm{s}^{-1}$, and occupies the cluster regime in plotted scaling relations.

Using our newly determined fossil cluster ICM X-ray properties, we combine our fossil sample with fossils in the literature to construct the largest fossil sample yet assembled. This sample is compared with a literature sample of normal groups and clusters, where significant effort has been made to homogenize the global $L_{\mathrm{X}}$, $T_{\mathrm{X}}$, $L_r$, and $\sigma_v$ data for the fossil and non-fossil samples.

Plotting the $L_{\mathrm{X}}$--$T_{\mathrm{X}}$, $L_{\mathrm{X}}$--$\sigma_v$, $L_{\mathrm{X}}$--$L_r$, and $T_{\mathrm{X}}$--$\sigma_v$ relations shows no difference between the properties of fossils and normal groups and clusters. Furthermore, we provide the first fits to three of these relations which reveals the relations of fossils systems agree within error to the relations of normal groups and clusters. Our work indicates that on the global scale, fossil systems are no different than non-fossil systems. However, the distinguishing large magnitude gap in the bright end of the fossil system luminosity function is still unexplained and thus further studies are necessary to characterize the properties of these objects and understand their nature.


\section*{Acknowledgements}

This research has made use of data obtained from the {\it Suzaku} satellite, a collaborative mission between the space agencies of Japan (JAXA) and the USA (NASA). We thank the anonymous referee for valuable comments, K. Hamaguchi and K. Pottschmidt at the {\it Suzaku} Helpdesk for useful advice on multiple aspects of our analysis, and D. Eckert for helpful discussions and for suggesting the flickering pixels issue as an explanation for the excess in the 0.5--0.7 keV range.

Support for this research was provided by NASA Grant No. NNX13AE97G, and by the University of Wisconsin-Madison Office of the Vice Chancellor for Research and Graduate Education with funding from the Wisconsin Alumni Research Foundation. FG acknowledges the financial contribution from contract PRIN INAF 2012 (`A unique dataset to address the most compelling open questions about X-ray galaxy clusters') and the contract ASI INAF NuSTAR I/037/12/0. ED gratefully acknowledges the support of the Alfred P. Sloan Foundation. MG acknowledges funding from MIUR PRIN2010-2011 (J91J12000450001). JALA has been partly funded from MINECO AYA2013-43188-P. EMC is partially supported by Padua University through grants 60A02-4807/12, 60A02-5857/13, 60A02-5833/14, and CPDA133894. JMA acknowledges support from the European Research Council Starting Grant (SEDmorph; PI V. Wild).


{\footnotesize
\bibliographystyle{mn2e}

}


\appendix

\section{Testing FGS24 for SWCX contamination}\label{appendix:swcxappendix}

\begin{figure}
\centering
\includegraphics[width=\linewidth]{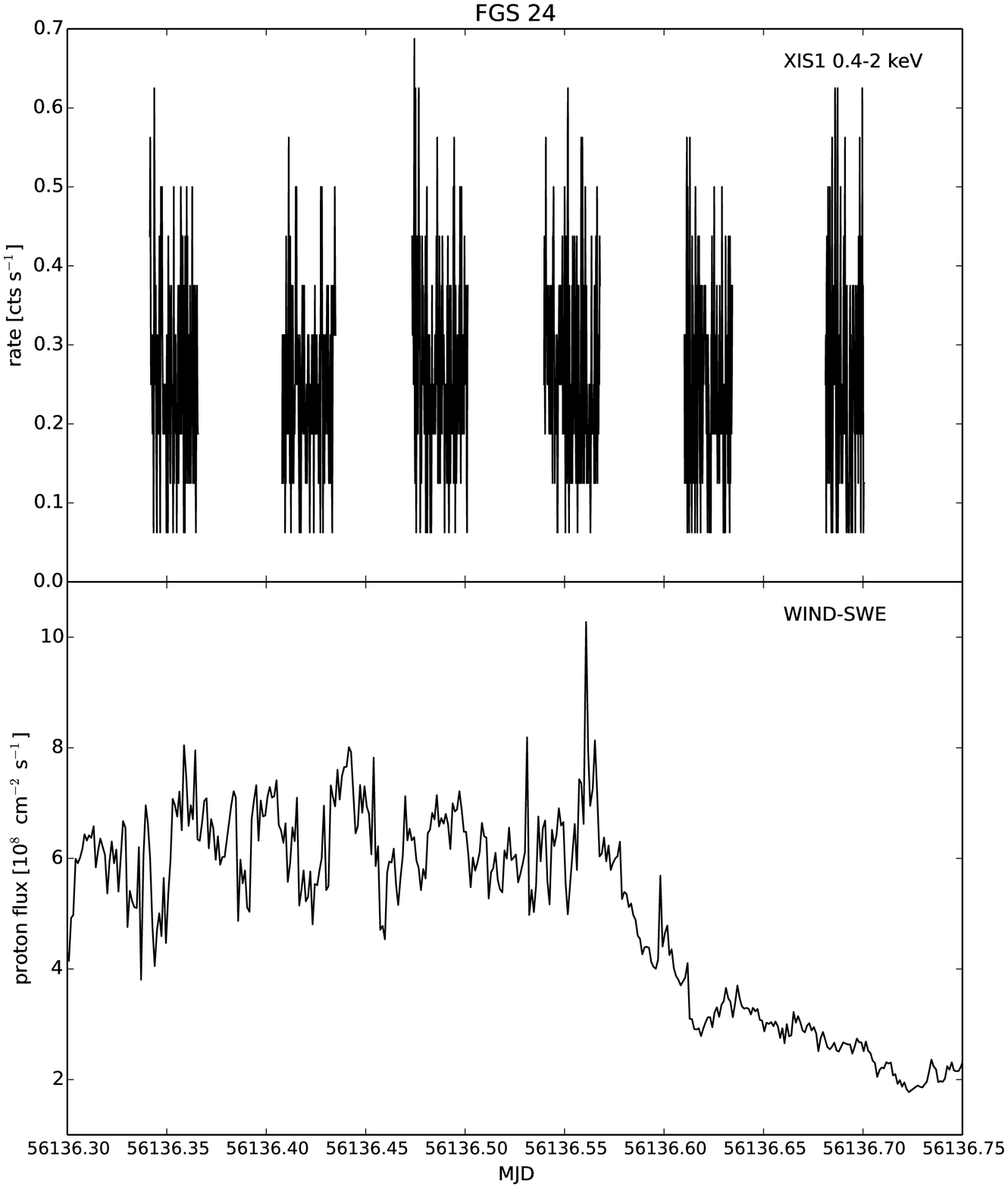}
\caption{Top: the observed XIS1 light curve for FGS24. Bottom: the WIND-SWE proton flux light curve plotted for the same time span. Proton flux has been found to be correlated to SWCX. The elevated proton flux levels during the FGS24 observation may potentially cause significant SWCX contaminating emission.}
\label{fig:807058010lc}
\end{figure}

The NASA WIND-SWE proton flux light curve displays elevated flux levels greater than $4\times10^8 \ \mathrm{cm}^{-2} \ \mathrm{s}^{-1}$ during a significant portion of the FGS24 observation (Fig.~\ref{fig:807058010lc}) which indicates SWCX photons may contaminate the lower $E<1$ keV region of the spectrum (see Section~\ref{sec:swcx}). To test for evidence of this contamination, we repeat the spectral analysis of Section~\ref{sec:optreg} for the time intervals where the flux was less than $4\times10^8 \ \mathrm{cm}^{-2} \ \mathrm{s}^{-1}$. These results are recorded in Table~\ref{table:alt58} and we find these results are consistent within error with those of using the full timespan of the observation (Table~\ref{table:bfp}).

\begin{table}
\begin{turn}{90}
\begin{minipage}[c][70mm]{235mm}
\caption{Best fit spectral parameters during a low proton flux time interval for FGS24.}
\begin{threeparttable}
\begin{tabular}{c c | c c c c | c c c | c c c c c c }
\hline
FGS & r$_{\mathrm{ap,src}}$ & \multicolumn{4}{c}{{\sc apec}} & \multicolumn{3}{c}{{\sc powerlaw}} & \multicolumn{6}{c}{{\sc apec}+{\sc powerlaw}} \\ 
& & $kT_{\mathrm{apec}}$ & $Z_{\mathrm{apec}}$ & norm$_{\mathrm{apec}}^{a}$ & $\chi^2$/d.o.f ($\chi^{2}_{r})$ & $\Gamma_{\mathrm{PL}}$ & norm$_{\mathrm{PL}}^{b}$ & $\chi^2$/d.o.f ($\chi^{2}_{r}$) & $kT_{\mathrm{apec}}$ & $Z_{\mathrm{apec}}$ & norm$_{\mathrm{apec}}^{a}$ & $\Gamma_{\mathrm{PL}}$ & norm$_{\mathrm{PL}}^{b}$ & $\chi^2$/d.o.f ($\chi^{2}_{r}$) \\
& & $[$keV$]$ & $[$Z$_{\odot}]$ & $[10^{-3}]$ & & & [$10^{-4}$] & & $[$keV$]$ & $[$Z$_{\odot}]$ & $[10^{-3}]$ & & [$10^{-4}$] & \\ [0.5ex]
\hline
24 & 2.7' & 4.96$_{-0.91}^{+1.59}$ & 0.5$_{-0.38}^{+0.48}$ & 4.4$_{-0.6}^{+0.6}$ & 68/35 (1.95) & 1.87$_{-0.13}^{+0.14}$ & 8.7$_{-1.0}^{+1.0}$ & 67/36 (1.86) & 0.19$_{-0.18}^{+6.67}$ & 0.3 & 9.6$_{-11.9}^{+1102.2}$ & 1.8 & 8.2$_{-5.0}^{+0.7}$ & 66/35 (1.9) \\
\hline
\end{tabular}
\begin{tablenotes}
\item [$a$] norm$_{\mathrm{apec}}=\frac{10^{-14}}{4 \pi [D_A (1+z)]^2} \int n_e n_H dV \mathrm{cm}^{-5} $
\item [$b$] norm$_{\mathrm{PL}}$ has units of photons keV$^{-1}$ cm$^{-2}$ s$^{-1}$ arcmin$^{-2}$ at 1 keV
\end{tablenotes}
\label{table:alt58}
\end{threeparttable}
\end{minipage}
\end{turn}
\end{table}


\section{Characterizing the Suzaku XRT PSF}\label{appendix:psffwhm}

We determine a radial model for the {\it Suzaku} XRT PSF to complete our image analysis in Section~\ref{sec:sbp}. Our PSF characterization employs archival observations of the X-ray point source SS Cyg observed for an effective 52 ks between 2005 November 18 and 19 ({\it Suzaku} sequence number 400007010). We clean the SS Cyg event files following the same procedure applied to our {\it Suzaku} observations (see Section~\ref{sec:observations}).

\begin{figure}
\centering
\includegraphics[width=.9\linewidth]{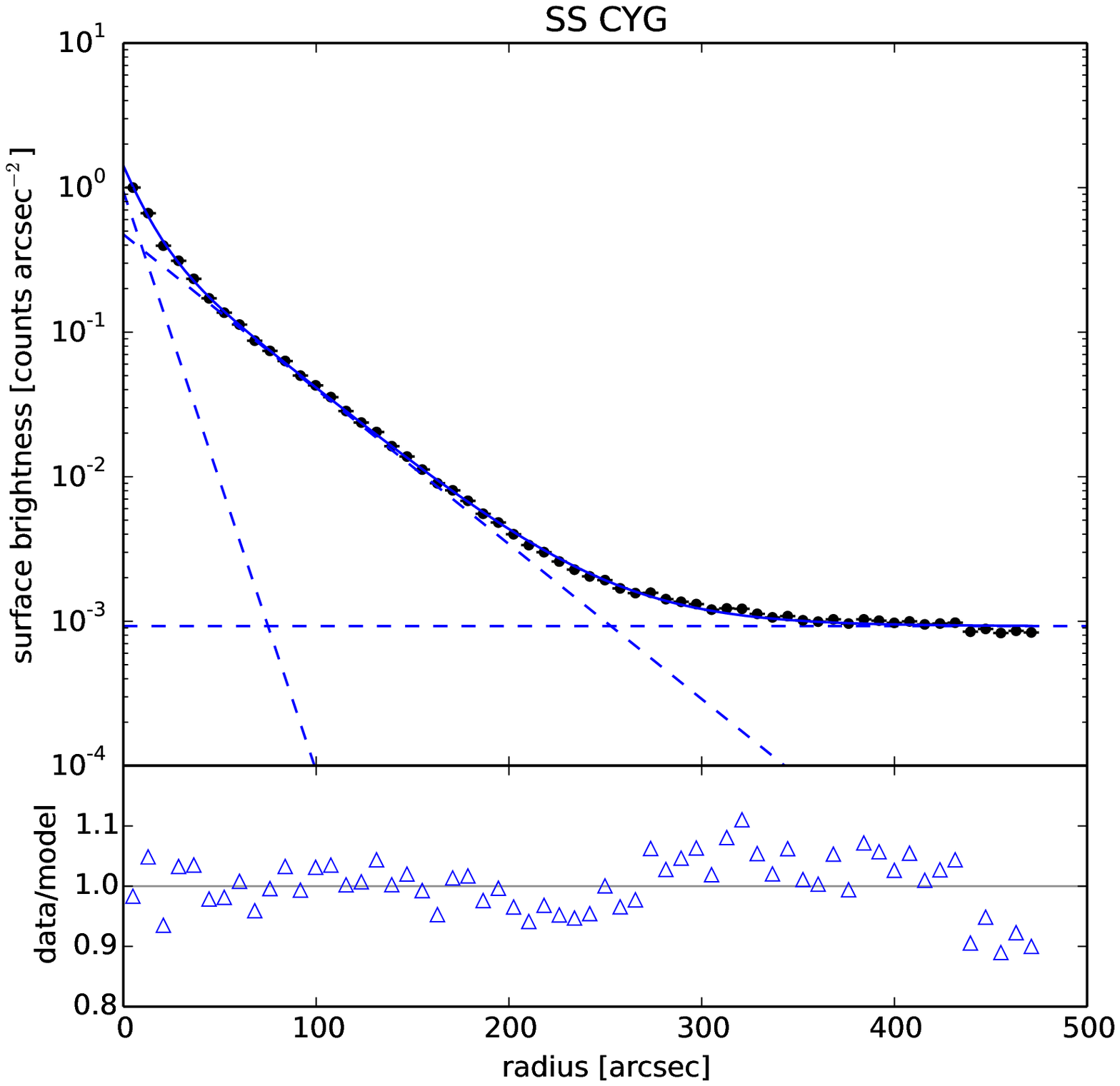}
\caption{Stacked and normalized XIS0+XIS1+XIS3 radial brightness profile for point-source SS Cyg in the 0.5--10 keV band. The best-fitting model, consisting of the sum of two exponentials and a background constant, is plotted in solid blue. Components of the model are plotted with dashed lines, and residuals are plotted as triangles. Best-fitting parameters for the model are recorded in Table~\ref{table:psfmodel}.}
\label{fig:psffit}
\end{figure}

The PSF is characterized using the radial profile of the stacked XIS0+XIS1+XIS3 image of SS Cyg that has been extracted in the 0.5--10 keV energy range and normalized to 1 (Fig.~\ref{fig:psffit}). The average PSF full width at half-maximum (FWHM) is found to be $\sim 35$ arcsec. Our PSF model consists of the sum of two exponentials, as recommended by \citet{sugizaki2009}, and thus the model fit to the SS Cyg brightness profile is:
\begin{equation}
S(r) = A_{1} \mathrm{e}^{c_{1}(r-r_{0,1})} + A_{2} \mathrm{e}^{c_{2}(r-r_{0,2})} + k,
\end{equation}
where the constant $k$ accounts for the background. The best-fitting parameters for this model are recorded in Table ~\ref{table:psfmodel}.

\begin{table}
\begin{center}
\caption{Best-fitting model to the radial brightness profile of SS Cyg.}
\begin{threeparttable}
\begin{tabular}{c c c c}
\hline
Component & Parameter & Value & Units\\ [0.5ex]
\hline
exp1 & $A_1$ & 0.46$_{-0.45}^{+0.16}$ & {\scriptsize counts arcsec$^{-2}$} \\
& $c_1$ & -2.5$_{-0.01}^{+0.01}$ & $10^{-2}$ {\scriptsize arcsec$^{-1}$} \\
& $r_{0,1}$ & 0.9$_{-61.6}^{+115.4}$ & {\scriptsize arcsec} \\
\hline
exp2 & $A_2$ & 0.54$_{-0.54}^{+2.57}$ & {\scriptsize counts arcsec$^{-2}$} \\
& $c_2$ & -9.2$_{-0.1}^{+0.1}$ & $10^{-2}$ {\scriptsize arcsec$^{-1}$} \\
& $r_{0,2}$ & 6.0$_{-174.7}^{+38.9}$ & {\scriptsize arcsec} \\
\hline
background & $k$ & 9.2$_{-0.1}^{+0.1}$ & $10^{-4}$ {\scriptsize counts arcsec$^{-2}$} \\
\hline
$\chi^2/\mathrm{d.o.f} (\chi^2_r)$ & & 405/53 (7.6) \\
\end{tabular}
\label{table:psfmodel}
\end{threeparttable}
\end{center}
\end{table}


\section{Notes on the Sample}\label{appendix:notes}

\textit{FGS03} is a \citetalias{zarattini2014} verified fossil system. The AGN (2MASX J07524421+4556576) associated with the BCG of this system is both confirmed in the optical \citep{vv2010} and radio. The radio emission from this object consists of strong bipolar jets extending 57 arcsec \citep{hess2012}. This AGN has also been identified as a Type I Seyfert \citep{stern2012}, and appears to dominate the X-ray emission observed from FGS03. The spectrum of this object is better fit by a power-law ($\chi_r^2=1.02$) than a thermal model ($\chi_r^2=1.17$), and no improvement in the fit occurs when a thermal component is added to the power-law model. Furthermore, our imaging analysis finds a $\beta$-model poorly describes the observed surface brightness profile. \citetalias{zarattini2014} find a velocity dispersion of $\sigma_v=259$ km s$^{-1}$, the smallest dispersion of the \citetalias{santos2007} catalogue. Such a low velocity dispersion is typically associated with a cool ICM temperature, which would explain why there appears to be very little thermal emission when compared to a very bright AGN.

\textit{FGS04} is a fossil candidate and has the coolest measured ICM of our sample ($T_{\mathrm{X}}$ = 2.81 keV). The BCG of this system contains the blazar NVSS J080730+340042 \citep{mgl2009} and in the radio, \citet{hess2012} find bipolar jets originating from this source. We do not see evidence of contribution from this object in the spectral analysis - the spectrum of FGS04 is fit significantly better by a thermal model than a power-law (compare a $\chi_r^2$ of 1.14 to 1.43).

\textit{FGS09} is a fossil candidate system at $z=0.125$. A background $z=0.73$ AGN (QSO B1040+0110; RA=10:43:03.84, Dec.=+00:54:20.42) is located 15 arcsec from the peak X-ray coordinates of FGS09. This AGN is confirmed in the optical \citep{vv2010} and the radio \citep{hess2012} bands. Based on our surface brightness profile and spectral analyses, this AGN is significantly contributing to the observed projected X-ray emission of FGS09. A large reduced chi-squared of $\chi_r^2$=5.7 is found for the $\beta$-model fit to the radial brightness profile. And, a power-law model ($\chi_r^2=0.92$) fits the spectrum of FGS09 much better than the thermal model ($\chi_r^2=1.08$).

\textit{FGS14} is a confirmed fossil system and is the largest, hottest, and most X-ray luminous cluster in our sample, with $r_{500}$ = 1 Mpc, $T_{\mathrm{X}}$ = 5.3 keV, and $L_{\mathrm{X}}$ = $7.7\times10^{44}$ erg s$^{-1}$. \citet{hess2012} detected radio-loud emission from two central sources; however, we did not see evidence of X-ray bright non-thermal emission in our spectral tests.

\textit{FGS15} is a rejected fossil candidate \citepalias{zarattini2014}. There are a number of contaminating sources in the XIS FOV of this source. A radio-loud AGN with an asymmetric jet is associated with the BCG of this system \citep{hess2012}. Within 40 arcsec of the peak system X-ray, the background ($z=0.45$) quasar [VV2010] J114803.2+565411 has been identified optically and in the radio \citep{vv2010,hess2012}. Of the two visually distinguishable point sources excluded in our analysis, the object closest to the centre of the system is spatially consistent with the QSO [VV2010] J114755.9+564948 at $z=4.32$ \citep{vv2010}. The further south removed point source is located at (RA=11:48:08.38, Dec.=+56:48:18.64). The closest known spatial match to this object is the radio source NVSS J114838+565327 located $\sim$2 arcmin away. Our surface brightness profile analysis reveals that a $\beta$-model ($\chi_r^2$=5.2) poorly fits the observed emission, and additionally the best-fitting spectral model of FGS15 is a power-law. For this object, it is possible multiple AGN are contributing to the observed emission; however, as noted by \citetalias{zarattini2014}, FGS15 could also be a filament due to its small number of constituent galaxies with large differences in velocity.

\textit{FGS24} is a rejected fossil candidate. No associated AGN were identified in the literature. However, the spectrum of FGS24 is better fit by a power-law than a thermal model (compare a $\chi_r^2$ of 1.33 to 1.38). FGS24 was observed during a period of potentially strong SWCX emission. While we found the best-fitting spectral parameters of the full observation match those of the isolated time interval of low proton flux, it is possible SWCX contamination is occurring even during this interval, obscuring the emission from FGS24.

\textit{FGS25} is a non-fossil galaxy cluster \citepalias{zarattini2014}. It is the second hottest cluster in our sample with $T_{\mathrm{X}}$ = 3.92 keV and a corresponding estimated mass of $M_{500} = 2.4\times10^{14} \ \mathrm{M}_{\odot}$. \citet{hess2012} find a radio-loud central point source in this cluster; however, our spectral analysis indicates no point source contribution as the FGS25 spectrum is much better described by a thermal model ($\chi_r^2=0.96$) than a power-law model ($\chi_r^2=1.26$).

\textit{FGS26} is a \citetalias{zarattini2014} confirmed fossil with $T_{\mathrm{X}}$ = 3.3 keV and $L_{\mathrm{X}}$ = $0.8\times10^{44}$ erg s$^{-1}$. We find no associated significant non-thermal signatures in the spectrum.

\textit{FGS27} is a confirmed fossil with measured global properties of $T_{\mathrm{X}}$ = 3.3 keV and $L_{\mathrm{X}}$ = $3.4\times10^{44}$ erg s$^{-1}$. Our spectral analysis does not indicate contribution of significant non-thermal emission.

\textit{FGS30} is a confirmed fossil with measured global properties of $T_{\mathrm{X}}$ = 3.4 keV and $L_{\mathrm{X}}$ = $3.06\times10^{44}$ erg s$^{-1}$. A radio-loud AGN (2MASX J17181198+5639563) is associated with its bright central galaxy \citep{hess2012}. The spectrum of FGS30 is better described by the thermal model ($\chi_r^2=1.05$) in comparison to the power-law model ($\chi_r^2=1.41$).


\label{lastpage}

\end{document}